\documentstyle[12pt,aaspp4,epsf]{article}
\def\fun#1#2{\lower3.6pt\vbox{\baselineskip0pt\lineskip.9pt
  \ialign{$\mathsurround=0pt#1\hfil##\hfil$\crcr#2\crcr\sim\crcr}}}
\def\lap{\mathrel{\mathpalette\fun <}}
\def\gap{\mathrel{\mathpalette\fun >}}

\begin{document}

\title{ELLIPTICAL GALAXY DYNAMICS}
\author{David Merritt}
\affil{Department of Physics and Astronomy}
\affil{Rutgers University}
\authoraddr{Serin Physics Laboratories, Piscataway, NJ 08855}

\vskip 1.truein

\clearpage

\tableofcontents

\clearpage

\section{INTRODUCTION}

Hubble (1936) divided the ``regular nebulae'' into two classes, 
the spirals and the ellipticals, defining the latter
as ``highly concentrated and [showing] no indications of 
resolution into stars.''
He emphasized the featureless appearance of most ``elliptical 
nebulae'' and noted that only two of their characteristics were
useful for further classification:
the shapes of their isophotal contours; and their luminosity gradients.
The latter were difficult to measure quantitatively at the time, 
and Hubble based his classification scheme entirely on
the ellipticity $(a-b)/a$, with $a$ and $b$ the major and minor 
axis lengths.
Hubble understood that the observed ellipticity was only a lower 
limit to the true elongation due to the unknown orientation 
of a galaxy's ``polar axis;'' he called this uncertainty 
``serious, but unavoidable.''
Nevertheless he was able to compute an estimate of the 
frequency function of intrinsic shapes by assuming that 
elliptical galaxies were oblate spheroids with random 
orientations (\cite{hub26}).

Hubble's remarks remain nearly as valid today as they were six 
decades ago.
Classification of elliptical galaxies is still based almost
entirely on their luminosity distributions; and although Hubble's 
analysis of the intrinsic shape distribution has been 
considerably refined, we still know little of a definite nature about 
the three-dimensional shapes of these systems.
An important shift in our understanding of elliptical galaxies  
took place in 1975,
following the discovery that most ellipticals rotate 
significantly more slowly than expected for a fluid body with the 
same flattening (\cite{bec75}).
Elliptical galaxies were revealed to be ``hot'' stellar systems,
in which most of the support against gravitational collapse comes 
from essentially random motions rather than from ordered 
rotation.
Two questions immediately arose from these observations: what 
produces the observed flattenings; and, given that rotation plays 
only a minor role, are elliptical galaxies axisymmetric or fully 
triaxial?
Binney (1978) suggested that the flattenings were due in large 
part to anisotropic velocity distributions and noted that
triaxial figures were no less likely than axisymmetric ones.
His suggestion was quickly followed up by Schwarzschild (1979, 1982) who 
showed that self-consistent triaxial models could be constructed by 
superposition of time-averaged orbits.
The phenomenon of triaxiality has since remained central 
to our understanding of elliptical galaxy dynamics.

Following Schwarzschild's pioneering work, 
a common theme of dynamical studies has been the
essentially regular character of the motion in triaxial 
potentials, a point of view reflected also in many 
review articles (\cite{bin82c}; \cite{def91}; \cite{ger94}; \cite{dez96}) 
and texts (\cite{frp84}; \cite{sas85}; \cite{bit87}).
Regular motion -- that is, motion that respects as many integrals as 
there are degrees of freedom -- is crucial for the success of the
self-consistency studies, since regular orbits have a variety of 
time-averaged shapes that make them well suited to reproducing 
the mass distribution of an elongated or triaxial galaxy
(\cite{sch81}).
However this view was challenged by the discovery, around 1993, 
that the luminosity profiles of real elliptical galaxies continue to rise, 
roughly as power laws, into the smallest observable radii
(\cite{cra93}; \cite{fer94}).
There is also growing evidence that most elliptical galaxies 
and bulges contain supermassive black holes at their centers, 
presumably relics of the quasar epoch (\cite{kor95}).
The orbital motion in triaxial models with 
central cusps or black holes
can be very different from the motion in models like 
Schwarzschild's, which had a large, constant-density core. 
Many orbits -- particulary the box-like orbits that visit the 
center -- are found to be chaotic, densely filling a 
region that is much rounder than an isodensity contour.
The non-integrability of realistic triaxial potentials is reflected 
also in the character of the regular orbits,
which are strongly influenced by resonances between the 
frequencies of motion in different directions.
A growing body of work supports the view that the dynamical 
influence of central density cusps and black holes can extend far 
beyond the nucleus of a triaxial galaxy, and may be responsible 
for many of the large-scale systematic properties of elliptical galaxies, 
including the fact that few of these systems exhibit strong evidence
for triaxiality.

Non-integrability and its consequences are therefore the major themes of 
the present review.
A number of other topics are highlighted here, 
both because of their intrinsic importance and because of their
relative neglect in recent reviews. 
These topics include torus construction; dynamical instabilities;
and mechanisms for collisionless relaxation.
Among the important topics not treated in detail 
here are intrinsic shapes (recently reviewed by Statler 1995); 
galaxy interactions and mergers (\cite{bar96});
and dynamical studies of the distribution of dark matter 
(\cite{bri99}; \cite{sac99}).

A number of standard formulae are in use for describing 
the density and potential of three-dimensional galaxies.
Several of the most common are defined here and referred to below.
Some of these formulae were intended to mimic the luminosity profiles of 
real galaxies; others are poor descriptions of real galaxies but have
features that make them useful from a computational point of view.
Models, like H\'enon's isochrone, that were first defined in 
the spherical geometry are often generalized to the 
ellipsoidal case by replacing the radial variable $r$ with $m$, 
where $m^2=(x/a)^2 + (y/b)^2 + (z/c)^2$ is constant on ellipsoidal 
shells.

1. The logarithmic potential:
\begin{equation}
\Phi(m^2) = {1\over 2}v_0^2\ln\left(m^2+r_c^2/a^2\right)
\label{logarithm}
\end{equation}
whose large-radius dependence corresponds to a density that falls off as 
$r^{-2}$.
The isodensity contours are peanut-shaped and the density falls 
below zero on the short axis when the elongation is sufficiently great.

2. The Perfect Ellipsoid:
\begin{equation}
\rho(m^2) = {\rho_0\over(1+m^2)^2}
\label{perfect}
\end{equation}
(\cite{kuz56}, 1973; \cite{dez85b}).
The Perfect Ellipsoid is the most general, ellipsoidally-stratified
mass model whose gravitational potential supports three isolating
integrals of the motion.

3. The ``imperfect ellipsoid'':
\begin{equation}
\rho(m^2) = {\rho_0m_0^2\over(1+m^2)(m_0^2+m^2)}, \ \ \ \ 0\le 
m_0\le 1
\label{imperfect}
\end{equation}
where $\rho_0=M(1+m_0)/(2\pi^2abcm_0^2)\ $ (\cite{mev96}).
For $m_0=1$, the Perfect density law is recovered, while for 
$m_0=0$ the density increases as $r^{-2}$ near the center.
The potential and forces can 
be efficiently calculated after a transformation to ellipsoidal 
coordinates.

4. H\'enon's isochrone, a spherical model with the potential:
\begin{equation}
\Phi(r) = {-GM\over r_0(u+2)},\ \ \ \ \ u = \sqrt{1+\left({r\over 
r_0}\right)^2} - 1
\end{equation}
(\cite{hen59a}, b).
The name ``isochrone'' refers to the independence of the radial
frequency of an orbit on its angular momentum.
The action-angle variables corresponding to quasiperiodic motion 
in the isochrone potential can be expressed in terms of simple functions 
(e.g. \cite{ges91}).

5. Dehnen's law:
\begin{equation}
\rho(m) = {(3-\gamma)M\over 4\pi 
abc}m^{-\gamma}(1+m)^{-(4-\gamma)}, \ \ \ \ 0\le\gamma<3
\label{dehnen}
\end{equation}
with $M$ the total mass (\cite{deh93}).
The potential in the triaxial geometry 
may be expressed in terms of one-dimensional 
integrals (\cite{mef96}).
Dehnen's law has a power-law central density dependence which 
approximates the observed luminosity profiles of early-type 
galaxies.
Its large-radius dependence is steeper than that of 
real elliptical galaxies.

\section{TORUS CONSTRUCTION}
Regular motion is defined as motion that respects at least $N$
isolating integrals, where $N$ is the number of degrees of freedom 
(DOF), i.e. the dimensionality of configuration space.
Regular motion can always be reduced to translation on a torus; 
that is, a canonical transformation $({\bf x}, {\bf 
v})\rightarrow({\bf J, \theta})$ can be found such that
\begin{equation}
\dot J_i = 0,\ \ \ \ \dot \theta_i = \omega_i, \ \ \ i = 1,...N.
\end{equation}
The constants $J_i$, called the actions, define the radii of 
the various cross-sections of the torus while the angles 
$\theta_i$ define the position on the torus (Figure 1).
The dimensionality of a torus is therefore equal to that
of an orbit in configuration space.
Each point on an orbit maps to $2N$ points on its torus; 
for $N=2$, these four points correspond to the four velocity 
vectors  $(\pm v_x, \pm v_y)$
through a given configuration space point.
The $\omega_i$ are called fundamental frequencies; 
in the generic case, they are incommensurable, i.e. 
their ratios can not be expressed as ratios of integers.
Orbits defined by incommensurable frequencies map onto the entire 
surface of the torus, filling it densely after a sufficiently long time.
For certain orbits, a resonance between the fundamental frequencies
occurs, i.e. 
${\bf n\cdot\omega}=0$ with ${\bf n}=(l,m)$ a nonzero integer vector.
In two degrees of freedom such a resonance implies $\omega_1/\omega_2=|m/l|$ 
and the orbit is periodic, closing on itself after $|m|$ revolutions in 
$\theta_1$ and $|l|$ revolutions in $\theta_2$.
When $N>2$, each independent resonance condition reduces the 
dimensionality of the orbit by one, and $N-1$ such conditions are required
for closure.
With the exception of special Hamiltonians like the spherical 
harmonic oscillator, resonant tori comprise a set of measure 
zero, although they are dense in the phase space.

In potentials that support $N$ global integrals, 
like that of the Perfect Ellipsoid, all trajectories 
lie on tori and the Hamiltonian can be written
\begin{equation}
H = H({\bf J})
\end{equation}
with
\begin{equation}
\omega_i=\dot\theta_i={\partial H\over\partial J_i}. 
\end{equation}
The tori of an integrable system are nested in a completely regular 
way throughout phase space.
According to the KAM theorem, these tori will survive under small 
perturbations of $H$ if their frequencies are 
sufficiently incommensurable (\cite{lil92}).
Resonant tori may be strongly deformed even under small 
perturbations, however, leading to a complicated phase-space structure of 
interleaved regular and chaotic regions.
Where tori persist, the motion can be characterized in terms of 
$N$ local integrals.
Where tori are destroyed, the motion is
chaotic and the orbits move in a space of higher 
dimensionality than $N$.

\begin{figure}
\plotone{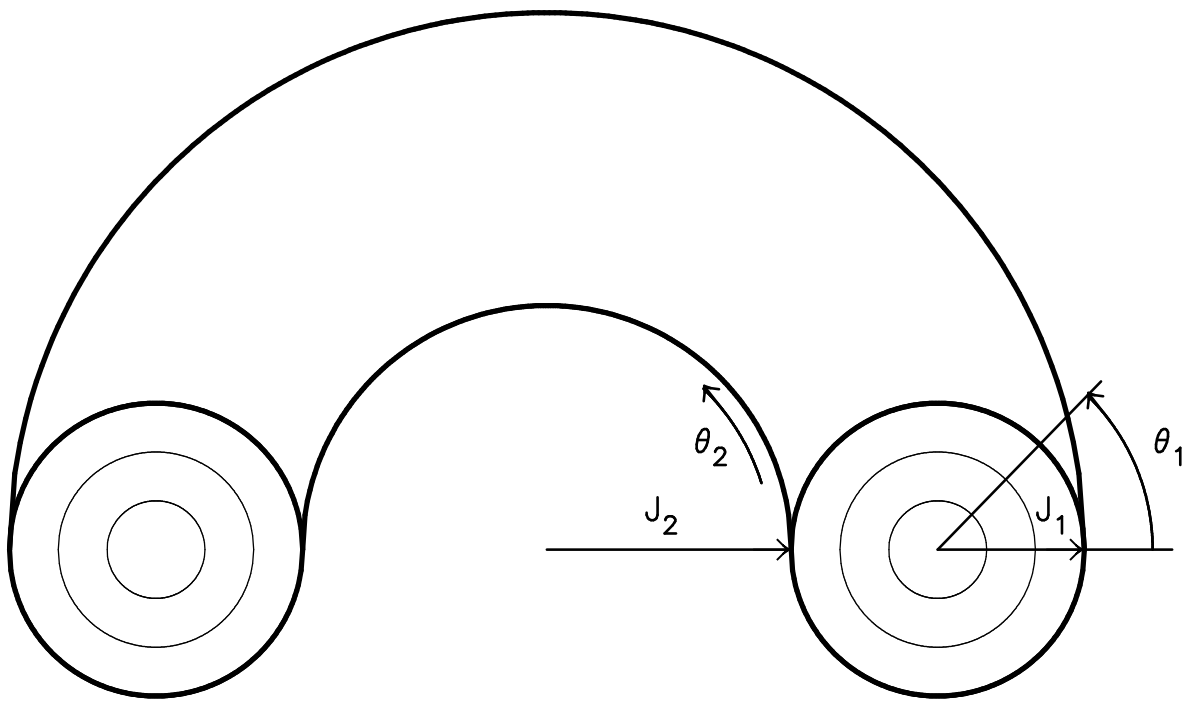}
\caption{
Invariant torus defining the motion of a regular orbit in a
two-dimensional potential.
The torus is determined by the values of the actions $J_1$ and
$J_2$; the position of the trajectory on the torus is defined by
the angles $\theta_1$ and $\theta_2$, which increase linearly
with time, $\theta_i = \omega_it + \theta_0$.
If the torus is resonant, $l\omega_1 + m\omega_2=0$, the
orbit is periodic, returning to its starting point after $|m|$
revolutions in $\theta_1$ and $|l|$ revolutions in $\theta_2$.
}
\end{figure}

While (local) action-angle variables are guaranteed to exist if the motion 
is regular, there is no general, analytic technique for calculating
the $({\bf J}, {\bf\theta})$ from the $({\bf x}, {\bf v})$.
This is unfortunate since the representation of an orbit in terms 
of its action-angle variables is maximally compact, reducing the 
$2N-1$ variables on the energy surface to just $N$ angles.
In addition, time-averaged quantities become trivial to compute 
since the probability 
of finding a star on the torus is uniform in the angle variables.
For instance, the configuration space density is
\begin{eqnarray}
\rho(x,y) &=&
\left|{\partial(\theta_1,\theta_2)\over\partial(x,y)}\right|
\rho(\theta_1,\theta_2) \nonumber
\\
&\propto&\left|{\partial x\over\partial\theta_1}{\partial y 
\over\theta_2}
- {\partial x\over\partial\theta_2}{\partial y\over\partial\theta_1}
\right|^{-1}
\end{eqnarray}
which can be trivially evaluated given ${\bf x}({\bf\theta})$.
Finally, the actions are conserved under slow deformations of the 
potential, a useful property if one wishes to compute the 
evolution of a galaxy that is subject to some gradual perturbing 
force.

Fortunately, a number of algorithms have been developed 
in recent years for numerically extracting the relations between 
the Cartesian and action-angle variables.
Most of this work has been directed toward 2 DOF systems,
making it applicable to axisymmetric potentials or to motion 
in a principal plane of a nonrotating triaxial galaxy.
Generalizations to three degrees of freedom are straightforward 
in principle, although in practice some of the algorithms 
described here become inefficient when $N>2$. 

\subsection{Iterative Approaches}

\subsubsection{Perturbative Methods}

In canonical perturbation theory, 
one begins by writing the Hamiltonian as the sum of two terms,
\begin{equation}
H = H_0({\bf J}) + \epsilon H_1({\bf J},{\bf\theta})
\label{cbt}
\end{equation}
where $H_0$ is integrable and $\epsilon$ is a 
(hopefully small) perturbation parameter.
One then seeks a canonical transformation to new action-angle
variables such that the Hamiltonian is independent of $\bf{\theta}$.
The transformation is found by expanding the generating function in 
powers of $\epsilon$ and solving the Hamilton-Jacobi 
equation successively to each order.
The KAM theorem states that such a transformation will 
sometimes exist, at least when $\epsilon$ is sufficiently small 
and when the unperturbed $\omega_i$ are sufficiently far 
from commensurability.
In a system with one degree of freedom, this approach is 
equivalent to Lindstedt's (1882) method in which both the 
amplitude and frequency of the oscillation are expanded as 
power series in $\epsilon$; the result is a uniformly convergent 
series solution for the motion.
Davoust (1983a,b,c) and Scuflaire (1995) presented applications
of Lindstedt's method to periodic motion in simple galactic potentials.

In systems with two or more degrees of freedom, 
the transformation to action-angle variables is not usually
expressible as a power series in $\epsilon$.
The reason is the ``problem of small denominators'': a Fourier 
expansion of $H_1$ in terms of the $\theta_i$ will generally
contain terms with ${\bf n}\cdot {\bf\omega} \approx 0$ which 
cause the perturbation series to blow up.
This phenomenon is indicative of a real change in the structure
of phase space near resonances: if the perturbation parameter
$\epsilon$ is small, $H_0+\epsilon H_1$ can be topologically
different from $H_0$ only if $H_1$ is large.
However, a number of techniques have been developed to formally 
suppress the divergence.
One approach, called secular perturbation theory, is 
useful near resonances in the unperturbed Hamiltonian, 
$\l\omega_1+m\omega_2=0$.
One first transforms to a frame that rotates with the resonant 
frequency; in this frame, the new angle variable measures 
the slow deviation from resonance.
The Hamiltonian is then averaged over the other, fast angle 
variable. 
To lowest order in $\epsilon$, the new action is $lJ_2+mJ_1$ and 
the motion in the slow variable can be found by quadrature.
Solutions obtained in this way are asymptotic approximations to the 
exact ones, i.e. they differ from them by at most $O(\epsilon)$ 
for times of order $\epsilon^{-1}$.
Whether they are good descriptions of the actual motion depends 
on the strength of the perturbation and the timescale of 
interest.

Verhulst (1979) applied the averaging technique to 
motion in the meridional plane of an axisymmetric galaxy.
He expanded the potential in a quartic polynomial about 
the circular orbit in the plane, thus restricting his results to 
epicyclic motion but at the same time guaranteeing a finite 
Fourier expansion for $H_1$.
He related the fundamental frequencies by setting
$\omega_z^2 = (r/s)^2\omega_x^2 + \delta(\epsilon)$ 
and found solutions to first order in $\epsilon$ for various 
choices of the integers $(r,s)$.
For $r=s=1$ he recovered Contopoulos's (1960) famous ``third 
integral,'' as well as Saaf's (1968) formal integral for a quartic 
potential.
De Zeeuw \& Merritt (1983) and de Zeeuw (1985a) applied Verhulst's 
formalism to motion near the center of a triaxial galaxy with an 
analytic core.
Here the unperturbed $\omega_i$ are the frequencies of harmonic 
oscillation near the center.
De Zeeuw \& Merritt showed 
that the choice $r=s=1$ produced a reasonable representation of
the orbital structure in the plane of rotation, including the 
important $1:1$ closed loop orbits that generate the 
three-dimensional tubes.
Robe (1985, 1986, 1987) used an exact technique to recover the 
periodic orbits in the case of $1:1:1$ resonance and investigated 
their stability.

Gerhard \& Saha (1991) compared the usefulness of three
perturbative techniques for reproducing the meridional-plane motion in
axisymmetric models.
They considered:  (1) Verhulst's (1979) averaging technique;
(2) a resonant method based on Lie transforms; and 
(3) a ``superconvergent'' method.
The latter was proposed originally by Kolmogorov (1954) 
and developed by Arnold (1963) and Moser (1967) in their proof of 
the KAM theorem.
The KAM method can only generate orbit families that are
present in $H_0$.
Gerhard \& Saha took for their unpertured potential the spherical 
isochrone and set $\Phi_1 = g(r) P_2^0(\cos\theta)$.
They were able to reproduce the main features of the 
motion using the resonant Lie transform method, including the 
box orbits that are not present when $\Phi_1=0$.
The KAM method could only generate loop orbits but did so 
with great accuracy when taken to high order.
Dehnen \& Gerhard (1993) used the approximate integrals generated 
by the resonant Lie-transform method to construct three-integral 
models of axisymmetric galaxies, as described in \S3.3.

\subsubsection{Nonperturbative Methods}

The failure of perturbation expansions to converge reflects the 
change in phase space topology that occurs near resonances in the 
unperturbed Hamiltonian.
As an alternative to perturbation theory, one can assume 
that a given orbit is confined to a torus and solve directly for 
the action-angle variables in terms of ${\bf x}$ and ${\bf v}$.
If the orbit is indeed regular, such a solution is 
guaranteed to exist and corresponds to a perturbation 
expansion carried out to infinite order (assuming the latter 
is convergent).
For a chaotic orbit, the motion is not confined to a torus
but one might still hope to derive an approximate torus that 
represents the average behavior of a weakly chaotic trajectory over 
some limited interval of time.

Ratcliff, Chang \& Schwarzschild (1984) pioneered this approach 
in the context of galactic dynamics.
They noted that the equations of motion of a 2D orbit,
\begin{equation}
\ddot x = -{\partial\Phi\over\partial x},\ \ \ \ 
\ddot y = -{\partial\Phi\over\partial y},
\label{rcs1}
\end{equation}
can be written in the form
\begin{eqnarray}
\left(\omega_1{\partial\over\partial\theta_1} + 
	\omega_2{\partial\over\partial\theta_2}\right)^2x & = &
	-{\partial\Phi\over\partial x}, \nonumber \\
\left(\omega_1{\partial\over\partial\theta_1} + 
	\omega_2{\partial\over\partial\theta_2}\right)^2y & = &
	-{\partial\Phi\over\partial y},
\label{rcs2}
\end{eqnarray}
where $\theta_1=\omega_1t$ and $\theta_2=\omega_2t$, the 
coordinates on the torus. 
If one specifies $\omega_1$ and $\omega_2$ and treats 
$\partial\Phi/\partial x$ and $\partial\Phi/\partial y$ as 
functions of the $\theta_i$, equations (\ref{rcs2}) can be  
viewed as nonlinear equations for $x(\theta_1,\theta_2)$ and 
$y(\theta_1,\theta_2)$.
No very general method of solution exists for such equations;
iteration is required and success depends on the stability of the 
iterative scheme.
Ratcliff et al. chose to express the coordinates as
Fourier series in the angle variables,
\begin{equation}
{\bf x}({\bf\theta}) = \sum_{\bf n}{\bf X}_{\bf n}e^{i{\bf 
n\cdot\theta}}.
\label{rcs3}
\end{equation} 
Substituting (\ref{rcs3}) into (\ref{rcs2}) gives
\begin{equation}
\sum_{\bf n}({\bf n\cdot\omega})^2{\bf X}_{\bf n}e^{i{\bf 
n\cdot\theta}}=\nabla\Phi
\label{rcs4}
\end{equation}
where the right hand side is again understood to be a function of the angles.
Ratcliff et al. truncated the Fourier series
after a finite number of terms and required equations (\ref{rcs4}) 
to be satisfied on a 
grid of points around the torus.
They then solved for the ${\bf X}_n$ by iterating from some initial guess.
Convergence was found to be possible if the initial guess 
was close to the exact solution.
Guerra \& Ratcliff (1990) applied a similar algorithm to orbits in the
plane of rotation of a nonaxisymmetric potential.

Another iterative approach to torus reconstruction was first 
developed in the context of semiclassical quantum theory by 
Chapman, Garrett \& Miller (1976). 
Binney and collaborators 
(\cite{mcb90}; \cite{bik93}; \cite{kab94a}, 1994b; 
\cite{kaa94}, 1995a, b) 
further developed the technique and applied it to galactic 
potentials.
One starts by dividing the Hamiltonian $H$ into separable and 
non-separable parts $H_0$ and $H_1$, as in equation (\ref{cbt}); 
however $\epsilon H_1$ is no longer required to be small.
One then seeks a generating function $S$ that maps the known tori 
of $H_0$ into tori of $H$.
For a generating function of the $F_2$-type (Goldstein 1980), 
one has
\begin{equation}
{\bf J}({\bf\theta},{\bf J}') = {\partial S\over\partial\theta},
\ \ \ \ {\bf\theta}'({\bf\theta},{\bf J}') = 
{\partial S\over\partial {\bf J}'}
\label{btr1}
\end{equation}
where $({\bf J}, {\bf\theta})$ and $({\bf J}', {\bf\theta}')$
are the action-angle variables of $H_0$ and $H$ respectively.
The generator $S$ is determined, for a specified ${\bf J}'$, 
by substituting the first of equations (\ref{btr1}) 
into the Hamiltonian $H({\bf J},{\bf\theta})$ and requiring the 
result to be independent of ${\bf\theta}$.
One then arrives at $H({\bf J}')$.
Chapman et al. showed that a sufficiently general form for $S$ is
\begin{equation}
S({\bf\theta},{\bf J}') = {\bf\theta}\cdot{\bf J}' - i\sum_{{\bf 
n}\ne 0} S_{\bf n}({\bf J}')e^{i{\bf n}\cdot{\bf\theta}},
\end{equation}
where the first term is the identity transformation, and they 
evaluated a number of iterative schemes for finding the 
$S_{\bf n}$.
One such scheme was found to recover the results of first-order
perturbation theory after a single iteration.

The generating function approach is useful for assigning energies 
to actions, $H({\bf J}')$, but most of the other quantities of 
interest to galactic dynamicists require additional effort.  
For instance, equation (\ref{btr1}) gives $\theta'(\theta)$ 
as a derivative of $S$, but since $S$ must be computed separately 
for every ${\bf J}'$ its derivative is likely to be 
ill-conditioned.
Binney \& Kumar (1993) and Kaasalainen \& Binney (1994a) discussed 
two schemes for finding $\theta'(\theta)$; the first 
required the solution of a formally infinite set of equations, while the 
latter required multiple integrations of the equations of motion 
for each torus.

Another feature of the generating function approach 
is its lack of robustness.
Kaasalainen \& Binney (1994a) noted that the success of the 
method depends somewhat on the choice of $H_0$.
For box orbits, which are most naturally described as coupled rectilinear 
oscillators, they found that a harmonic-oscillator 
$H_0$ gave poor results unless an additional point 
transformation was used to deform the rectangular orbits of $H_0$ 
into narrow-waisted boxes like those in typical galactic potentials.
Kaasalainen (1995a) considered orbits belonging to higher-order 
resonant families and found that it was generally necessary to
define a new coordinate transformation for each family.

Warnock (1991) presented a modification of the Chapman et al. 
algorithm in which the generating funtion $S$ was 
derived by numerically integrating an orbit from appropriate 
initial conditions, transforming the coordinates to $({\bf J, 
\theta})$ of $H_0$ and interpolating ${\bf J}$ on a regular grid 
in ${\bf\theta}$.
The values of the $S_{\bf n}$ then follow from the 
first equation of (\ref{btr1}) after a discrete Fourier 
transform.
Kaasalainen \& Binney (1994b) found that Warnock's scheme could be 
used to substantially refine the solutions found via their iterative 
scheme.

Having computed the energy on a grid of ${\bf J}'$ values, one can 
interpolate to obtain the full Hamiltonian $H(J_1',J_2')$.
If the system is not in fact completely integrable, this $H$ may 
be rigorously interpreted as smooth approximation to the true $H$
(\cite{wru91}, 1992) and can be taken as 
the starting point for secular perturbation theory. 
Kaasalainen (1994, 1995b) developed this idea and showed how to recover 
accurate surfaces of section in the neighborhood of low-order
resonances in the planar logarithmic potential.

Percival (1977) described a variational principle for constructing
tori.  
His technique has apparently not been implemented in the context
of galactic dynamics.

\subsection{Trajectory-Following Approaches}

A robust and powerful alternative to the generating function 
approach is to construct tori by Fourier decomposition of the 
trajectories.
Trajectory-following algorithms are based on the fact that 
integrable motion is quasiperiodic; in other words, in any 
canonical coordinates $({\bf p,q})$, the motion can be expressed as 
\begin{eqnarray}
{\bf p}(t) & = &\sum_{\bf n}{\bf p_n}({\bf J})\exp\left[i{\bf 
n}\cdot(\omega t + {\bf\phi}_0)\right], \nonumber \\ 
{\bf q}(t) & = &\sum_{\bf n}{\bf q_n}({\bf J})\exp\left[i{\bf 
n}\cdot(\omega t + {\bf\phi}_0)\right] 
\label{qp1}
\end{eqnarray}
where the $\omega_i$ are the $N$ fundamental frequencies on the 
torus.
It follows from equations (\ref{qp1}) that the Fourier transform 
of ${\bf x}(t)$ or ${\bf v}(t)$ will consist of a set of spikes at 
discrete frequencies $\omega_k$ that are linear combinations of the 
fundamental frequencies.
An analysis of the frequency spectrum yields both the fundamental 
frequencies and the integer vectors $\bf n$ associated with each spike.
The relation between the Cartesian coordinates and the angles 
follows immediately from equation (\ref{qp1}).
For instance, the $x$ coordinate in a 2 DOF system becomes
\begin{eqnarray}
x(t) & = & \sum_kX_ke^{i\omega_kt} 
\nonumber \\ 
& = & \sum_{lm}X_{l,m}e^{i(l\theta_1 + m\theta_2)} 
\nonumber \\ 
& = & x(\theta_1, \theta_2)
\label{qp2}
\end{eqnarray}
and similarly for $y(t)$.
The actions can be computed from Percival's (1974) formulae,
\begin{eqnarray}
J_1 = \sum_{l,m}l\left(l\omega_1+m\omega_2\right)
\left(X_{lm}^2 + Y_{lm}^2\right), \nonumber \\ 
J_2 = \sum_{l,m}m\left(l\omega_1+m\omega_2\right)
\left(X_{lm}^2 + Y_{lm}^2\right),
\label{per1}
\end{eqnarray}
thus yielding the complete map $({\bf x}, {\bf v}) \rightarrow 
({\bf J}, {\bf\theta})$.
Binney \& Spergel (1982) pioneered the trajectory-following 
approach in galactic dynamics, using a least-squares algorithm to 
compute the ${\bf X}_k$.
They were able to recover the fundamental frequencies in a 2 DOF 
potential with a modest accuracy of $\sim 0.1\%$ after $\sim 25$ 
orbital periods.
Binney \& Spergel (1984) used Percival's formula to construct the 
action map for orbits in a principal plane of the logarithmic 
potential.

A major advance in trajectory-following algorithms was made by 
Laskar (1988, 1990), who developed a set of tools, the 
``numerical analysis of fundamental frequencies'' (NAFF), for 
extracting the frequency spectra of quasiperiodic systems with 
very high precision.
The NAFF algorithm consists of the following steps:

1. Integrate an orbit for a time $T$ 
and record the phase space variables at $M$ equally spaced intervals.
Translate each time series $f(t)$ to an interval $[-T/2,T/2]$ 
symmetric about the time origin.

2. Using a discrete Fourier transform, construct an approximation to 
the frequency spectrum of $f(t)$ and identify the peaks.
The location of any peak $\omega_k$ will be defined to an 
accuracy of $\sim 1/MT$.

3. Refine the estimate of the location of the 
strongest peak by finding the maximum of the function
\begin{equation}
\phi(\omega) = {1\over T}\int_{-T/2}^{T/2} f(t) e^{-i\omega 
t}\chi(t) \ dt
\label{naff1}
\end{equation}
where $\chi(t) = 1 + \cos(2\pi t/T)$ is the Hanning window 
function.
The integral can be approximated by interpolating the 
discretely-sampled $f(t)$.
The Hanning filter broadens the peak but greatly 
reduces the sidelobes, allowing a very precise determination of 
$\omega_k$.

4. Compute the amplitude $X_k$ by projecting $e^{i\omega_kt}$ onto 
$f(t)$, and subtract this component from the time series.

5. Repeat steps 3 and 4 until the residual function does not significantly 
decrease following subtraction of another term.
Since subsequent components $e^{i\omega_jt}$ will not be mutually 
orthogonal, a Gram-Schmidt procedure is used to construct orthonormal basis 
functions before carrying out the projection in step 4.

6. Identify the integer vector ${\bf n}_k=(l_k,m_k)$ associated with each 
$\omega_k$.

\noindent Laskar's algorithm recovers the fundamental frequencies with an 
error that falls off as $T^{-4}$ (\cite{las96}), compared with 
$\sim T^{-1}$ in algorithms like Binney \& Spergel's (1982).
Even for modest integration times of $\sim 10^2$ orbital 
periods, the NAFF algorithm is able to recover fundamental frequencies 
with an accuracy of $\sim10^{-8}$ or better.
Such extraordinary precision allows the extraction of a large number of 
components from the frequency spectrum, hence a very precise
representation of the torus.

Papaphilippou \& Laskar (1996) applied the NAFF algorithm to 
2DOF motion in a principal plane of the logarithmic potential.
They experimented with different choices for the quantity $f$ whose
time series is used to compute the frequency spectrum.
Ideally, one would choose $f$ to be an angle variable, in terms of 
which the frequency spectrum reduces to a single peak, but 
the angles are not known a priori and the best one can do is to use the angle
variables corresponding to some well-chosen Hamiltonian.
However Papaphilippou \& Laskar found that the convergence of the
quasiperiodic expansion was only weakly affected by the choice of
$f$; for most orbits, Cartesian coordinates (or polar coordinates in the
case of loop orbits) were found to work almost as well as other choices.
This result implies that trajectory-following
methods are more easily automated than generating function 
methods which require a considerable degree of cleverness in the 
choice of coordinates.

Papaphilippou \& Laskar (1996) focussed on the fundamental frequencies 
rather than the actions
for their characterization of the tori,
in part because the frequencies can be obtained with more precision
than the actions,
but also because KAM theory predicts that the structure of phase 
space is determined in large part by resonances between the 
$\omega_i$.
They defined the ``frequency map,'' the curve of $\omega_2/\omega_1$ 
values determined by a set of orbits of a given energy;
this curve is discontinuous whenever the initial conditions pass 
over a resonance associated with chaotic motion.
The important resonances, and the sizes of their associated 
chaotic regions, are immediately apparent from the frequency map.
Papaphilippou \& Laskar showed that most of the chaos in the
logarithmic potential was 
associated with the unstable short-axis orbit, a $1:1$ resonance,
but they were also able to identify significant chaotic zones 
associated with higher-order resonances like the $1:2$ banana 
orbit.

One shortcoming of trajectory-following algorithms is that they 
must integrate long enough for the orbit to adequately sample its
torus.
When the torus is nearly resonant, 
${\bf n}\cdot{\bf\omega}\approx 0$, the orbit is restricted 
for long periods of time to a subset of the torus and the required 
integration interval increases by a factor 
$\sim 1/{\bf n}\cdot{\bf\omega}$.
Another problem is the need to sample the time series with very high 
frequency in order to minimize the effects of aliasing.

\section{MODELLING AXISYMMETRIC GALAXIES}

Motion in an axisymmetric potential is qualitatively simpler than 
in a fully triaxial one due to conservation of angular 
momentum about the symmetry axis.
Defining the effective potential
\begin{equation}
\Phi_{\rm eff} \equiv \Phi(R,z) + {L_z^2\over 2R^2},
\label{axi1}
\end{equation}
where $(R,z,\phi)$ are cylindrical coordinates and 
$L_z=R^2\dot\phi={\rm constant}$,
the equations of motion are
\begin{equation}
\ddot R = -{\partial\Phi\over\partial R} - {L_z^2\over R^3}
 = -{\partial\Phi_{\rm eff}\over\partial R} ,\ \ \ \ \ \ddot z 
= -{\partial\Phi\over\partial z}
= -{\partial\Phi_{\rm eff}\over\partial z},
\label{axi2}
\end{equation}
and $\dot\phi=L_z/R^2$.
These equations describe the two-dimensional motion of a star in 
the $(R,z)$, or meridional, plane which rotates non-uniformly about 
the symmetry axis.
Motion in axisymmetric potentials is therefore a 2 DOF problem.

Every trajectory in the meridional plane is constrained 
by energy conservation to lie within the zero-velocity curve, the 
set of points satisfying $E=\Phi_{\rm eff}(R,z)$.
While the equations of motion (\ref{axi2}) can not be solved in closed
form for arbitrary $\Phi(R,z)$, numerical integrations demonstrate that 
most orbits do not densely fill the zero-velocity curve but instead 
remain confined to narrower, typically wedge-shaped regions 
(\cite{oll62}); in three dimensions, the orbits are tubes around
the short axis.
\footnote{Because of their boxlike shapes in the meridional plane, 
such orbits were originally called ``boxes'' even though their 
three-dimensional shapes are more similar to doughnuts.}
The restriction of the motion to a subset of the region defined 
by conservation of $E$ and $L_z$ is indicative of the existence of 
an additional conserved quantity, or third integral $I_3$, for the 
majority of orbits.
Varying $I_3$ at fixed $E$ and $L_z$ is roughly 
equivalent to varying the height above and below
the equatorial plane of the orbit's
intersection with the zero velocity curve.
In an oblate potential, extreme values of $I_3$ correspond either 
to orbits in the equatorial plane, or to ``thin tubes,'' 
orbits which have zero radial action and which reduce to precessing 
circles in the limit of a nearly spherical potential.
In prolate potentials, two families of thin tube orbits may exist: 
``outer'' thin tubes, similar to the thin tubes in oblate 
potentials,
and ``inner'' thin tubes, orbits similar to helices that wind 
around the long axis (\cite{kuz73}).

The area enclosed by the zero velocity curve tends to zero as
$L_z$ approaches $L_c(E)$, the angular momentum of a circular 
orbit in the equatorial plane.
In this limit, the orbits may be viewed as perturbations of
the planar circular orbit, and an additional isolating integral 
can generally be found (\cite{ver79}).
As $L_z$ is reduced at fixed $E$, the amplitudes of allowed 
motions in $R$ and $z$ increases and resonances between the 
two degrees of freedom begin to appear.
Complete integrability is unlikely in the presence of resonances,
and in fact one can find often small regions of stochasticity at 
sufficiently low $L_z$ in axisymmetric potentials.
However the fraction of phase space associated with chaotic
motion typically remains small unless $L_z$ is close to zero
(\cite{ric82}; \cite{les92}; \cite{eva94}).
The most important resonances at low $L_z$ in oblate potentials 
are $\omega_z/\omega_R=1:1$, which produces the banana orbit
in the meridional plane, and
$\omega_z/\omega_R=3:4$, the fish orbit.
The banana orbit bifurcates from the $R$-axial (i.e. planar) 
orbit at high $E$ and low $L_z$, causing the latter to lose its stability; 
the corresponding three-dimensional orbits are shaped like saucers with 
central holes.
The fish orbit bifurcates from the thin tube orbit typically
without affecting its stability.
In prolate potentials, the banana orbit does not exist and 
higher-order bifurcations first occur from the 
thin, inner tube orbit (\cite{eva94}).

Once the orbital families in an axisymmetric potential have been 
identified, one can search for a population of orbits that 
reproduces the kinematical data from some observed galaxy.
In practice, this procedure is made difficult by lack of information
about the distribution of mass that determines the 
gravitational potential and about the intrinsic elongation or orientation 
of the galaxy's figure.
Faced with these uncertainties, galaxy modellers have often
chosen to tackle simpler problems with well-defined solutions.
One such problem is the derivation of the two-integral 
distribution function $f(E,L_z)$ that self-consistently 
reproduces a given mass distribution $\rho(R,z)$. 
Closely related is the problem of finding three-integral $f$'s for 
models based on integrable, or St\"ackel, potentials.
These approaches make little or no use of kinematical data and 
hence are of limited applicability to real galaxies.
More sophisticated algorithms can construct the family of 
three-integral $f$'s that reproduce an observed luminosity 
distribution in any assumed potential $\Phi(R,z)$, in addition 
to satisfying an additional set of constraints imposed by the 
observed velocities.
Most difficult, but potentially most rewarding, are approaches that attempt 
to simultaneously infer $f$ and $\Phi$ in a model-independent way
from the data.
These different approaches are discussed in turn below.

\subsection{Two-Integral Models}

One can avoid the complications associated with resonances and 
stochasticity in axisymmetric potentials by simply postulating 
that the phase space density is constant on hypersurfaces of constant 
$E$ and $L_z$, the two classical integrals of motion.
Each such piece of phase space generates a configuration-space 
density
\begin{equation}
\delta\rho = f(E,L_z)d^3{\bf v} = 2\pi f(E,L_z) v_m dv_m 
dv_{\phi} = {2\pi\over R} f(E,L_z) dE dL_z
\end{equation}
where $v_m=\sqrt{v_R^2+v_z^2}$, the velocity in the meridional 
plane; $\delta\rho$ is defined to be nonzero only at points 
$(R,z)$ reached 
by an orbit with the specified $E$ and $L_z$.
The total density contributed by all such phase-space pieces is
\begin{equation}
\rho(R,z) = {4\pi\over R}\int_{\Phi}^0 dE 
\int_0^{R\sqrt{2(E-\Phi)}} f_+(E,L_z) dL_z,
\label{fel1}
\end{equation}
where $f_+$ is the part of $f$ even in $L_z$, $f_+(E,L_z) 
= {1\over 2} \left[f(E,L_z) + f(E,-L_z)\right]$; the odd part of
$f$ affects only the degree of streaming around the symmetry 
axis.
Equation (\ref{fel1}) is a linear relation between 
known functions of two variables,
$\rho(R,z)$ and $\Phi(R,z)$, and an unknown function of two variables, 
$f_+(E,L_z)$; hence one might expect the solution for $f_+$ to be 
unique.
Formal inversions were presented by Lynden-Bell (1962a), Hunter (1975) 
and Dejonghe (1986) using integral transforms; however these 
proofs impose fairly stringent conditions on $\rho$.
Hunter \& Qian (1993) showed that the solution
can be formally expressed as a path integral in the complex 
$\Phi$-plane and calculated a number of explicit solutions.
Even if a solution may be shown to exist, finding it is rarely 
straightforward since one must invert a 
double integral equation.
Analytic solutions can generally only be found for 
potential-density pairs such that the ``augmented density,''
$\rho(R,\Phi)$, is expressible in simple form (\cite{dej86}).

An alternative approach is to represent $f$ and $\rho$ discretely 
on two-dimensional grids; the double integration then becomes a 
matrix operation which can be inverted to give $f$.
Results obtained in this straightforward way tend to be extremely noisy 
because of the strong ill-conditioning of the inverse operation,
however (e.g. \cite{kui95}, Figure 2).
Regularization of the inversion can be achieved via a number of 
schemes.
The functional form of $f$ can be restricted by representation in a
basis set that includes only low order, i.e. slowly varying, terms 
(\cite{deg94}; \cite{mag95}), or by truncated
iteration from some smooth initial guess (\cite{deh95}).
Neither of these techniques deals in a very flexible way with the 
ill-conditioning.
An alternative approach is suggested by modern techniques for function 
estimation: one recasts equation (\ref{fel1}) as a 
penalized-likelihood problem, the solution to which is
smooth without being otherwise restricted in functional form
(Merritt 1996).

The two-integral $f$'s corresponding to a large number of 
axisymmetric potential-density pairs have been found using these 
techniques; compilations are given by Dejonghe (1986) and by 
Hunter \& Qian (1993).
A few of these solutions may be written in closed form 
(\cite{lyn62a}; \cite{lak81}; \cite{bad93}; \cite{eva93}, 
1994) but most can be expressed only as infinite series or as numerical 
representations on a grid.
Since the existence of such solutions is not in question, 
the most important issue addressed by these studies is the 
positivity of the derived $f$'s.
If $f$ falls below zero for some $E$ and $L_z$, one may 
conclude either that no self-consistent distribution function exists 
for the assumed mass model or (more securely) that any such function 
must depend on a third integral.
For example, Batsleer \& Dejonghe (1993) 
derived analytic expressions for $f(E,L_z)$ corresponding to the 
Kuzmin-Kutuzov (1962) family of mass models, whose density profile 
matches that of the isochrone in the spherical limit.
They found that $f$ becomes negative when the (central) axis 
ratio of a prolate model exceeds the modest value of $\sim 1.3$.
A similar result was obtained by Dejonghe (1986) for the prolate branch of
Lynden-Bell's (1962a) family of axisymmetric models.
By contrast, the two-integral $f$s corresponding to oblate 
mass models typically remain non-negative for all 
values of the flattening.

The failure of two-integral $f$'s to describe prolate 
models can be understood most simply in terms of the tensor 
virial theorem.
Any $f(E,L_z)$ implies isotropy of motion in the meridional 
plane, since $E$ is symmetric in $v_R$ and $v_z$ and $L_z$ 
depends only on $v_{\phi}$.
Now the tensor virial theorem states that the mean square velocity 
of stars in a steady-state galaxy must be highest in the direction 
of greatest elongation. 
In an oblate galaxy, this can be accomplished by making either 
$\sigma_R^2$ or $\langle v^2_{\phi}\rangle$ large compared to 
$\sigma_z^2$.
But $\sigma_R=\sigma_z$ in a two-integral model,
hence the flattening must come from large $\phi$- velocities,
i.e. $f$ must be biased toward orbits with large $L_z$.
Such models may be physically unlikely but will never 
require negative $f$'s.
In a prolate galaxy, however, the same argument implies that the 
number of stars on nearly-circular orbits must be reduced as the 
elongation of the model increases.
This strategy eventually fails when the population of
certain high-$L_z$ orbits falls below zero.
The inability of two-integral $f$'s to reproduce the density of 
even moderately elongated prolate spheroids suggests that barlike 
or triaxial galaxies are generically dependent on a third integral.

The ``isotropy'' of two-integral models allows one to infer a great 
deal about their internal kinematics without even deriving $f(E,L_z)$.
The Jeans equations that relate the potential of an axisymmetric 
galaxy to gradients in the velocity dispersions 
are
\begin{eqnarray}
\nu{\partial\Phi\over\partial z} & = & 
-{\partial(\nu\sigma^2)\over\partial
z} ,\label{jeans1a} \\
\nu{\partial\Phi\over\partial R} & = &
-{\partial(\nu\sigma^2)\over\partial R} -
{\nu\over R}\left(\sigma^2 - \overline{v_{\phi}^2} \right),
\label{jeans1b}
\end{eqnarray}
with $\nu$ the number density of stars and 
$\sigma=\sigma_R=\sigma_z$ the velocity dispersion in the 
meridional plane.
If $\nu$ and $\Phi$ are specified, these equations have solutions
\begin{eqnarray}
\nu\sigma^2 & = & \int_z^{\infty} \nu{\partial\Phi\over\partial z} 
dz, \label{jeans2a} \\
\nu\overline{v_{\phi}^2} & = & \nu\sigma^2 + R\int_z^{\infty} 
\left( {\partial\nu\over\partial R}{\partial\Phi\over\partial z} 
- {\partial\nu\over\partial z}{\partial\Phi\over\partial R}\right) dz. 
\label{jeans2b}
\end{eqnarray}
The uniqueness of the solutions is a consequence of the 
uniqueness of the even part of $f$; the only remaining freedom 
relates to the odd part of $f$, i.e.  
the division of $\overline{v_{\phi}^2}$ into mean motions 
and dispersions about the mean, $\overline{v_{\phi}^2} = 
\overline{v_{\phi}}^2 + \sigma^2_{\phi}$.
A model with streaming motions adjusted such that 
$\sigma_{\phi}=\sigma_R=\sigma_z$ everywhere is called an
``isotropic oblate rotator'' since the model's flattening may be 
interpreted as being due completely to its rotation.
The expressions (\ref{jeans2a}, \ref{jeans2b}) 
have been evaluated for a number 
of axisymmetric potential-density pairs (\cite{fil86}; 
\cite{ded88}; \cite{deg94}; \cite{evd94}); 
the qualitative nature of the solutions 
is only weakly dependent on the choices of $\nu$ and $\Phi$.

The relative ease with which $f(E,L_z)$ and its moments can be 
computed given $\nu$ and $\Phi$ has tempted a number of workers 
to model real galaxies in this way.
The approach was pioneered by Binney, Davies \& Illingworth 
(1990) and has been very widely applied (\cite{vdm90};  
\cite{vdm91}; \cite{dej93}; \cite{vdmm94}; \cite{kui95}; \cite{deh95}; 
\cite{qia95}).
Typically, a model is fit to the luminosity density and the 
potential is computed assuming that mass follows light, often
with an additional central point mass representing a black hole. 
The even part of $f$ or its moments are then uniquely determined, 
as discussed above.
The observed velocities are not used at all in the construction 
of $f_+$ except insofar as they determine the normalization of 
the potential.
Models constructed in this way have been found to reproduce the 
kinematical data quite well in a few galaxies, notably M32 
(\cite{deh95}; \cite{qia95}).
The main shortcoming of this approach is that it gives no insight into 
how wide a range of three-integral $f's$ could fit the same data.
Furthermore, if the model fails to reproduce the observed
velocity dispersions, one does not know whether the two-integral 
assumption or the assumed form for $\Phi$ (or both) are incorrect.
 
\subsection{Models Based on Special Potentials}

The motion in certain special potentials is simple enough that 
the third integral can be written in closed form, allowing one to 
derive tractable expressions for the (generally non-unique)
three-integral distribution functions that reproduce $\rho(R,z)$.
Such models are mathematically motivated 
and tend to differ in important ways from real galaxies, 
but the hope is that they may give insight into more realistic models.
Dejonghe \& de Zeeuw (1988) pioneered this approach by 
constructing three-integral $f$'s for the Kuzmin-Kutuzov (1962) 
family of mass models, which have a potential of St\"ackel form and 
hence a known $I_3$.
They wrote $f = f_1(E,L_z) + f_2(E,L_z,I_3)$ and chose a simple
parametric form for $f_2$, $f_2=|E|^lL_z^m(L_z+I_3)^n$.
The contribution of $f_2$ to the density was then 
computed and the remaining part of $\rho$ was required to come from
$f_1$.

Bishop (1987) pointed out that the mass density of any oblate 
St\"ackel model can be reconstructed from the thin short-axis 
tube orbits alone.
The density at any point in Bishop's ``shell'' models is 
contributed by a set of thin tubes that differ in only one parameter, 
their turning point.
The distribution of turning points that reproduces the density 
along every shell in the meridional plane can be found by solving an 
Abel equation.
If all the orbits in such a model are assumed to circulate in the 
same direction about the symmetry axis, the result is the 
distribution function with the highest total angular momentum consistent
with the assumed distribution of mass.
Bishop constructed shell-orbit distribution functions 
corresponding to a number of oblate St\"ackel models.
De Zeeuw \& Hunter (1990) applied Bishop's algorithm to the 
Kuzmin-Kutuzov models, and Evans, de Zeeuw \& Lynden-Bell (1990) 
derived shell models based on flattened isochrones.
Hunter et al. (1990) derived expressions analogous to Bishop's 
for the orbital distribution in prolate shell models in which 
the two families of thin tube orbits permit a range of different
solutions for a given mass model. 

The ease with which thin-orbit distribution functions can be 
derived has motivated a number of schemes in which $f$ is assumed 
to be close to $f_{\rm shell}$, i.e. in which the orbits have a 
small but nonzero radial thickness.
Robijn \& de Zeeuw (1996) wrote $f=f_{\rm shell} \times g(E,L_z,I_3)$
with $g$ a specified function and described an iterative scheme
for finding $f_{\rm shell}$.
They used their algorithm to derive a number of three-integral 
$f$'s corresponding to the Kuzmin-Kutuzov models. 
De Zeeuw, Evans \& Schwarzschild (1996) noted that, in models where the 
equipotential surfaces are spheroids with fixed axis ratios (the 
``power-law'' galaxies), one can write an approximate third 
integral that is nearly conserved for tube orbits with small 
radial thickness.
This ``partial integral'' reduces to the total angular momentum 
in the spherical limit; its accuracy in non-spherical models is 
determined by the degree to which thin tube orbits deviate from 
precessing circles.
Evans, H\"afner \& de Zeeuw (1997) used the partial integral to 
construct approximate three-integral distribution functions for 
axisymmetric power-law galaxies.

The restriction of the potential to St\"ackel form implies that 
the principal axes of the velocity ellipsoid are aligned with the 
same spheroidal coordinates in which the potential is separable 
(\cite{edd15}).
This fact allows some progress to made in finding solutions to 
the Jeans equations.
Dejonghe \& de Zeeuw (1988) and Evans \& Lynden-Bell (1989) 
showed that specification of a single kinematical function, e.g.
the velocity anisotropy, over the complete meridional plane is 
sufficient to uniquely determine the second velocity moments 
everywhere in a St\"ackel potential.
In the limiting case $\sigma_z=\sigma_R$, their result reduces to 
equations (\ref{jeans2a}, \ref{jeans2b}).
Evans (1992) gave a number of numerical solutions to the Jeans 
equations based on an assumed form for the radial dependence 
of the anisotropy.
Arnold (1995) showed that similar solutions could be found 
whenever the velocity ellipsoid is aligned with a separable 
coordinate system, even if the underlying potential is not 
separable.

\subsection{General Axisymmetric Models}

In all of the studies outlined above, restrictions were placed 
on $f$ or $\Phi$ for reasons of mathematical convenience alone.
One would ultimately like to infer both functions in an unbiased
way from observational data, a difficult problem for which no very general 
solution yet exists.
An intermediate approach consists of writing down physically-motivated 
expressions for $\Phi(R,z)$ and $\nu(R,z)$, then deriving 
a numerical representation of $f(E,L_z,I_3)$ that reproduces $\nu$ 
as well as any other observational constraints in the assumed potential.
For instance, $\nu$ might be derived from the observed luminosity 
density and $\Phi$ obtained via Poisson's equation under the 
assumption that mass follows light.
The primary motivation for such an approach is that the relation 
between $f$ and the data is linear once $\Phi$ has been 
specified, which means that solutions for $f$ can be 
found using standard techniques like quadratic programming 
(Dejonghe 1989).
Models so constructed are free of the biases that result from 
placing arbitrary restrictions on $f$; furthermore, if the 
expression for $\Phi$ is allowed to vary over some set of parameters, 
one can hope to assign relative likelihoods to different 
models for the mass distribution.

Most observational constraints take the form of moments of 
the line-of-sight velocity distribution, and it is appropriate 
to ask how much freedom is allowed in these moments once $\Phi$ 
and $\nu$ have been specified.
The Jeans equations for a general axisymmetric galaxy are similar 
to the ones given above for two-integral models, except that 
$\sigma_z$ and $\sigma_R$ are now distinct functions and 
the velocity ellipsoid can have nonzero $\overline{v_Rv_z}$, 
corresponding to a tilt in the meridional plane:
\begin{eqnarray}
\nu{\partial\Phi\over\partial z} & = &-{\partial(\nu\sigma_z^2)\over\partial z}
 - {\partial (\nu\overline{v_{R}v_z}R)\over R\partial R}, \\
\nu{\partial\Phi\over\partial R} & = &
-{\partial(\nu\sigma_{R}^2)\over\partial R} -
{\partial(\nu\overline{v_{R}v_z})\over\partial z} -
{\nu\over R}\left(\sigma_{R}^2-\overline{v}_{\phi}^2-\sigma_{\phi}^2
\right).
\label{jeans3}
\end{eqnarray}
Unlike the two-integral case, the solutions to these equations are 
expected to be highly nonunique since the shape and orientation of 
the velocity ellipsoid in the meridional plane are free to vary -- 
a consequence of the dependence of $f$ on a third integral.
Fillmore (1986) carried out the first thorough investigation of 
the range of possible solutions; he considered oblate spheroidal 
galaxies with de Vaucouleurs density profiles, and 
computed both internal and projected velocity moments for 
various assumed elongations and orientations of the models.
Fillmore forced the velocity ellipsoid to have one of two, fixed 
orientations: either aligned with the coordinate axes 
($\overline{v_Rv_z}=0$), or radially aligned, i.e. oriented such 
that one axis of the ellipsoid was everywhere directed toward the center.
He then computed solutions under various assumptions about 
the anisotropies.
Solutions with large $\sigma_{\phi}$ tended to produce large 
line-of-sight velocity dispersions $\sigma_p$ along the major axis, 
and contours of $\sigma_p$ that were more flattened than the 
isophotes.
Solutions with large $\sigma_R$ had more steeply-falling major 
axis profiles and $\sigma_p$ contours that were rounder than the
isophotes, or even elongated in the $z$-direction.
These differences were strongest in models seen nearly edge-on.
Fillmore suggested that the degree of velocity anisotropy could be 
estimated by comparing the velocity dispersion gradients along 
the major and minor axes.

Dehnen \& Gerhard (1993) carried out an extensive study in which 
they constructed explicit expressions for $f(E,L_z,I_3)$;
in this way they were able to avoid finding solutions of the 
moment equations that corresponded to negative $f$'s.
They approximated $I_3$ using the first-order 
resonant perturbation theory of Gerhard \& Saha (1991) described 
above; their mass model was the same one used in that study, a 
flattened isochrone.
Dehnen \& Gerhard made the important point that the 
mathematically simplest integrals of motion are not necessarily 
the most useful physically.
They defined new integrals $S_r$ and $S_m$, 
called ``shape invariants,'' as algebraic functions of $E$, $L_z$ 
and $I_3$.
The radial shape invariant $S_r$ is an approximate measure of the 
radial extent of an orbit, while the meridional shape invariant 
$S_m$ measures the extent of the orbit above and below the equatorial plane.
Two-integral distribution functions of the form $f=f(E,S_m)$ are 
particularly interesting since they assign equal phase space 
densities to orbits of all radial extents $S_r$,
leading to roughly equal dispersions in the $R-$ and $\phi-$ 
directions.
Classical two-integral models, 
$f=f(E,L_z)$, accentuate the nearly circular orbits to an extent 
that is probably unphysical. 
Dehnen \& Gerhard also investigated choices for $f$ that produced 
radially-aligned velocity ellipsoids with anisotropies that varied from 
pole to equator.

The most general, but least elegant, way to construct $f$ in a 
specified potential is to superpose individual orbits, integrated 
numerically.
Richstone (1980, 1982, 1984) pioneered this approach by building 
scale-free oblate models with $\nu\sim r^{-2}$ in a 
self-consistent, logarithmic potential.
Levison \& Richstone (1985a,b) generalized the algorithm to 
models with a logarthmic potential but a more realistic 
luminosity distribution, $\nu\sim r^{-3}$.
Fillmore \& Levison (1989) carried out a survey of 
highly-flattened oblate models with a de Vaucouleurs 
surface brightness distribution and with two choices for the 
gravitational potential, self-consistent and logarithmic.
They found that the range of orbital shapes was sufficient to 
produce models in either potential with similar observable 
properties; for instance, models could be constructed in both 
potentials with velocity dispersion profiles that increased or 
decreased along either principal axis over a wide range of radii.
Hence they argued that it would be difficult to infer the 
presence of a dark matter halo based on the observed slope of 
the velocity dispersion profile alone.

\begin{figure}
\plotone{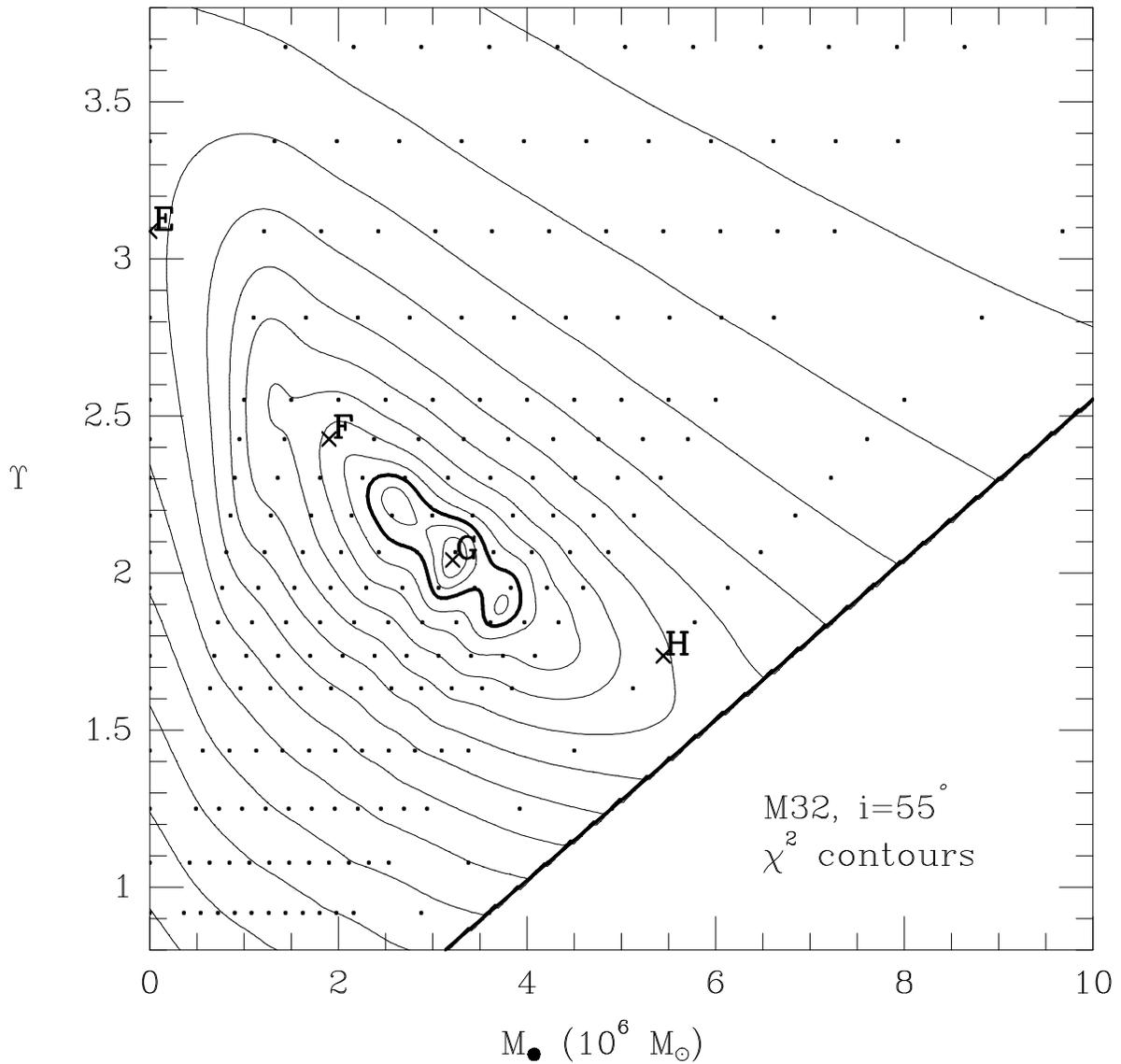}
\caption{
Contours of constant $\chi^2$ that measure the goodness-of-fit
of axisymmetric models to kinematical data for M32 (van der Marel
et al. 1998).
The abscissa is the mass of a central point representing a nuclear
black hole; the ordinate is the mass-to-light ratio of the stars.
These two parameters together specify the potential $\Phi(R,z)$.
Each dot represents a single model constructed by varying the orbital
population so as to best reproduce the observations
in the specified potential.
The plateau of nearly-constant $\chi^2$ corresponds to models
in which changes in $\Phi$ can be compensated for by changes in
the orbital population, leaving the goodness-of-fit essentially
unchanged.
}
\end{figure}

Orbit-based algorithms like Fillmore \& Levison's have now 
been written by a number of groups (\cite{geb98}; \cite{vdm98}; 
\cite{val98}).
In spite of Fillmore \& Levison's discouraging conclusions about the
degeneracy of solutions,
the most common application of these algorithms is to 
potential estimation, i.e. inferring the form of $\Phi(R,z)$ 
based on observed rotation curves and velocity dispersion 
profiles.
A standard approach is to represent $\Phi$ in terms of a small set 
of parameters; for every choice of parameters, the $f$ is found that
best reproduces the kinematical data, and the optimum 
$\Phi$ is defined in terms of the parameters for which the derived 
$f$ provides the best overall fit.
For instance, $\Phi$ may be written 
\begin{equation}
\Phi(R,z) = \left({M\over L}\right)\Phi_L(R,z) - {GM_h\over r}
\end{equation}
where $M/L$ is the mass-to-light ratio of the stars, $\Phi_L$ is 
the ``potential'' corresponding to the observed luminosity 
distribution, and $M_h$ is the mass of a central black hole.
An example is given in Figure 2 which shows $\chi^2$ contours in 
$(M_h,M/L)$-space 
derived from ground-based and HST data for M32 
(\cite{vdm98}).
The expected degeneracy appears as a plateau of nearly constant 
$\chi^2$; this plateau reflects the freedom to adjust
a three-integral $f$ in response to changes in $\Phi$ such that 
the goodness-of-fit to the data remains precisely unchanged.
When the potential is represented by just two parameters, this 
non-uniqueness appears as a ridge line in parameter space, since 
the virial theorem implies a unique relation between the two 
parameters that define the potential (\cite{mer94}).
Imperfections in the data or the modelling algorithm broaden this 
ridge line into a plateau, often with spurious local minima.
The extreme degeneracy of models derived from such data means
that is usually impossible to learn much about the potential that
could not have been inferred from the virial theorem alone.

\subsection{The Axisymmetric Inverse Problem}

Modelling of elliptical galaxies has evolved in a very different way 
from modelling of disk galaxies, where it was recognized early 
on that most of the information about the mass distribution is contained 
in the velocities, not in the light.
By contrast, most attempts at elliptical galaxy modelling have 
used the luminosity as a guide to the mass, with the velocities 
serving only to normalize the mass-to-light ratio.
One could imagine doing much better, going from the observed 
velocities to a map of the gravitational potential.
The difficulties in such an ``inverse problem'' approach are 
considerable, however.
The desired quantity, $\Phi$, appears implicitly as a 
non-linear argument of $f$, which itself is unknown and
must be determined from the data.
There exist few uniqueness proofs that would even justify 
searching for an optimal solution, much less 
algorithms capable of finding those solutions.

A notable attempt was made by Merrifield (1991), who asked 
whether it was possible to infer a dependence of $f$ on a third 
integral in a model-independent way.
Merrifield pointed out that the velocity dispersions along either 
the major or minor axes of an edge-on, two-integral axisymmetric 
galaxy could be independently used to evaluate the kinetic energy 
term in the virial theorem.
A discrepancy between the two estimates might be taken as 
evidence for a dependence of $f$ on a third integral.
Merrifield's test may be seen as a consequence of the fact that 
$f(E,L_z)$ is uniquely determined in an axisymmetric galaxy with 
known $\Phi$ and $\nu$.
However, as Merrifield emphasized, a spatially varying $M/L$ 
could mimic the effects of a dependence of $f$ on a third integral.

An algorithm for simultaneously recovering
$f(E,L_z)$ and $\Phi(R,z)$ in an edge-on galaxy, without any 
restrictions on the relative distribution of mass and light,
was presented by Merritt (1996).
The technique requires complete information about the 
rotational velocity and line-of-sight velocity dispersion 
over the image of the galaxy.
One can then deproject the data to find unique expressions for 
$\sigma(R,z)$, $\sigma_{\phi}(R,z)$ and $\overline{v_{\phi}}(R,z)$.
Once these functions are known, the potential follows immediately
from either of the Jeans equations (\ref{jeans1a}, \ref{jeans1b}); 
$f_+(E,L_z)$ is 
also uniquely determined, as described above.
The odd part of $f$ is obtained from the
deprojected $\overline{v_{\phi}}(R,z)$.
This work highlights the impossibility of ruling out
two-integral $f$'s for axisymmetric galaxies based on observed 
moments of the velocity distribution, since the potential can 
always be adjusted in such a way as to reproduce the data without 
forcing $f$ to depend on a third integral.

The algorithm just described may be seen as the generalization to 
edge-on axisymmetric systems of algorithms that infer $f(E)$ and 
$\Phi(r)$ in spherical galaxies from the velocity dispersion profile 
(e. g. \cite{gef95}).
The spherical inverse problem is highly degenerate if 
$f$ is allowed to depend on $L^2$ as well as $E$ (e.g. \cite{dem92}), 
and one expects a similar degeneracy in the axisymmetric inverse problem 
if $f$ is allowed to depend on $I_3$.
Thus the situation is even more discouraging than envisioned by Fillmore \& Levison 
(1989), who assumed that the data were restricted to the major or minor 
axes: even knowledge of the velocity moments over the full image 
of a galaxy is likely to be consistent with a large number of 
$(f,\Phi)$ pairs.
Distinguishing between these possible solutions clearly requires
additional information, and one possible source is line-of-sight velocity 
distributions (LOSVD's), which are now routinely measured with 
high precision (\cite{cal94}).
In the spherical geometry, LOSVD's have been shown to be 
effective at distinguishing between different $f(E,L^2)-\Phi(R,z)$ 
pairs that reproduce the velocity dispersion data equally well 
(\cite{mes93}; \cite{ger93}; \cite{mer93}).
A second possible source of information is proper 
motions, which in the spherical geometry allow one to infer the 
variation of velocity anisotropy with radius (\cite{lem89}); 
however most elliptical galaxies are too distant for 
stellar proper motions to be easily measured.
A third candidate is X-ray gas, from which the potential can in 
principle be mapped using the equation of hydrostatic equilibrium 
(\cite{sar88}).

All of the techniques described above begin from the assumption 
that the luminosity distribution $\nu(R,z)$ is known.
Rybicki (1986) pointed out the remarkable fact that $\nu$ is 
uniquely constrained by the observed surface brightness 
distribution of an axisymmetric galaxy
only if the galaxy is seen edge-on, or if some other 
restrictive condition applies, e.g. if the isodensity contours 
are assumed to be coaxial ellipsoids with known axis ratios.
Gerhard \& Binney (1996) constructed axisymmetric density 
components that are invisible when viewed in projection and
showed how the range of possible $\nu$'s increases as the 
inclination varies from edge-on to face-on.
Kochanek \& Rybicki (1996) developed methods to produce families 
of density components with arbitrary equatorial density 
distributions; such components typically look like disks.
Romanowsky \& Kochanek (1997) explored how uncertainties in 
deprojected $\nu$'s affect computed values of the kinematical 
quantities in two-integral models with constant mass-to-light ratios.
They found that large variations could be produced in the 
meridional plane velocities but that the projected profiles were 
generally much less affected.

These studies suggest that the dynamical inverse 
problem for axisymmetric galaxies is unlikely to have a unique 
solution except under fairly restrictive conditions.
This fact is useful to keep in mind when evaluating 
axisymmetric modelling studies, in which conclusions about the 
preferred dynamical state of a galaxy are usually affected 
to some degree by restrictions placed on the models for reasons of 
computational convenience only.

\section{TRIAXIALITY}

Motion in triaxial potentials differs in three important ways 
from motion in axisymmetric potentials. 
First, the lack of rotational symmetry means that no component of an 
orbit's angular momentum is conserved.
While tube orbits that circulate about the symmetry axes 
still exist in triaxial potentials, 
other orbits are able to reverse their sense of rotation and 
approach arbitrarily closely to the center.
The box orbits of St\"ackel potentials are prototypical examples.
Second, triaxial potentials are 3 DOF systems, and the objects 
that lend phase space its structure are the resonant tori that 
satisfy a condition between the three fundamental frequencies of 
the form $l\omega_1+m\omega_2+n\omega_3=0$.
Unlike the case of 2 DOF systems, where a resonance between
two fundamental frequencies implies commensurability and
hence closure, the resonant trajectories in 3 DOF systems
are not generically closed; instead, they densely fill a thin, 
two-dimensional surface.
Third, much of the phase space in realistic triaxial potentials is 
chaotic, particularly in models where the gravitational 
force rises rapidly toward the center.

The original motivation for studying triaxial models came from 
the observed slow rotation of elliptical galaxies (\cite{bec75}; 
\cite{ill77}), 
which effectively ruled out ``isotropic oblate rotator'' models.
The low rotation could have been explained without invoking 
triaxiality, since any axisymmetric model can be made nonrotating by 
requiring equal numbers of stars to circulate in the two senses 
about the symmetry axis.
But Binney (1978) argued that it was 
more natural to grant the existence of a global third integral, 
hence to assume that the flattening was due in
part to anisotropy in the meridional plane.
Binney argued further that two non-classical integrals were no 
less natural than one, and therefore that one might 
be able to build galaxies without rotational symmetry whose elongated 
shapes were supported primarily by the extra integrals.
This suggestion was confirmed by Schwarzschild (1979, 1982) who 
showed that most of the orbits in triaxial potentials with large smooth 
cores do respect three integrals and that self-consistent 
triaxial models could be constructed by superposition of such orbits.
Subsequent support for the triaxial hypothesis came from $N$-body 
simulations of collapse in which the final configurations
were often found to be non-axisymmetric 
(\cite{wij82}; \cite{val82}).

The case for triaxiality is perhaps less compelling now than it 
was ten or fifteen years ago.
$N$-body simulations of galaxy formation that include a dissipative 
component often reveal an evolution to axisymmetry in the stars once 
the gas has begun to collect in the center
(\cite{udr93}; \cite{dub94}; \cite{bah96}).
A central point mass, representing a nuclear black hole, has a similar
effect (\cite{nmv85}; \cite{meq98}). 
A plausible explanation for the evolution toward axisymmetry in
these simulations is stochasticity of the box orbits 
resulting from the deepened central potential.
Similar conclusions have been drawn from self-consistency studies
of triaxial models with realistic, centrally-concentrated density 
profiles: the shortage of regular box orbits is often found to 
preclude a stationary triaxial solution
(\cite{sch93}; \cite{mef96}; \cite{mer97}).
Observational studies of minor-axis rotation suggest
that few if any elliptical galaxies are strongly triaxial 
(\cite{fiz91}), and detailed modelling of 
a handful of nearby ellipticals, as discussed above, reveals that 
their kinematics can often be very well reproduced by assuming
axisymmetry.
The assumption that an oblate galaxy with
counter-rotating stars would be unphysical has also been weakened 
by the discovery of a handful of such systems
(\cite{rgk92}; \cite{mek94}).

The phenomenon of triaxiality nevertheless remains a topic of 
vigorous study, for a number of reasons.
At least some elliptical galaxies and bulges exhibit clear 
kinematical signatures of non-axisymmetry (s.g. \cite{scg79};
\cite{fih89}), and the 
observed distribution of Hubble types is likewise inconsistent
with the assumption that all ellipticals are precisely axisymmetric 
(\cite{trm95}, 1996; \cite{ryd96}).
Departures from axisymmetry (possibly transient) are widely 
argued to be necessary for 
the rapid growth of nuclear black holes during the quasar 
epoch (\cite{shb90}), for the fueling of starburst galaxies 
(\cite{sam96}),
and for the large radio luminosities of some ellipticals (\cite{bic97}).
These arguments suggest that most elliptical galaxies or bulges
may have been triaxial at an earlier epoch, and perhaps
that triaxiality is a recurrent phenomenon induced by 
mergers or other interactions.

Almost all of the work reviewed below deals with nonintegrable 
triaxial models.
Integrable triaxial potentials do exist -- the Perfect Ellipsoid 
is an example -- but the integrable models always have features, like
large, constant-density cores, that make 
them poor representations of real elliptical galaxies.
More crucially, integrable potentials are ``non-generic'' in the 
sense that their phase space lacks many of the features that are 
universally present in non-integrable potentials.
For instance, the box orbits in realistic triaxial 
potentials are strongly influenced by resonances between the 
three degrees of freedom, while in integrable potentials these 
resonances (although present) have no effect and all the box 
orbits belong to a single family.
Integrable triaxial models are reviewed by de Zeeuw (1988) and 
Hunter (1995).

The standard convention is adopted here in which the long and 
short axes of a triaxial figure are identified with the $x$ 
and $z$ axes respectively.

\subsection{The Structure of Phase Space}

The motion in smooth triaxial potentials has many features 
in common with more general dynamical systems.
Some relevant results from Hamiltonian dynamics are reviewed here
before discussing their application to triaxial potentials.

Motion in non-integrable potentials is strongly influenced by 
resonances between the fundamental frequencies.
A resonant torus is one for which the frequencies $\omega_i$ 
satisfy a relation 
${\bf n\cdot\omega}=0$ with ${\bf n}=(l,m,n)$ 
an integer vector. 
Resonances are dense in the phase space of an integrable 
Hamiltonian, in the sense that every torus lies near to a torus
satisfying ${\bf n\cdot\omega}=0$ for some (perhaps very large)
integer vector $\bf n$.
However, most tori are very non-resonant in the sense that
${\bf n}\cdot\omega$ is large compared with $|{\bf n}|^{-(N+1)}$,
with $N$ the number of degrees of freedom.
As one gradually perturbs a Hamiltonian away from integrable form, 
the KAM theorem guarantees that the very non-resonant tori will 
retain their topology, i.e. that the motion in their vicinity will remain 
quasiperiodic and confined to (slightly deformed) invariant tori.
Resonant tori, on the other hand, can be strongly affected by even a 
small perturbation.

In a 2 DOF system, motion on a resonant torus is closed, 
since the resonance condition implies that the two frequencies 
are expressible as a ratio of integers, $\omega_1/\omega_2=|m/l|$.
In three dimensions, a single relation 
${\bf n\cdot\omega}=0$ between the three frequencies
does not imply closure; instead, the trajectory is confined to a 
two-dimensional submanifold of its 3-torus.
The orbit in configuration space lies on a thin sheet. 
(In the special case that one of the elements of ${\bf n}$ is 
zero, two of the frequencies will be commensurate.)
For certain tori, two independent resonance conditions 
${\bf n\cdot\omega}=0$ will apply; in this case, 
the fundamental frequencies 
can be written $\omega_i=n_i\omega_0$, i.e. there is 
commensurability for each frequency pair and
the orbit is closed.

On a resonant torus in an integrable Hamiltonian, 
every trajectory satisfies the resonance conditions (of number 
$K\le N-1$) regardless of its phase.
Among this infinite set of resonant trajectories, only a finite 
number persist under perturbation of the Hamiltonian, as resonant 
tori of dimension $N-K$.
The character of these remaining tori alternates from stable 
to unstable as one varies the phase around the original torus.
Motion in the vicinity of a stable resonant torus is regular and 
characterized by $N$ fundamental frequencies.
Motion in the vicinity of an unstable torus is generically
stochastic, even for small perturbations of the Hamiltonian.
Furthermore, in the neighborhood of a stable resonant torus,
higher-order resonances occur which lead to secondary regions of 
regular and stochastic motion, etc., down to finer and finer scales.

In a weakly perturbed Hamiltonian, the stochastic regions tend 
to be isolated and the associated orbits are often found to mimic 
regular orbits for many oscillations.
\footnote{Strictly speaking, the different stochastic regions in 
a 3 DOF system are always interconnected, but the time scale for 
diffusion from one such region to another (Arnold diffusion) is
very long if the potential is close to integrable.}
As the perturbation increases,
the stochastic regions typically grow at the expense of the 
invariant tori.
Eventually, stochastic regions associated with different 
unstable resonances overlap, producing large regions of phase 
space where the motion is interconnected.
One often observes a sudden transition to large-scale 
or ``global'' stochasticity as some perturbation parameter is varied.
In a globally stochastic part of phase space, different orbits 
are effectively indistinguishable and wander in a few 
oscillations over the entire connected region.
Such orbits rapidly visit the entire configuration space region 
within the equipotential surface, giving them a 
time-averaged shape that is approximately spherical.

The way in which resonances affect the phase space structure of 
nonrotating triaxial potentials has recently been clarified 
by a number of studies 
(\cite{caa98}; \cite{pal98}; \cite{vam98}) 
that used trajectory-following algorithms 
to extract the fundamental frequencies and to identify stochastic 
orbits.
The discussion that follows is based on this work and on 
the earlier studies of Levison \& Richstone (1987), Schwarzschild 
(1993) and Merritt \& Fridman (1996). 

At a fixed energy, the phase space of a triaxial galaxy is 
five-dimensional.
One expects most of the regular orbits to be recoverable by
varying the initial conditions over a space of lower dimensionality; 
for instance, in an integrable potential, all the orbits 
at a given energy can be specified via the 2D set of 
actions.
Levison \& Richstone (1987) advocated a 4D initial 
condition space consisting of coordinates $(x_0,z_0,y_0=0)$ 
and velocities $(\dot x_0,\dot z_0)$, with $\dot y_0$ 
determined by the energy.
Schwarzschild (1993) argued that most orbits could be recovered 
from just two, 2D initial condition spaces.
His first space consisted of initial conditions with zero velocity
on one octant of the equipotential surface; these initial conditions 
generate box orbits in St\"ackel potentials.
The second space consisted of initial conditions in the $(x,z)$ plane 
with $\dot x_0=\dot z_0 =0$ and $\dot y_0$ determined by the energy.
This space generates the four families of tube orbits in St\"ackel 
potentials.
Papaphilippou \& Laskar (1998) suggested that boxlike orbits 
could be generated more simply by setting all coordinates to zero 
and varying the two velocity components parallel to a principal plane.
This choice for box-orbit initial condition space is not 
precisely equivalent to Schwarzschild's; it excludes 
resonant orbits (e.g. the banana) that avoid the center, while 
Schwarzschild's excludes orbits (e.g. the anti-banana) that 
have no stationary point.

Figure 3a illustrates stationary
(box-orbit) initial condition space at one energy in a triaxial
Dehnen model 
with $c/a=0.5$, $b/a=0.79$ and $\gamma=0.5$, a ``weak cusp.''
The figure was constructed from $\sim 10^4$ orbits with starting 
points lying on the equipotential surface; the greyscale was adjusted 
in proportion to the logarithm of the 
stochasticity, measured via the rate of diffusion in frequency 
space.
Initial conditions associated with regular orbits are white.
Figure 3b shows the frequency map defined by the same orbits,
i.e. the ratios $\omega_x/\omega_z$ and $\omega_y/\omega_z$ 
between the fundamental frequencies associated with motion along 
the three coordinate axes.
The most important resonances, $l\omega_x+m\omega_y+n\omega_z=0$,
are indicated by lines in the frequency map and by their integer
components $(l,m,n)$ in Figure 3a.
Intersections of these lines correspond to closed orbits,
$\omega_i=n_i\omega_0$.
In order to keep the frequency map relatively simple, all orbits 
associated with a stable resonance have been plotted precisely on the 
resonance line; the third fundamental frequency (associated with 
slow rotation around the resonance) has been ignored.
Unstable resonances appear as gaps in the frequency map, since
the associated orbits do not have well-defined frequencies.

\begin{figure}
\plotfiddle{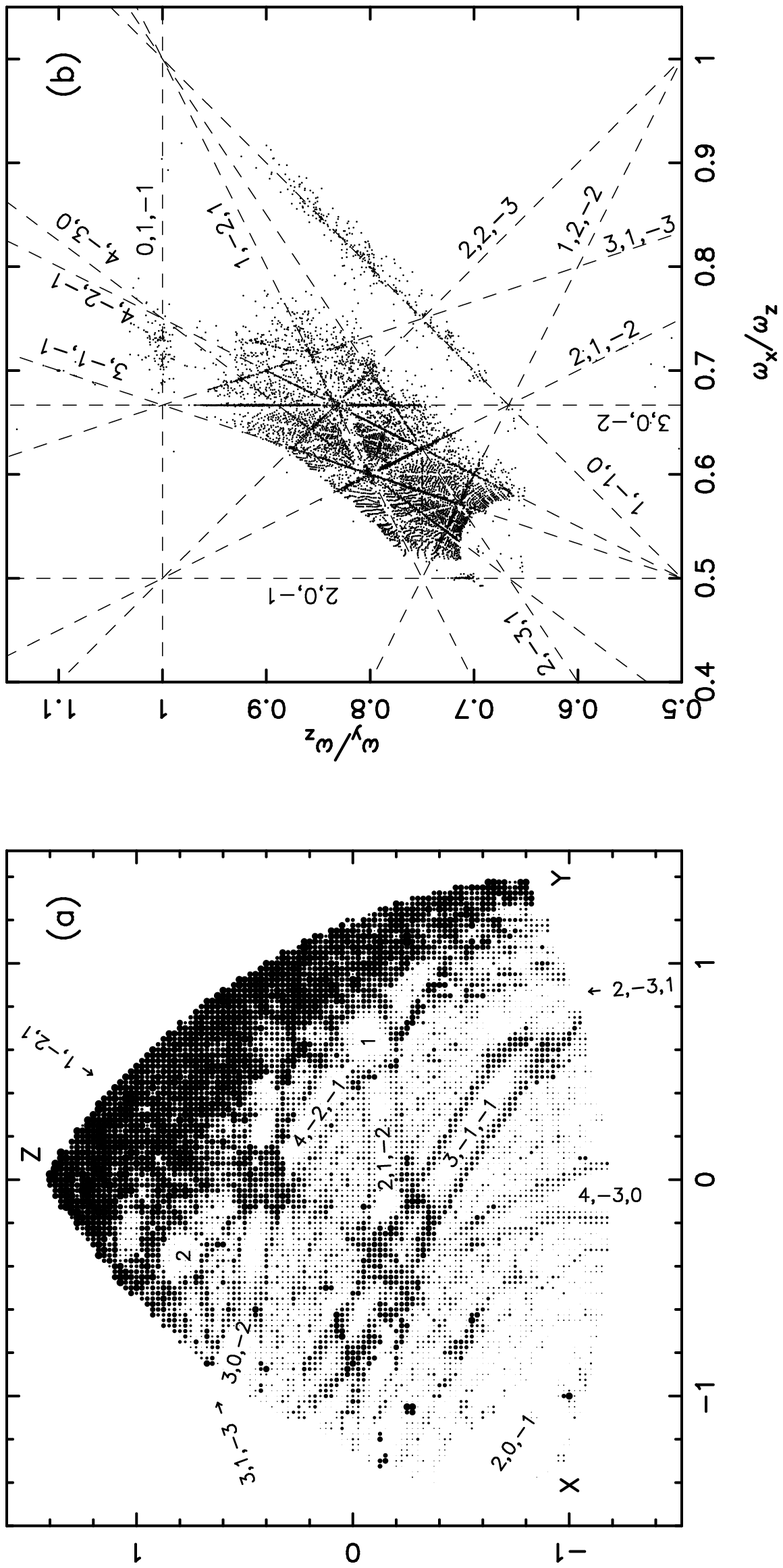}{6.in}{270.}{75}{75}{-300}{+450}
\caption{
Box-orbit phase space at roughly the half-mass energy in a
triaxial Dehnen model with cusp slope $\gamma=0.5$
(Valluri \& Merritt 1998).
(a) Initial condition space: one octant of the equipotential surface
has been projected onto a plane.
Each orbit begins on this surface with zero velocity.
The top, left and right corners correspond to the $z$ (short),
$x$ (long) and $y$ (intermediate) axes.
The grey scale is proportional to the logarithm of the
diffusion rate of orbits in frequency space; initial conditions
corresponding to regular orbits are white.
The regions labelled ``1'' and ``2'' are starting points
associated
with the $5:6:8$ and $7:9:10$ periodic orbits, respectively.
Other regions are labelled with the integers $(l,m,n)$ that
define resonance zones.
(b) Frequency map: the fundamental frequencies
are plotted as ratios $\omega_x/\omega_z$ and $\omega_y/\omega_z$
for each of the orbits in (a).
The most important resonances are labelled.
Stable resonances produce solid lines; gaps correspond
to unstable (chaotic) resonances.
}
\end{figure}

Following the discussion above, one expects the regular orbits 
to come in three families corresponding to the number 
$K=(0,1,2)$ of independent resonance conditions that define their 
associated phase space regions.
The three families are in fact apparent in Figure 3.
First are the regular orbits that lie in the regions between the 
resonance zones, $K=0$.
Orbits in these regions are regular for the same reason that box 
orbits in a St\"ackel potential are regular, i.e. because the 
region of phase space in which they are located is locally 
integrable.
\footnote{The regularity of the box orbits 
in St\"ackel potentials is sometimes erroneously attributed 
to the stability of the long-axis orbit.}
In configuration space, these orbits densely fill 
a three-dimensional region centered on the orgin.
A second set of regular orbits lie in zones associated with a 
single resonance, $K=1$; examples are the $(2,1,-2)$, $(3,-1,-1)$ and 
$(4,-2,-1)$ resonance zones.
The stable resonant orbits that define these regions are 
``thin boxes'' that fill a sheet in configuration space
(Figure 4); the regular 
orbits around them have a small but finite thickness.
The planar $x-z$ banana $(2,0,-1)$ and fish $(3,0,-2)$, and the
$x-y$ pretzel $(4,-3,0)$, also generate families of thin orbits.
The third set of regular orbits, the ``boxlets,'' 
surround periodic orbits that
lie at the intersection of two resonances, $K=2$.
Two such regions are apparent in Figure 3a, associated with 
the $5:6:8$ and $7:9:10$ periodic orbits (marked 1 and 2).

\begin{figure}
\plotone{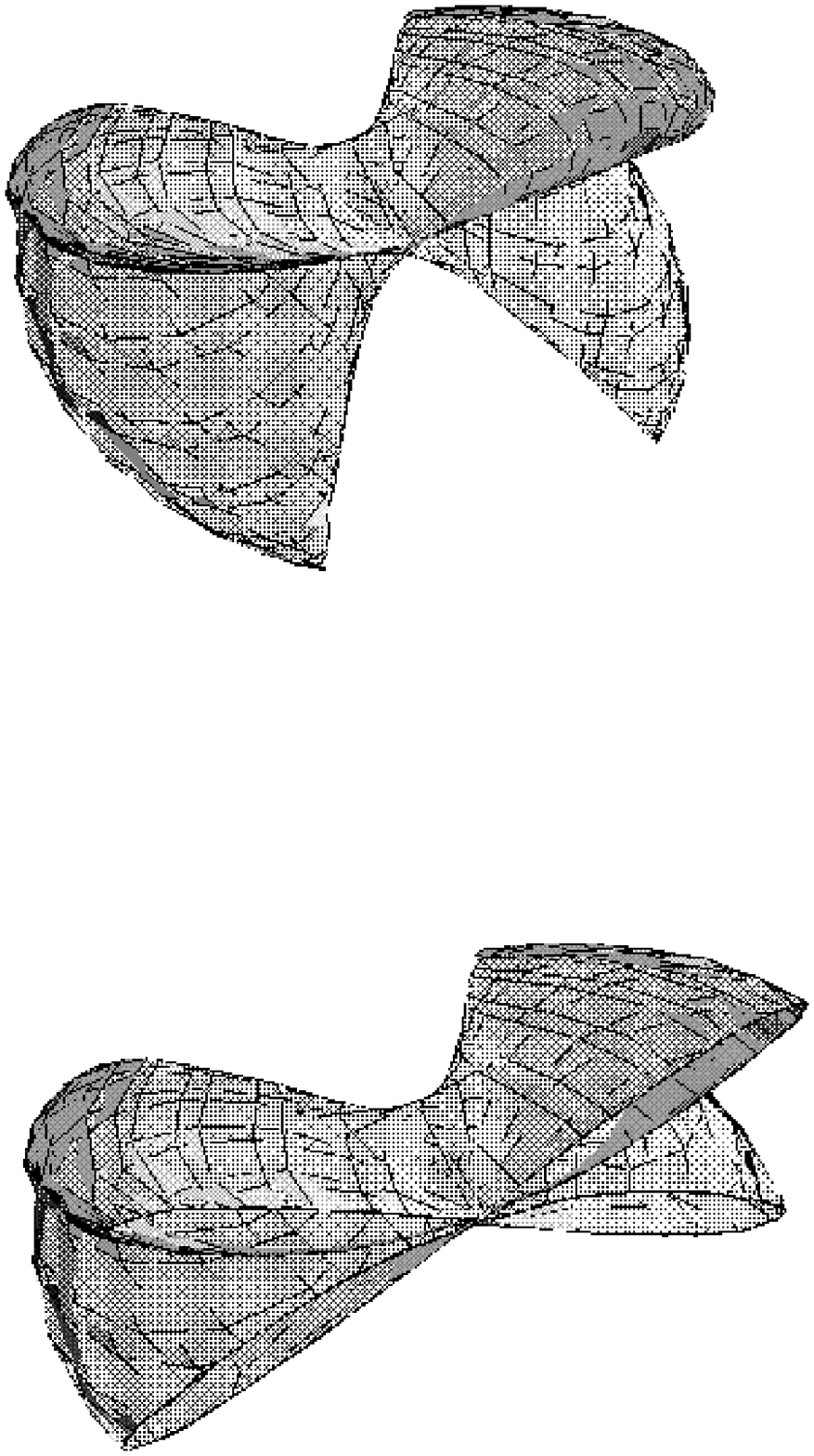}
\caption{A thin box orbit associated with the $(2,1,-2)$ resonance
in Figure 3.
The lower view is a cutaway showing that the orbit is confined
to a membrane.
}
\end{figure}

Close inspection of Figure 3a suggests that even the 
``integrable'' ($K=0$) regions are threaded with 
high-order resonances and their associated chaotic zones.
One expects to find such structure since resonant tori 
are dense in the phase space of the unperturbed Hamiltonian.
However the stochasticity associated with the high-order 
resonances is very weak and for practical purposes the orbits 
in these regions are regular throughout.

All of the box orbits in a St\"ackel potential belong to a single 
family with smoothly varying actions.
While resonances like those shown in Figure 3 are present 
in St\"ackel potentials, their effect on the motion
is limited to the resonant tori themselves.
By contrast, in the phase space of tube 
orbits, even St\"ackel potentials contain (up to) three important 
resonances at each energy corresponding to the three families of 
thin tube orbits.
These primary resonances persist in nonintegrable potentials 
and generate families of regular tube orbits similar to those in 
St\"ackel potentials.
The transition zones between the various tube orbit families, 
which are occupied by unstable orbits in a St\"ackel potential, 
are now stochastic, although the stochastic zones are typically
narrow.
Valluri \& Merritt (1998) illustrated the time evolution of a stochastic
tube orbit.

The structure in Figure 3 is characteristic of 
mildly non-integrable 
triaxial potentials.
As the perturbation of the potential away from integrability increases 
-- due to an increasing central density, for instance --
the parts of phase space corresponding to tube orbits are only 
moderately affected.
However the boxlike orbits are sensitively dependent on the form 
of the potential near the center. 
Papaphilippou \& Laskar (1998) studied ensembles of orbits in the 
logarithmic triaxial potential at fixed energy; they took as 
their perturbation parameter the axis ratios of the model.
At extreme elongations, high-order resonances became 
important in box-orbit phase space; for instance, in a 
triaxial model with $c/a=0.18$, significant stochastic regions 
associated with the $(3,-1,0)$ and $(6,-1,-1)$ resonances were 
found.
Valluri \& Merritt (1998) varied the cusp slope $\gamma$ and the 
mass of a central black hole in a family of triaxial Dehnen 
models.
They found that the relative contributions from the three types 
of regular orbit (satisfying 0, 1, or 2 resonance conditions) 
in box-orbit phase space
shifted as the perturbation increased, from $K=0$ (box orbits) 
at small perturbations,
to $K=1$ (thin boxes) at moderate perturbations,
to $K=2$ (boxlets) at large perturbations.

\subsection{Periodic Orbits}

As discussed above, the objects that give phase space its 
structure in 3 DOF systems are the resonant tori which satisfy a 
condition $l\omega_1+m\omega_2+n\omega_3=0$ between the three 
fundamental frequencies.
The corresponding orbits are thin tubes or thin boxes.
However most studies of the motion in triaxial potentials have 
focussed on the principal planes, in which resonances 
(i.e. $l\omega_1 + m\omega_2=0$) are equivalent to closed orbits.
Figure 3a suggests that most of the regular orbits in a 
triaxial potential are associated with thin orbits rather than 
with closed orbits.
The closed orbits are nevertheless worth studying for a 
number of reasons.
Gas streamlines must be non-self-intersecting, which restricts 
the motion of gas clouds to closed orbits like the $1:1$ loops.
Elongated boxlets like the bananas have shapes that make them 
very useful for reproducing a barlike mass distribution, and 
one expects such orbits to be heavily populated in 
self-consistent models.
The fraction of regular orbits associated with closed 
orbits (as opposed to thin orbits) 
also tends to increase as the phase space becomes more and 
more chaotic, as discussed below.

\subsubsection{Nonrotating Potentials}

Periodic orbits in the principal planes of triaxial models often 
first appear as bifurcations from the axial orbits.
Figure 5 is a representative bifurcation diagram for axial orbits 
in nonrotating triaxial models with finite central forces, based on the 
studies of Heiligman \& Schwarzschild (1979), Goodman \& Schwarzschild (1981), 
Magnenat (1982a), Binney (1982a), Merritt \& de Zeeuw (1983), 
Miralda-Escud\'e \& Schwarzschild (1989), Pfenniger \& de Zeeuw (1989), 
Lees \& Schwarzschild (1992), Papaphilippou \& Laskar (1996, 1998), 
and Fridman \& Merritt (1997).
When the central force is strongly divergent, as would be the case in a 
galaxy with a central black hole, the axial orbits are unstable 
at all energies but many of the other periodic orbits in 
Figure 5 still exist.

\begin{figure}
\plotfiddle{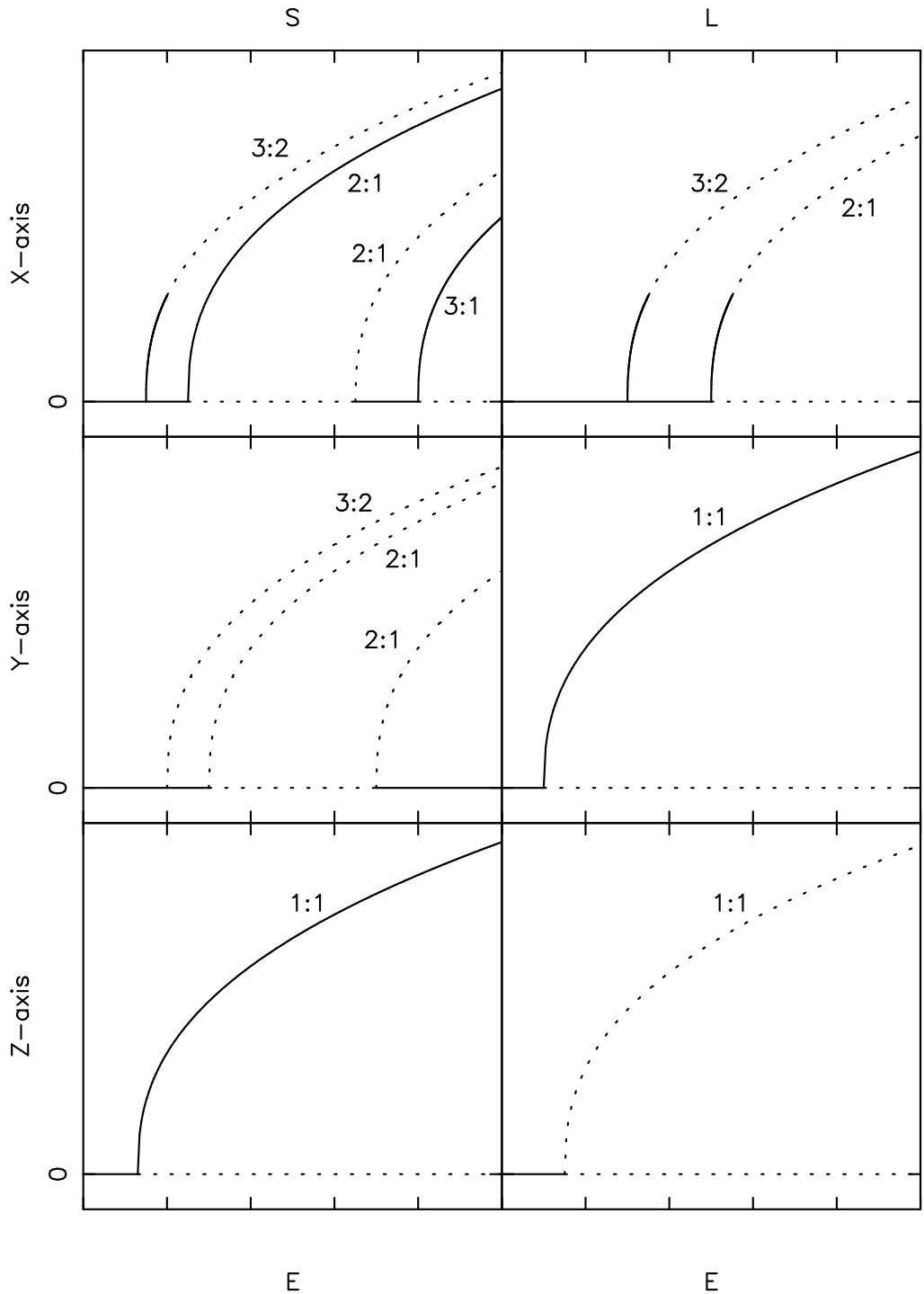}{7.in}{0.}{75}{75}{-220}{-40}
\caption{
Representative bifurcation diagram for planar periodic orbits in
non-axisymmetric potentials with finite central forces.
The three rows show bifurcations from the $x$ (long), $y$ (intermediate)
and $z$ (short) axes.
The left column (S) shows bifurcations in the plane containing the shorter
of the two remaining axes, while the right column (L) shows bifurcations
in the plane containing the longer axis.
Solid lines correspond to stable orbits and dotted lines to unstable orbits,
where stability is defined with respect to motion in the appropriate plane.
}
\end{figure}

At low energies in a harmonic core, all three axial orbits are 
stable.
The axial orbits typically change their stability properties at 
the $n:1$ bifurcations where the frequency of oscillation equals 
$1/n$ times the frequency of a transverse perturbation.
This typically first occurs first along the $y$ (intermediate) axis when the 
frequency of $y$ oscillations falls to the frequency of an 
$x$ perturbation, producing a $1:1$ bifurcation.
The $y$-axis orbit becomes unstable and a closed $1:1$ 
loop orbit appears in the $x-y$ plane. 
The $x-y$ loop is initially elongated in the direction of the $y$ 
axis but becomes rapidly rounder with increasing energy.
Similar $1:1$ bifurcations occur at slightly higher energies from the 
$z$ axis: first in the $y-z$ plane, then in the $x-z$ plane, 
producing two more planar loop orbit families.
The $x-z$ loop is typically unstable to vertical perturbations; the 
other two loops are generally stable and generate the long- and 
short-axis tube orbit families.

The $x$-axis orbit does not experience a $1:1$ bifurcation since its 
frequency is always less than that of perturbations along either 
of the two shorter axes.
However at sufficiently high energies, the $x$ oscillation frequency 
falls to $1/2$ the frequency of a $z$ perturbation and the $2:1$ 
$x-z$ banana orbit appears (Figure 6).
At still higher energies, a second $2:1$ 
bifurcation produces the $x-y$ banana; following this bifurcation, 
the $x$-axis orbit is typically unstable in both directions.
This second bifurcation occurs most readily in nearly 
prolate models where the $y$ and $z$ axes are nearly equal in 
length.
The energy of the first, $x-z$ bifurcation is a strong function of 
the degree of central concentration of the model, as measured for 
instance by the cusp-slope parameter $\gamma$ in Dehnen's model.
For $\gamma\gap 1$, the $x$-axis orbit is unstable at most 
energies in all but the roundest triaxial models.

\begin{figure}
\plotfiddle{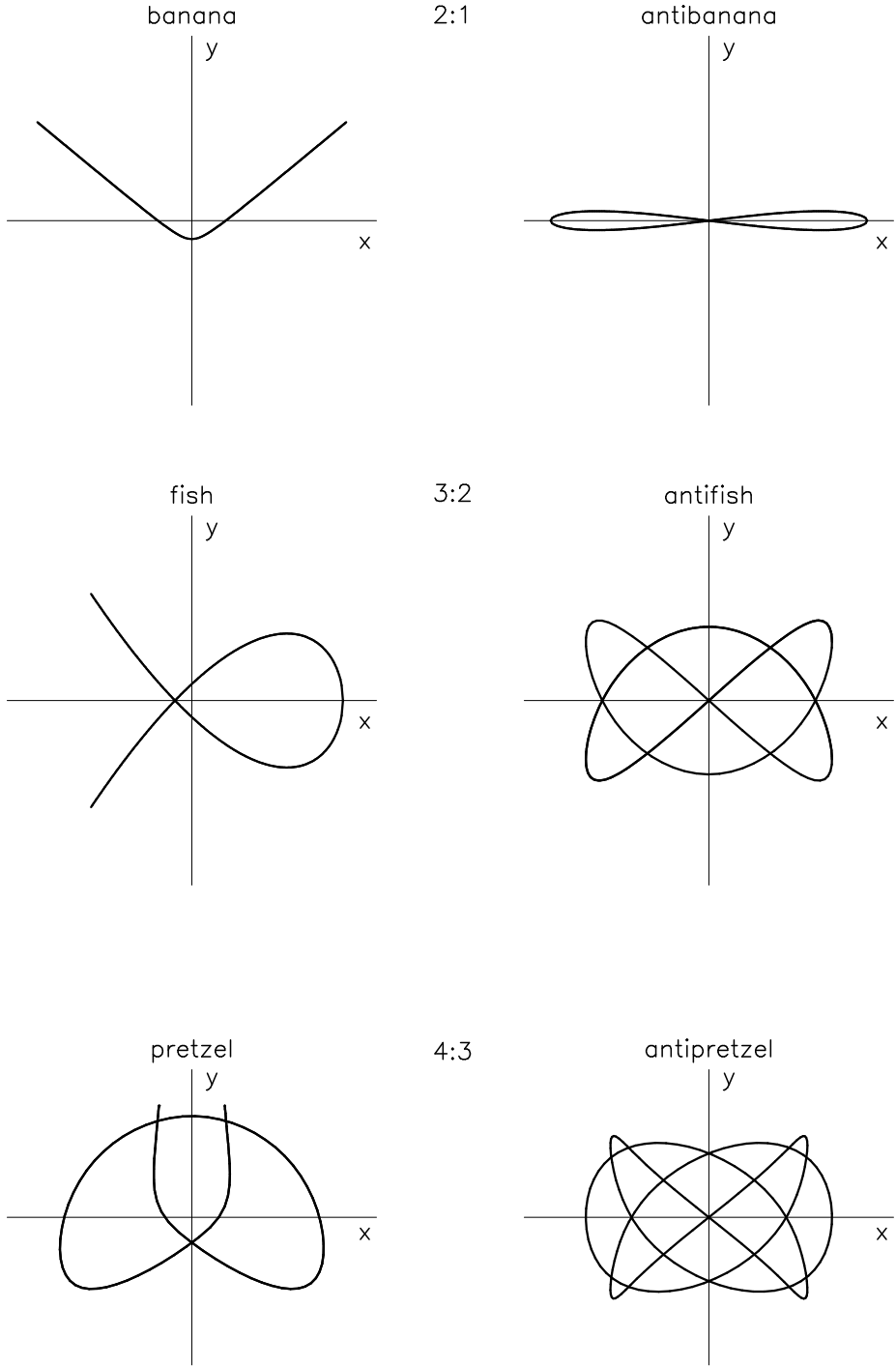}{7.5in}{0.}{80}{80}{-260}{-50}
\caption{
Resonant box orbits in a principal plane of a triaxial potential.
Left: centrophobic boxlets; these are typically stable.
Right: centrophilic boxlets; these are typically unstable.
}
\end{figure}

In highly elongated models, $c/a\lap 0.45$, the $x$ axis orbit 
can return to stability at the bifurcation of the ``antibanana'' 
orbit, a $2:1$ resonant orbit that passes through the 
center (Figure 6).
Miralda-Escud\'e \& Schwarzschild (1989) call such orbits 
``centrophilic;'' resonant orbits that avoid the center, like the 
$x-z$ banana, are ``centrophobic.''
In nearly oblate models, this return to stability in the $x-z$ 
plane causes the $x$-axis orbit to become stable in both 
directions.
The $x$-axis orbit can become unstable once more at still higher 
energies when $c/a$ is sufficiently small, through the appearance of 
a $3:1$ resonance in the direction of the $z$ axis. 

Both the $y$- and $z$-axis orbits are unstable at all energies 
above the $1:1$ bifurcations.
The $y$-axis orbit can become unstable in both directions 
following the $2:1$ bifurcation that produces the $y-z$ banana.
In highly elongated models, the $y$-axis orbit returns briefly to 
stability in the $z$-direction through the appearance of the 
$y-z$ anti-banana.

Additional bifurcations of the form $n:m$, with both $n$ and $m$ 
greater than one, can occur from the axial orbits.
These bifurcations typically do not affect the
stability of the axial orbits but they are nevertheless 
important because they generate additional families of periodic orbits.
The name ``boxlet'' was coined by Miralda-Escud\'e \& Schwarzschild (1989)
for these orbits (Figure 6).
The $3:2$ boxlets are ``fish,'' the $4:3$ boxlets are ``pretzels,''
etc.
Only the fish orbits have been extensively studied; they first
appear as $3:2$ bifurcations from the $x$-axis orbit 
($x-z$ and $x-y$ fish) or the $y$-axis orbit ($y-z$ fish),
typically at energies below the banana
bifurcation. 
The $x-y$ fish is only important in highly prolate models.

The boxlets can themselves become unstable, either to 
perturbations in the plane of the orbit or to vertical 
perturbations.
The $x-z$ banana exists and is stable over a wide range of model 
parameters; it becomes unstable only in nearly prolate models, 
through a vertical $2:1$ bifurcation. 
The $x-y$ and $y-z$ bananas are usually vertically unstable; the 
$x-y$ banana returns to stability only in highly elongated, 
nearly prolate models.
In nearly oblate models, the $x-z$ fish first becomes unstable to 
perturbations in the orbital plane, while for strongly triaxial 
and prolate models instability first appears in the vertical 
($y$) direction.
The $x-y$ fish is only important in strongly prolate models; in
strongly triaxial or oblate models, it either does not exist, 
or is generally unstable to vertical perturbations.
The $y-z$ fish is almost always vertically unstable.

As the central concentration of a triaxial model increases, the 
axial orbits become unstable over a progressively wider range of 
energies.
In models with steep central cusps or central black holes, the 
axial orbits are unstable at all energies and the (centrophobic)
boxlets may extend all the way to the center.
Little systematic work has been done on the properties 
of boxlets in such models.
Miralda-Escud\'e \& Schwarzschild (1989) investigated the orbit 
structure in the principal planes of two triaxial models with 
$\Phi=\log\left(R_c^2+m^2\right)$ as $R_c$ was varied; for $R_c=0$, 
their models have an $r^{-2}$ central density cusp.
Miralda-Escud\'e \& Schwarzschild found that the lowest-order boxlets 
continued to exist, typically with no change in their stability 
properties, as $R_c$ was reduced to zero.
The centrophobic boxlets were also found to be not strongly affected 
by the addition of a central point mass.
Pfenniger \& de Zeeuw (1989) investigated the banana boxlets in a 
second family of models with $\gamma=2$ cusps.
They found that the bananas became strongly bent for model axis 
ratios $c/a\lap 0.65$, making them less useful for reconstructing 
the model shape.
In their survey of orbits in two triaxial models with Dehnen's 
density law, Merritt \& Fridman (1996) 
found that the $x-z$ fish was vertically unstable at most energies for 
$\gamma=1$ and at all energies for $\gamma=2$.
Only the $x-z$ fish, of all the planar boxlet families, remained 
stable over an appreciable energy range in the two models 
investigated by them.
A number of higher-order resonances outside of the principal 
planes were also found to be important, including the $4:5:7$, 
$5:6:8$ and $6:7:9$ boxlets.

\subsubsection{Rotating Models}

A great deal of work has been done on periodic orbits in models 
of barred galaxies 
(\cite{cog89}; \cite{sew93}).
Most of this work has been restricted to motion in the equatorial
plane; 
furthermore, models of barred galaxies typically have low central 
concentrations, mild departures from axisymmetry (``weak bars''), 
and rapid rotation, making them poor representations of elliptical galaxies.  
The smaller number of studies of periodic orbits in rotating 
elliptical galaxy models have focussed on two low-order resonant
families:
the $1:1$ orbits that bifurcate from the axial orbits, 
giving rise to the tubes; and the $1:2$ banana orbits.

An early study of the closed orbits in Schwarzschild's (1979) 
triaxial model (\cite{mer79}) revealed that the $1:1$ orbits 
circling the long axis continue to exist when the figure rotates, 
but are tipped by the Coriolis force out of the $y-z$ plane.
The tip angle increases with energy and the family terminates 
when it mergers with the retrograde closed orbits in the equatorial plane.
Heisler, Merritt \& Schwarzschild (1982) called these tipped 
orbits ``anomalous'' and noted the existence of a second, 
unstable family of anomalous orbits circling the intermediate 
axis. 
A large number of additional studies have 
traced the existence and stability of these orbits in a variety 
of rotating triaxial potentials (\cite{bin81}; \cite{tod82};
\cite{mag82a}, b; \cite{dem83}; \cite{dur83}; \cite{mez83};
\cite{mul83}; \cite{muh84}; \cite{maz88}; \cite{cle89}; \cite{paz90};
\cite{hpn93}).
One motivation for this work has been the expectation that 
gas or dust falling into a rotating triaxial galaxy might dissipatively 
settle onto the anomalous orbits before finding its way into the 
center (\cite{tod82}; \cite{vks82}).

The existence of the anomalous orbits is tied to the vertical 
instability of the orbits in the equatorial plane.
There are four major families of planar orbits: 
two of these (the ``$x_1$'' and ``$x_4$'' families in 
Contopoulos's convention), both prograde, remain close to the $x$ 
and $y$ axes respectively; the other two ($x_2$ and $x_3$) are 
prograde and retrograde counterparts of the $1:1$ orbit that 
bifurcates from the $y$ axis in nonrotating models.
One of these -- either the retrograde orbit, if figure rotation 
is about the short axis, or the prograde orbit, if rotation is 
about the long axis -- is unstable to perturbations out of the 
plane of rotation over a wide radial range.
In either case, a family of anomalous orbits exists that is 
inclined to the plane of rotation; the family has two branches 
that are tipped in opposite directions with respect to the equatorial 
plane but which circle the short axis in the same sense.
The anomalous orbits first appear at low energies as $1:1$ 
bifurcations of the $z-$ axis orbit.
In models without cores, the $z$-axis orbit may be unstable at 
all energies and the anomalous orbits extend into the center.
In models with sufficiently high rotation, on the other hand,
the anomalous orbits can
bifurcate twice from the family in the plane, never reaching the 
$z$-axis.
The unstable anomalous orbits that circle the intermediate axis 
first appear at the bifurcation of the $1:1$ orbit in the $x-z$ plane.

Miller \& Smith (1979) investigated the motion of particles in 
a 3D, rotating $N$-body bar.
They found that nearly half of the orbits could be 
described as respecting a $2:1$ resonance between 
the $x$ and $z$ motions.
Pfenniger \& Friedli (1991) likewise found a large population of
$2:1$ orbits in their rotating $N$-body models.
This family of orbits, which reduces to the $x-z$ bananas in the 
case of zero figure rotation, has been studied by a number of 
other authors including Mulder \& Hooimeyer (1984), Cleary (1989), 
Martinet \& Udry (1990), Patsis \& Zachilas (1990), Udry (1991),
and Hasan, Pfenniger \& Norman (1993).
The family bifurcates from the prograde $x_1$ family at low 
energies and so retains the $x-y$ commensurability of that 
orbit -- either $1:1$ (low energies) or $3:1$ (high energies).
The radial range over which it exists decreases with increasing 
rotation of the figure.
At large energies, the $x_1$ orbit returns to vertical stability 
through bifurcation of the $x-z$ anti-banana.
At even greater energies, both banana families may merge again 
with the $x_1$ orbits.
Udry (1991) traced the existence and stability of the $2:1$ 
orbits that bifurcate from the $y$ and $z$ axes and from the 
other $1:1$ families in the equatorial plane.
None of these families is likely to be as important as the $x-z$ 
bananas in self-consistent models.

\subsection{Stochasticity}

\subsubsection{Origin}

Stochasticity always occurs near an unstable resonant torus.
Such tori exist and are dense even in integrable systems, but 
the motion in their vicinity is (by definition) regular; 
under perturbation of the potential, however, resonant tori are destroyed 
and the motion undergoes a qualitative change.
The reason for this change is suggested by Figure 7, which shows 
an orbit that lies close to the short- ($y$-) axis orbit in the potential 
of a 2D bar.
Such an orbit can be represented as the superposition of a rapid 
radial oscillation and a slow angular rotation, with frequencies 
$\omega_r$ and $\omega_{\theta}$ respectively; $\omega_{\theta}$ tends to zero 
as the orbit approaches the axial orbit.
Resonances between the fast and slow motions satisfy a condition
$l\omega_{\theta} - m\omega_r = 0$.
As $\omega_{\theta}$ tends to zero, the separation between neighboring 
resonances, defined by $l$ and $l+1$, also becomes small, 
resulting in motion that is extremely complex.
In fact one can show that the motion 
is generically no longer confined to tori; instead the trajectory moves 
more-or-less randomly through a region of dimensionality $2N-1$, 
where $N$ is the number of degrees of freedom.
Among the properties of the motion in these stochastic regions is 
exponential divergence of initially close trajectories 
(\cite{mil64}; \cite{ose68}), resulting in extreme
sensitivity to small changes in the initial conditions.

\begin{figure}
\plotone{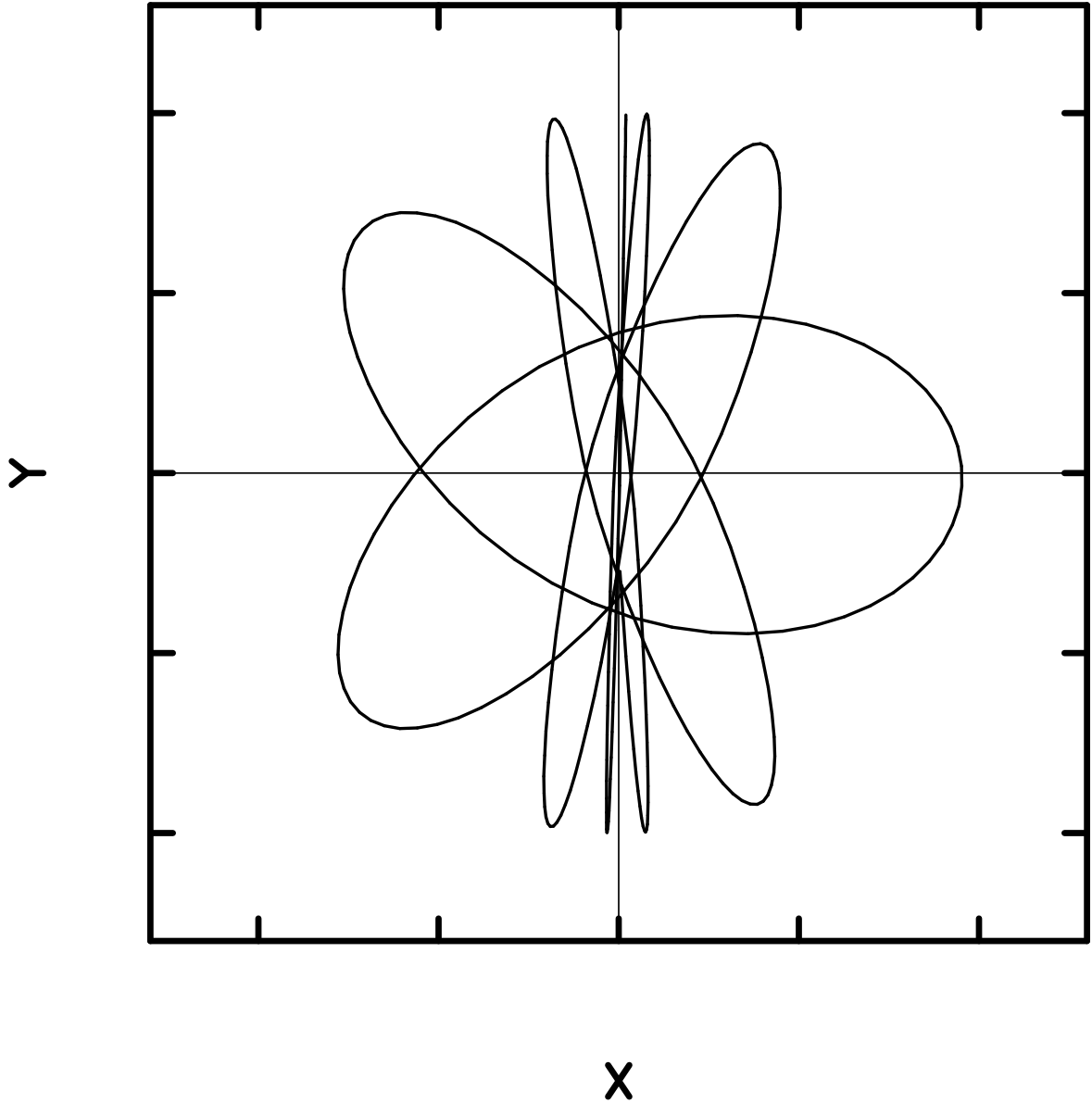}
\caption{
An orbit that lies close to the unstable, short- ($y$-) axis orbit in a
2D barred potential.
The motion consists of a rapid radial oscillation combined with a slow
rotation.
Resonances between the fast and slow motions produce a stochastic
zone where the motion is exponentially unstable.
}
\end{figure}

An early indication of the importance of stochasticity in triaxial 
potentials was the discovery that some of the orbits in 
Schwarzschild's (1979) triaxial model yielded different 
occupation numbers when integrated using a different computer
(\cite{mer80}),
a consequence of the exponential divergence just mentioned.
Goodman \& Schwarzschild (1981) computed rates of divergence 
between pairs of box-orbit trajectories at one energy in this 
potential; they found that about a fourth of the orbits were stochastic.
Most of the unstable orbits had their starting points near the 
$z$ axis, producing a narrow strip of instability
in initial condition space; a similar stochastic strip can be 
seen in Figure 3a.
Goodman \& Schwarzschild showed  
that both the $y$ and $z$ axis orbits were linearly unstable at 
this energy while the $x$ axis orbit was stable.
They speculated that the stochasticity was connected with
this instability; 
in particular, they noted that the $z$-axis orbit 
was strongly unstable to perturbations in both directions,
following the bifurcation of the $1:1$ loop 
orbits in the $y-z$ and $x-z$ planes.

The stochasticity of the motion near the short axis orbit
(as opposed to the instability of the axial orbit, a weaker 
condition)
was established by Gerhard (1985) using a method due to Melnikov (1963).
Melnikov's method is based on the qualitative change that occurs 
in the shape of the phase curve, or ``separatrix,'' of an 
unstable 2D orbit in the surface of section when the motion in its 
vicinity becomes stochastic.
In a two-dimensional integrable potential, 
the separatrix is a smooth curve that 
intersects itself at the fixed (``homoclinic'') point 
corresponding to an unstable orbit like the one in Figure 7.
Under perturbation, the two branches of the separatrix
corresponding to motion toward and away from the fixed point 
need not join smoothly; instead they can oscillate wildly, producing an 
infinite set of additional homoclinic points.
The stochasticity may be interpreted as a consequence of this 
complex behavior.
Melnikov defined an integral quantity that measures the 
infinitesimal separation between the two branches of the separatrix
that approach and depart from the unstable fixed point.
If this distance changes sign as one proceeds around the separatrix, 
nonintegrability has been rigorously established.
Gerhard took as his integrable potential a 2D St\"ackel 
model in which the total density fell off as $r^{-2}$ and the 
non-axisymmetric component as $r^{-4}$.
He showed that certain perturbations, e.g. density terms varying as 
$\cos(m\phi)$ with $m$ odd, produced a rapid breakdown in the 
regularity of the motion near the short-axis orbit.

Martinet \& Udry (1990) noted that most of the stochasticity in 
slowly-rotating planar models appeared to be associated with the $x_3$ 
family, which reduces to the unstable short-axis orbit 
in the limit of zero rotation.
They traced the interactions of higher-order resonances in the 
vicinity of the $x_3$ orbit and argued that the development of a large 
stochastic layer in the surface of section could be attributed to 
the simultaneous action of these resonances.
Martinet \& Udry found that the stochastic region associated with this orbit 
tended to contract as the rate of figure rotation increased.

The $1:1$ bifurcation that first induces instability in the $z$-axis 
orbit occurs at lower and lower energies as the central 
concentration of a triaxial model is increased.
At the same time, the sensitivity of the axial orbits to small 
perturbations increases as the central force 
steepens.
These two facts imply a greater role for stochasticity 
in triaxial models with high central densities.
\footnote{
By carefully adjusting the shape of the equipotential curve
as a function of radius, planar non-axisymmetric models can be 
constructed in which the motion is fully regular even when the central 
density is divergent (\cite{srt97}).
These models are probably too special to be relevant to real 
galaxies.}
Gerhard \& Binney (1985) investigated 2 DOF motion in a 
bar with a central density cusp, $\rho\sim r^{-\gamma}$, or a 
central point mass $M_h$ representing a supermassive black hole.
Using surfaces of section, they found that the fraction of phase space 
associated with stochastic motion increased with increasing 
$\gamma$ or $M_h$.
However, regular boxlike orbits persisted in families associated with 
low-order resonances like the $2:1$ banana.
Miralda-Escud\'e \& Schwarzschild (1989) studied the orbits in a 
planar logarithmic potential as a function of core radius $R_c$.
Even for $R_c=0$, corresponding to a $\rho\sim r^{-2}$ density 
cusp, the stochastic regions were found to be confined to narrow 
filaments in the surface of section between the resonant families.
Papaphilippou \& Laskar (1996) likewise found only a modest 
degree of stochasticity in their study of 2D logarithmic 
potentials with cores.
Touma \& Tremaine (1997) obtained similar results in the planar 
logarithmic potential using an approximate mapping technique.

Studies like these of 2 DOF motion tend to underestimate the 
importance of stochasticity in triaxial potentials,
for at least three reasons.
Periodic orbits that generate families of regular trajectories 
in the principal planes are often unstable to perturbations out of 
the plane; examples are the $x-y$ and $y-z$ bananas discussed 
above.
Second, there are unstable periodic orbits that exist 
outside of the principal planes; some examples can be seen
in Figure 3.
Third, much of the stochasticity in 3 DOF systems is 
associated with resonances between the three degrees of freedom
which have no analog in 2 DOF systems.

Techniques like surfaces of section that work well in 2 DOF 
systems can become cumbersome in three dimensions.
An alternative technique for identifying an orbit as stochastic is 
to compute its Liapunov exponents, the mean exponential
rates of divergence of a set of trajectories surrounding it.
In a 3 DOF system there are six Liapunov
exponents for every trajectory, corresponding to the six
dimensions of phase space; the exponents come in pairs of opposite sign, 
a consequence of the conservation of phase space volume.
Of the three independent exponents, one -- corresponding to
displacements in the direction of the motion -- is always zero.
The two remaining exponents, $\sigma_1$ and
$\sigma_2$, may be seen as defining the time-averaged
divergence rates in two directions orthogonal to the trajectory.
For a regular orbit, $\sigma_1=\sigma_2=0$; for a stochastic
orbit, at least one (and typically two) of these exponents is nonzero.

Udry \& Pfenniger (1988) computed all six Liapunov exponents for 
ensembles of orbits in triaxial potentials based on a modified 
Hubble law, which has a constant-density core and a large-radius 
density dependence of $r^{-3}$.
To this model they added perturbing forces of various kinds, 
corresponding to central mass concentrations and angular 
distortions of the density.
Udry \& Pfenniger presented their results in terms of
histograms of Liapunov exponents without distinguishing between 
different types of orbit or between orbits of different 
energies.
Nevertheless they found that the average degree of stochasticity 
for a given potential always increased with increasing strength of the 
perturbation.
Schwarzschild (1993) computed the largest Liapunov exponent for 
the orbits in his scale-free, $\rho\propto r^{-2}$ triaxial models 
discussed below.
He found that most of the boxlike orbits, and a modest but significant 
fraction of the tube orbits, were stochastic; the $y-z$ 
instability strip first discussed by Goodman \& Schwarzschild 
(1981) was typically greatly enlarged in the scale-free models.
Merritt \& Fridman (1996) demonstrated that the boxlike orbits in a 
triaxial model with a cusp as shallow as $\rho\sim r^{-1}$ were 
also mostly stochastic, though many of these orbits mimicked regular 
orbits for tens or hundreds of orbital periods.
They noted that much of the stochasticity was generated by 
resonances not associated with the $z$-axis orbit.
Merritt \& Valluri (1996) showed that the histogram of Liapunov 
exponents of box orbits at a given energy tended toward a narrow 
spike as the integration time increased, consistent with the 
expectation that the stochastic parts of phase space are 
interconnected via the Arnold web.
They found that both $\sigma_1$ and $\sigma_2$ were significantly 
nonzero for all of their orbits, suggesting that no isolating integrals
existed aside from the energy.

\subsubsection{Transition to Global Stochasticity}

The KAM theorem guarantees that motion in the vicinity of a ``very 
nonresonant'' torus in an integrable system will remain regular under
small perturbations of the potential.
Since the majority of tori are very nonresonant -- in the 
sense that rational numbers are rare compared to irrational ones
-- the KAM theorem says that a small perturbation of an 
integrable potential will leave most of the trajectories regular.
At the same time, the resonant tori are dense in the phase space, 
and even an infinitesimal perturbation can cause the motion in their 
vicinity to become chaotic.
When the perturbation is small, these chaotic regions are narrow 
and the motion within them tends to mimic regular motion for many
oscillations.
As the perturbation is increased, the chaotic zones from 
different resonant tori become wider and neighboring 
zones begin to overlap.
Eventually one expects to find large regions of interconnected
phase space where the 
motion is fully stochastic (\cite{heh64}).

\begin{figure}
\plotfiddle{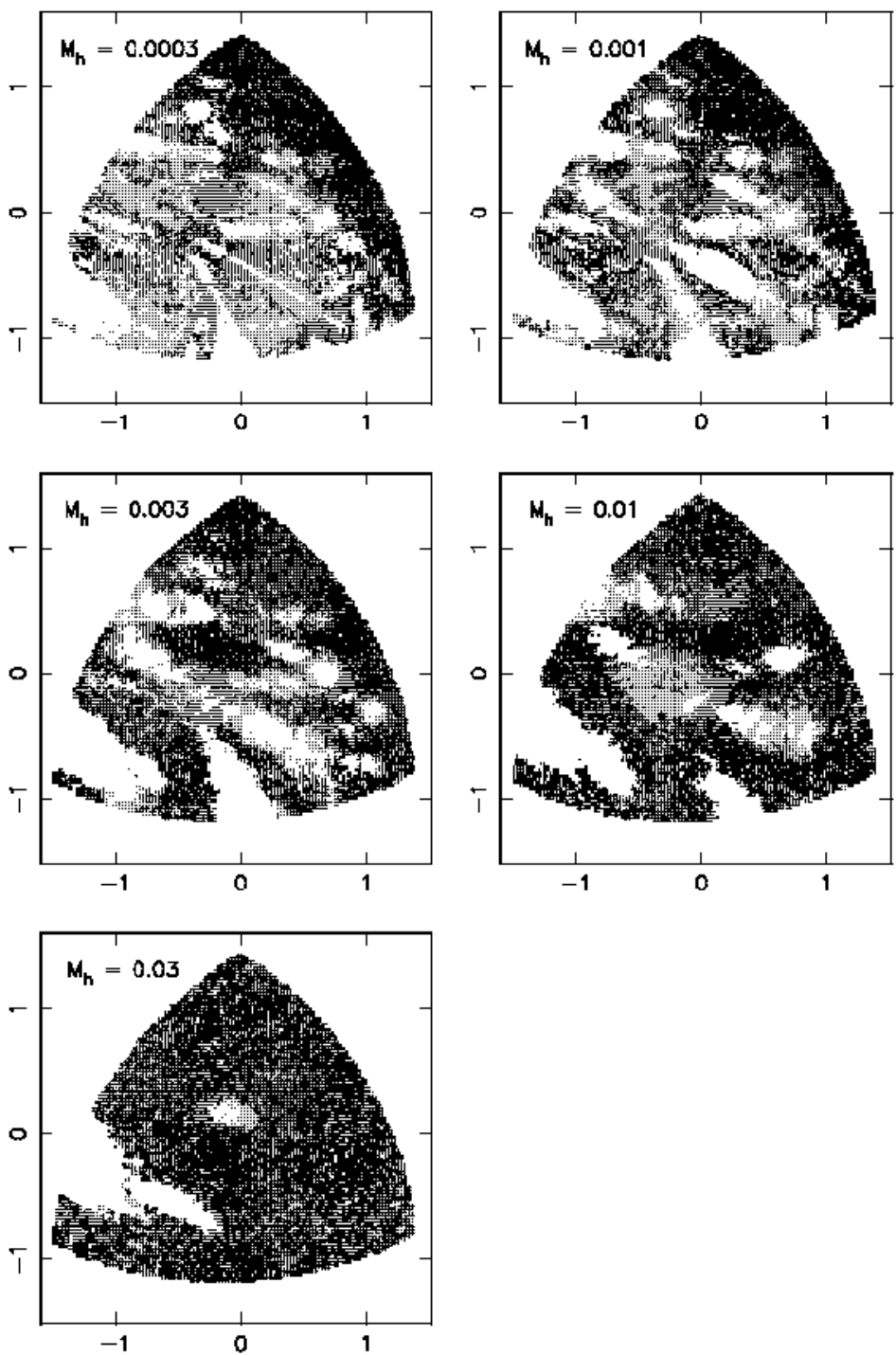}{7.in}{0.}{75}{75}{-220}{-30}
\caption{
Transition to global stochasticity in stationary (box-orbit) initial
condition space (Valluri \& Merritt 1998).
Each frame shows an octant of the equipotential surface, at roughly the
half-mass radius, in a triaxial Dehnen model with $\gamma=0.5$
and with a central point containing a fraction $M_h$ of the total mass of
the model.
The $z$- (short) axis is vertical and the $x$- (long) axis is to the left.
The grey scale is proportional to the logarithm of the diffusion rate
in frequency space; white regions correspond to regular orbits.
When $M_h> 0.02$, the motion is essentially fully stochastic at this
energy.
}
\end{figure}

The studies summarized above suggest that the behavior of box
orbits in triaxial potentials can be strongly influenced by the
steepness of the central force law, and it is natural to ask
how centrally concentrated a triaxial model must be
before the parts of phase space associated with boxlike orbits 
become globally stochastic.
Valluri \& Merritt (1998) investigated this question in triaxial 
Dehnen models, taking as their perturbation parameters the slope 
$\gamma$ of the central density cusp and the mass $M_h$ of a 
central point representing a nuclear black hole.
They measured stochasticity via the change $\Delta\omega$ of the 
fundamental frequencies of an orbit computed over a fixed interval 
of time; for a regular orbit $\Delta\omega=0$, while for a 
stochastic orbit the ``fundamental frequencies'' gradually change 
with time (\cite{las93}).
A quantity like $\Delta\omega$ is a more useful diagnostic of 
stochasticity than the Liapunov exponents since it measures a
finite, rather than infinitesimal, displacement of an orbit due 
to stochastic diffusion.
Valluri \& Merritt found a fairly well-defined transition to global 
stochasticity in the phase space of boxlike orbits 
as $M_h$ increased past $\sim 2\%$ of the galaxy mass
(Figure 8).
In models without a central point mass, the transition appeared 
to begin when the central cusp slope $\gamma$ reached $\sim 2$, 
close to the largest value seen in real galaxies.
Valluri \& Merritt constructed histograms of diffusion rates in 
frequency space and found that weakly chaotic potentials 
exhibited a $\sim 1/\Delta\omega$ distribution, extending over at 
least six decades in $\Delta\omega$.
As $M_h$ or $\gamma$ was increased, this distribution flattened, 
although every model investigated by them contained a significant 
number of slowly-diffusing stochastic orbits.
Valluri \& Merritt estimated that stochastic orbits would induce significant 
changes in the shape of a triaxial galaxy over its lifetime if 
$\gamma\gap 2$, or if $M_h$ exceed $\sim 0.5\%$ times the galaxy mass.

Papaphilippou \& Laskar (1998) used the same technique 
to compute diffusion rates in frequency space for ensembles of 
orbits in the logarithmic triaxial potential.
They chose a fixed energy and core radius $R_c$ such that the maximum 
amplitude of the motion was approximately $8R_c$;
as perturbation parameters they chose the axis ratios of the 
model.
They found a striking increase in the diffusion rates of boxlike 
orbits in frequency space as the models became flatter.

\subsection{Self-Consistent Models}

Following Schwarzschild (1979, 1982), a standard technique for 
constructing stationary triaxial models has been to 
integrate large numbers of orbits for $\sim 10^2$ 
periods, store their time-averaged densities in a set of cells, 
and then reproduce the known mass of the model in each cell via 
orbital superposition.
The technique is best suited to integrable potentials 
in which time-averaged densities correspond to uniformly 
populated tori (\cite{van84}; \cite{sta87}).
In non-integrable potentials, a decision must be made about how 
to populate the stochastic parts of phase space.
At one extreme, if stochastic diffusion rates are high, 
it is appropriate to assume that the stochastic parts of phase 
space at each energy are uniformly populated.
In a weakly chaotic potential, on the other hand, stochastic 
orbits should be treated more like regular orbits since they remain 
confined to restricted regions of phase space over astronomically 
interesting time scales.
Since orbits with a wide variety of shapes are useful for 
reproducing the density of a triaxial model, 
one expects the range of self-consistent solutions to be strongly 
dependent on how much of phase space is stochastic and on how
the stochastic orbits are treated.

Schwarzschild (1979, 1982) emphasized the regular appearance of most of 
the orbits in his modified Hubble model, a consequence of its 
relatively low degree of central concentration.
He found that self-consistency required appreciable numbers of 
orbits from both the box and tube families, a conclusion reached
also by Statler (1987) who constructed models based on the Perfect 
Ellipsoid.
The most strongly triaxial models in Statler's survey were 
dominated by box orbits; tube orbits appeared in significant numbers
only when the geometry was favorable, i.e. when the figure was nearly 
oblate or prolate.
Levison \& Richstone (1987) likewise found a predominance of regular or 
nearly-regular orbits in their model-building study based on the 
logarithmic potential with a finite-density core.
Levison \& Richstone noted that many of the box orbits failed to exhibit 
reflection symmetry, a likely indication that they were 
associated with resonances.

Kuijken (1993) carried out a self-consistency study of 
scale-free, non-axisymmetric disks.
His models had a surface density
\begin{equation}
\Sigma(x,y) = \left[x^n + (y/q)^n\right]^{-1/n}
\end{equation}
with $n=2$ or $4$; the asymptotic dependence is 
$\Sigma\propto r^{-1}$, corresponding to a logarithmic density 
law, and the parameter $n$ fixes the shapes of the isodensity 
contours, which are elliptical for $n=2$ and become more boxy 
with increasing $n$.
Kuijken found, in agreement with the earlier studies discussed 
above, that the 2D motion in triaxial potentials was mostly 
regular; however almost all of the boxlike orbits could be 
identified with one of the low-order resonant boxlets in Figure 6.
He found that only axis ratios $q\gap 0.7$ could be 
self-consistently reproduced for $n=2$, and setting $n=4$ 
restricted the allowed range of shapes even more.
Syer \& Zhao (1998) carried out a study similar to Kuijken's 
based on the scale-free disk models of Sridhar \& Touma (1997), 
which have $\Sigma(R,\theta) = R^{\alpha-1}S(\theta)$ for 
$0\le\alpha\le 1$.
The shape function $S(\theta)$ in Sridhar \& Touma's models 
is determined, for a given $\alpha$, by the requirement that the 
motion be fully regular; the corresponding models are 
peanut-shaped and highly elongated, with short-to-long axis 
ratios of $\sim 1/3$.
Syer \& Zhao found that no choice of the parameter $\alpha$ 
permitted a self-consistent solution, a result which they 
ascribed to the restricted range of shapes of the boxlike orbits.

\begin{figure}
\plotfiddle{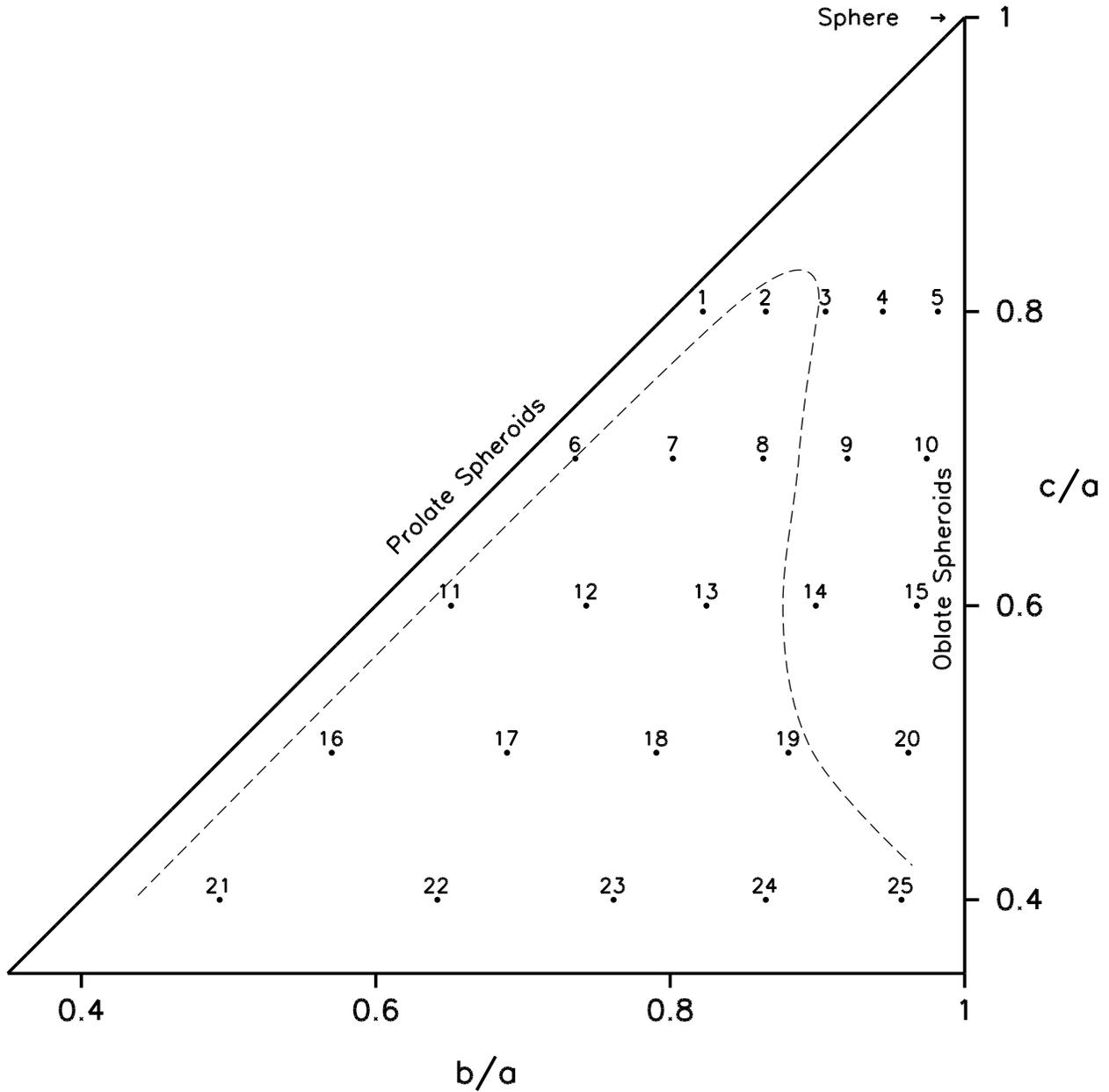}{7.in}{0.}{95}{95}{-290}{-150}
\caption{
Allowed axis ratios for self-consistent triaxial models with
Dehnen's density law and $\gamma=2$, a ``strong cusp''
(Merritt 1997).
Only regular orbits were allowed in the solutions.
The dotted curve denotes the approximate limit of solution space;
points below correspond to models for which no self-consistent
solution was found.
Since many of the boxlike orbits in these models are stochastic,
and since box orbits are necessary for maintaining triaxial figures,
only nearly oblate and prolate solutions were found.
}
\end{figure}

Schwarzschild (1993) carried out a study similar to Kuijken's in
three dimensions.
He chose the scale-free logarithmic potential
with six different values for the axis ratios of the figure, 
from nearly oblate to nearly prolate; four of these were
highly flattened.
Stochasticity was identified via the largest Liapunov exponent 
$\sigma_1$; as discussed above, a large fraction of the boxlike 
orbits were found to be stochastic in all of the model 
potentials.
Schwarzschild was able to find self-consistent equilibria for 
each of his six models when all the orbits -- both regular and 
stochastic -- were included.
In these models, the stochastic orbits were treated precisely 
like regular orbits, i.e. each was assigned a unique 
distribution of mass defined as the population of a set 
of grid cells averaged over $\sim 50$ orbital periods.
When the stochastic orbits were excluded, only the rounder models 
could be reconstructed.
Schwarzschild then estimated how rapidly the shapes of the models 
containing stochastic orbits would change due to their
non-uniform population of phase space.
He re-integrated the stochastic orbits for three times the 
interval of the original integration and recorded the change in 
their spatial distribution.
This evolution was found to result in modest but significant
changes in the overall shapes of the models.

Schwarzschild's scale-free models were designed to represent galactic 
halos, regions of low stellar density where a typical star will 
complete only a few tens of oscillations over the age
of the universe.
Closer to the centers of elliptical galaxies, orbital periods 
fall to 1\% or less of a galaxy lifetime; since the rates of stochastic 
diffusion scale approximately with orbital frequencies, one expects the 
central parts of a triaxial galaxy to be much more strongly affected 
by stochasticity than the outer parts.
Merritt \& Fridman (1996) investigated the self-consistency of 
non-scale-free triaxial models obeying Dehnen's density law with 
$\gamma=1$ (``weak cusp'') and $\gamma=2$ (``strong cusp'') and
$c/a=0.5$.
Treating the stochastic orbits like regular orbits yielded 
self-consistent solutions for both models; the regular 
orbits alone failed to reproduce either mass model.
Merritt \& Fridman then attempted to construct more nearly stationary 
solutions in which the stochastic parts of phase space were 
populated in a uniform way, especially at low energies,
by combining all of the stochastic orbits at a given energy into a 
single ensemble.
They found that no significant fraction of the mass could 
be placed on these ``fully mixed'' orbits in the strong-cusp 
model.
Merritt (1997) extended this study to strong-cusp models
with a range of shapes.
He found (Figure 9) that only fairly oblate, prolate or 
spherical models could be constructed using the regular orbits 
alone.

\section{COLLISIONLESS RELAXATION}
Elliptical galaxies are collisionless systems, in the 
sense that the forces between individual stars are 
unimportant compared with the mean gravitational field.
Most elliptical galaxies are nevertheless smooth and relaxed in 
appearance, a puzzle first noted by Zwicky (1939). 
In statistical mechanics, the approach of macroscopic systems 
to a relaxed state is tied to the exponential 
instability of the motion of individual particles, i.e. to chaos 
(\cite{wig71}; \cite{for75}).
In the case of a hard sphere gas, for instance, the chaos results from 
collisions between molecules which rapidly erase memory of the 
initial state (\cite{sin63}).
In galaxies, where close encounters between stars are 
rare, relaxation must be driven by some other mechanism that
eliminates correlations.
One candidate, discussed above, is the stochasticity associated with motion in 
non-axisymmetric potentials.
Motion in rapidly-varying potentials is also likely to be 
complex and one expects relaxation under such conditions 
to be especially efficient.
The term ``mixing'' is used by statistical physicists to describe all such
forms of relaxation, both collisional and collisionless, in fixed as well
as time-dependent potentials.
Mixing can take place at various speeds and can even be 
arbitrarily slow; 
an example of slow mixing in galaxies is the 
persistence over many crossing times of narrow features like
``shells'' (\cite{qui84}).
However, in statistical mechanics, mixing is almost always assumed to 
be associated with chaos: in part because stochasticity is the 
norm in complex systems, 
and in part because quasiperiodic motion 
tends to preserve correlations for long times.
The fact that most stars in elliptical galaxies are not associated 
with narrow features like shells suggests that chaotic mixing, 
or something similar, was active during galaxy formation.

The association of relaxation with exponential instability
is not often made in the astronomical literature,
where the emphasis has historically been on quasi-periodic motion.
For instance, Jeans's (1915) theorem states that the distribution function of 
a relaxed galaxy must be expressible in terms of the integrals of motion, 
and his theorem has been interpreted as according a priveledged status to 
integrable potentials (e.g. \cite{bin82b}).
Proofs of Jeans's theorem are sometimes based on 
the assumption that the potential is fully integrable
(e.g. \cite{bit87}, Appendix 4a).
Statistical physicists, on the other hand, view integrals of the motion as 
impediments to relaxation and routinely associate chaos with a 
steady state.
Similarly, discussions in the astronomical literature of relaxation in 
time-dependent potentials, or ``violent relaxation'' 
(\cite{lyn67}; \cite{shu78}), rarely focus on the mixing properties
of the flow, even though such properties are considered basic for 
understanding the evolution of other many-particle systems like
gases (Sinai 1979).

While there is merit in both points of view,
the astronomer's traditional preoccupation with regular motion 
has been something of an impediment toward 
understanding collisionless relaxation.
Regular orbits are the most useful building blocks for constructing
self-consistent models, but it does not follow that 
integrability is a necessary condition for a steady state to exist;
in fact, the redistribution of stars in phase space that 
drives relaxation is much easier to arrange when the motion is 
not quasi-periodic.
Furthermore, to the extent that galactic potentials are 
integrable -- and the work reviewed in the previous sections 
suggests that most elliptical galaxies may be close to axisymmetric,
hence nearly integrable --
a mechanism like chaotic mixing is probably required in order
to explain how they got that way.
It is likely too that our understanding of ``violent 
relaxation'' will be sharpened once attention is 
focussed on the properties of the 
phase-space flow that are responsible for driving 
evolution toward a steady state in time-dependent potentials.

\subsection{Jeans's Theorem}

Jeans (1915) noted that the phase space density of a 
collisionless stellar system satisfies the continuity equation
\begin{equation}
0={Df\over Dt}={\partial f\over\partial t} + {\bf v}\cdot\nabla f 
- \nabla\Phi\cdot{\partial f\over\partial {\bf v}},
\end{equation}
which states that $f$ remains constant as one follows the motion of any star.
In a steady state, $\partial f/\partial t=0$, $f$ must therefore 
be constant along trajectories at a {\it given} time, and
one way to achieve this is by writing $f$ as a function of the 
integrals of motion.

However not all integrals are valid arguments of $f$, as a number of 
authors (\cite{kur55}; \cite{con60}; \cite{lyn62b})
subsequently pointed out.
The relative phase, $\theta_i/\omega_i-\theta_j/\omega_j$, 
between motion in 
different directions around an invariant torus is a bona fide 
integral, but 
allowing $f$ to depend on it would yield a phase space density 
with multiple values at a single point since trajectories 
with different phases all fill the torus densely 
(assuming incommensurability of the frequencies).
Integrals that do not have this property are called ``isolating'': an 
isolating integral is one that, in some transformed coordinate 
system, makes $\partial H/\partial p_i$ a function only of the 
coordinate conjugate to $p_i$.
In an integrable system with $N$ degrees of freedom, the 
isolating integrals are the $N$ actions or any set of $N$ 
independent functions of the actions.
Jeans's theorem for a galaxy in which the motion is fully 
integrable becomes:

\begin{quote}
{\sl The phase space density of a stationary stellar system 
with a globally integrable potential
can be expressed in terms of the isolating integrals
in that potential.}
\end{quote}

\noindent
An equivalent statement is that $f$ is constant on every 
invariant torus.

A troublesome feature of Jeans's theorem when stated in this way 
is its apparent restriction to integrable systems, 
which must be vanishingly rare in nature.
But as a number of authors, notably Pfenniger (1986) and Kandrup 
(1998b), 
have emphasized, 
regularity of the motion is in no way a prerequisite for 
a steady state to exist.
The special status that Jeans's theorem seems to accord to 
integrable potentials disappears when one recognizes that 
the theorem is really a statement about phases, not 
about integrals: in a steady state,
stars must be uniformly distributed with respect to angle
over each invariant torus.
In a region of phase space that is not integrable, stationarity 
can likewise be achieved by requiring $f$ to be constant throughout the 
region: since $f$ is conserved following the flow, an 
initially constant value will remain constant forever.
The generalization of Jeans's theorem to general, non-integrable 
potentials is then:

\begin{quote}
{\sl The phase space density of a stationary stellar system 
must be constant within every well-connected region.}
\end{quote}

\noindent
A well-connected region is one that can not be decomposed into two 
finite regions such that all trajectories lie in either one 
or the other (what mathematicians call ``metric 
transitivity.'')
Invariant tori are such regions, but so are the more complex 
parts of phase space associated with stochastic trajectories.

An example of a well-connected region that is not an invariant 
torus is the ``Arnold web,'' the phase space region
accessible to every stochastic trajectory at a given energy in 
a 3 DOF system (\cite{lil92}).
Every point in the Arnold web has the same energy, hence a 
constant $f$ within such a region could formally be written as 
$f=f(E)$ as in Jeans's original formulation.
However there will generally exist invariant tori 
at the same energy on which the motion respects two additional 
integrals and where $f$ has different values.
The phase space density in such a system is therefore
likely to be an extremely complicated function of the coordinates.

In weakly chaotic potentials, the time required for a single 
trajectory to visit the entire Arnold web may be very long,
\footnote{The more relevant time scale is the chaotic mixing time,
defined below.}
and one would not expect the distribution of stars in stochastic
phase space to have reached a uniform value in a galaxy lifetime.
In such potentials, chaotic trajectories mimic 
regular orbits with effective integrals for many oscillations.
In more strongly chaotic potentials, 
one can define time-independent mass components that 
correspond to an approximately uniform filling of stochastic 
phase space at every energy.
Such components have shapes that are not very useful for 
reconstructing a galaxy's figure; nevertheless they can constitute
a significant fraction of the mass of a self-consistent model
(\cite{kac96}; \cite{mef96}), 
and there is every reason to believe that nature uses 
such components as well.

In a 2 DOF system, the stochastic regions at a given energy
are separated by the invariant tori.
This is the case in axisymmetric potentials, where $f$ can have 
different values in different stochastic regions at a given $E$ 
and $L_z$.
Axisymmetric models with this level of 
generality have yet to be constructed.
The classical approach of writing $f=f(E,L_z)$ for an axisymmetric 
galaxy assigns a constant density to every point on
hypersurfaces of constant $E$ and $L_z$.
Since some of these points correspond to stochastic orbits
in general, 
the two-integral approach to axisymmetric modelling 
actually depends on the more general form of Jeans's theorem 
given above for its justification.

\subsection{Approach to a Steady State}

\subsubsection{Ergodicity and Mixing}

The evolution of a collisionless ensemble of stars is never 
toward the uniform population of phase space demanded by Jeans's 
theorem.
An initially compact group of phase points gets drawn out into a 
filament of ever decreasing width, a consequence of the 
conservation of phase space volume implied by Liouville's theorem.
Observed with infinite resolution, the region 
occupied by these points becomes increasingly striated, not more 
uniform.
The most that can be hoped for is that the coarse-grained phase 
space density will approach a constant value within some region.
However even this outcome is not guaranteed by any very 
general property of Hamilton's equations.
For instance, an ensemble of points on an invariant torus does 
not evolve toward a coarse-grained steady state -- it simply translates, 
unchanged, around the torus.

The simplest model for collisionless relaxation is ``phase 
mixing,'' the gradual shearing of points in a fixed, 
integrable potential (Figure 10a).
Since the fundamental frequencies of regular motion are generally 
different on different tori, two points that lie on adjacent
tori will separate linearly with time.
After many revolutions, the density of stars in this filament, 
averaged over a finite phase-space volume, will be 
independent of angle.
Thus the coarse-grained density evolves asymptotically to a state 
that satisfies Jeans's theorem, in spite of the fact that the 
fine-grained density never reaches a steady state.

\begin{figure}
\plotfiddle{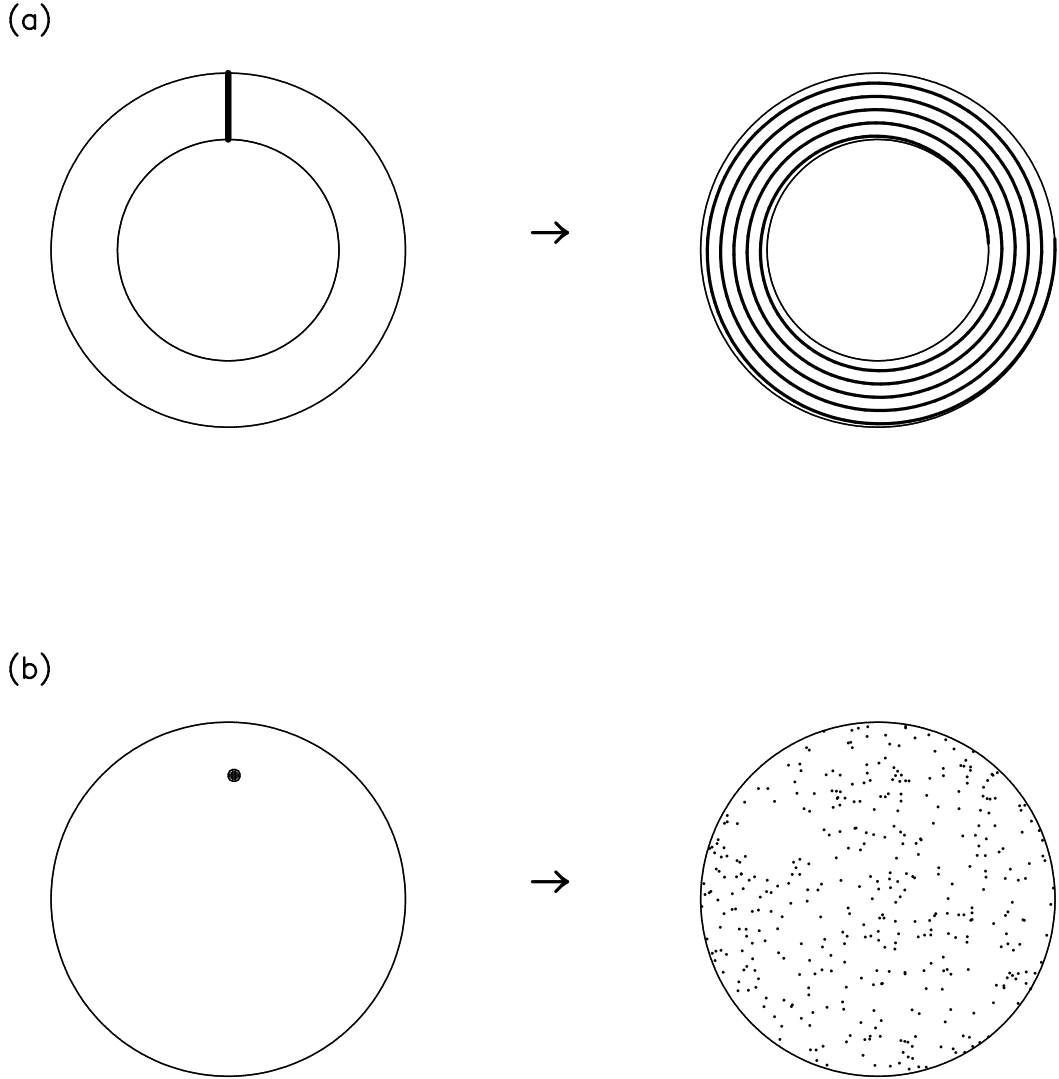}{6.in}{0.}{85}{85}{-280}{-200}
\caption{
(a) Phase mixing vs. (b) chaotic mixing.
Both types of mixing produce evolution toward a state
such that the coarse-grained phase space density is uniform
within some region.
In phase mixing, the separation between two initially nearby
points increases approximately linearly with time,
while in chaotic mixing the separation increases exponentially on average.
Although the two types of mixing are normally defined in
fixed potentials, qualitatively similar evolution of both sorts
occurs in time-dependent potentials as well; an example is shown in
Figure 12.
}
\end{figure}

This coincidence is a consequence of the ergodicity of 
regular motion.
Ergodic motion spends equal amounts of time, on average, in equal 
phase-space volumes (defined with respect to the invariant 
measure).
Ergodicity is a non-trivial property 
which can only be rigorously 
proved for certain classes of motion.
Regular motion is ergodic due to its quasi-periodicity: 
since the angles on the torus increase linearly with time, 
any trajectory fills the torus uniformly and densely in a 
time-averaged sense.
This time-averaged uniformity with respect to angle guarantees that 
phase mixing will tend toward a coarse-grained
state in which no angle is preferred.

Ergodicity by itself is a condition only on the time-averaged 
behavior of a trajectory.
Relaxation implies more in general: not only should the entire 
past of a given phase point cover the phase space uniformly, but 
so also should the present of any neighborhood of the original 
point.
In other words, a small patch of phase space should evolve in 
such a way that it uniformly covers, at a single later time, a 
much larger region (Figure 10b).
This process, which dynamicists call simply ``mixing,'' 
can occur in many ways and at various speeds,
but it is most commonly associated with the exponential instability of
stochastic motion, and the term ``chaotic mixing'' will be used here 
where there is danger of confusion with phase mixing.
Krylov (1979) was the first to emphasize the role of 
chaotic mixing in explaining relaxation processes in statistical 
mechanics, and nowadays chaotic mixing is regarded by statistical 
physicists as essentially equivalent to relaxation (\cite{sin79}).

Since chaotic mixing is driven by the exponential instability of the 
motion, it occurs locally at a rate that is determined by the 
Liapunov exponents.
It is also irreversible, in the sense that an infinitely precise 
fine-tuning of the velocities would be required in order 
to undo its effects.
Phase mixing {\it is} reversible, and it
has no well-defined time scale:
the rate at which a group of phase points shears depends on the 
range of orbital frequencies in the group.
If the maximum and minimum frequencies are $\omega_1$ and 
$\omega_2$ respectively, one expects phase mixing to take place on 
a time scale of order $(\omega_1-\omega_2)^{-1}$.
This time scale is never less than a dynamical time and can be much 
longer; for points restricted to a single torus, the time scale 
is infinite and no phase mixing occurs.

Phase mixing and chaotic mixing are idealizations of the mixing that
takes place in real galaxies, whose potentials may vary with time
in complex ways.
Nevertheless one often sees, in time-dependent numerical simulations, 
examples of mixing that are readily identified with these two basic types.
An example, discussed below, is illustrated in Figure 12.

\subsubsection{Mixing in Fixed Potentials}

Stochastic motion is nearly random in the sense that the likelihood 
of finding a particle anywhere in the stochastic region tends 
toward a constant value after a sufficiently long time.
An initially compact group of stars should therefore spread out 
until it covers the accessible phase space uniformly in a 
coarse-grained sense.
Kandrup and collaborators (\cite{kam94}; \cite{mah95}) 
first investigated this process in the context of galactic 
dynamics.
They focussed on motion in two-dimensional potentials, 
integrating ensembles of initially localized points until they 
had reached nearly invariant distributions.
They found that the coarse-grained distribution function 
typically exhibited an exponential approach toward equilibrium 
at a rate that correlated well with the mean Liapunov exponent 
for the ensemble.
When evolved for much longer times, $f$ was found to slowly
change as orbits diffused into regions that, although accessible, 
were avoided over the shorter time interval.
Kandrup et al. suggested that one could construct approximately 
steady-state models of galaxies using the near-invariant 
distributions, since the time scale for the slower evolution was 
typically much longer than a galaxy lifetime.

\begin{figure}
\plotfiddle{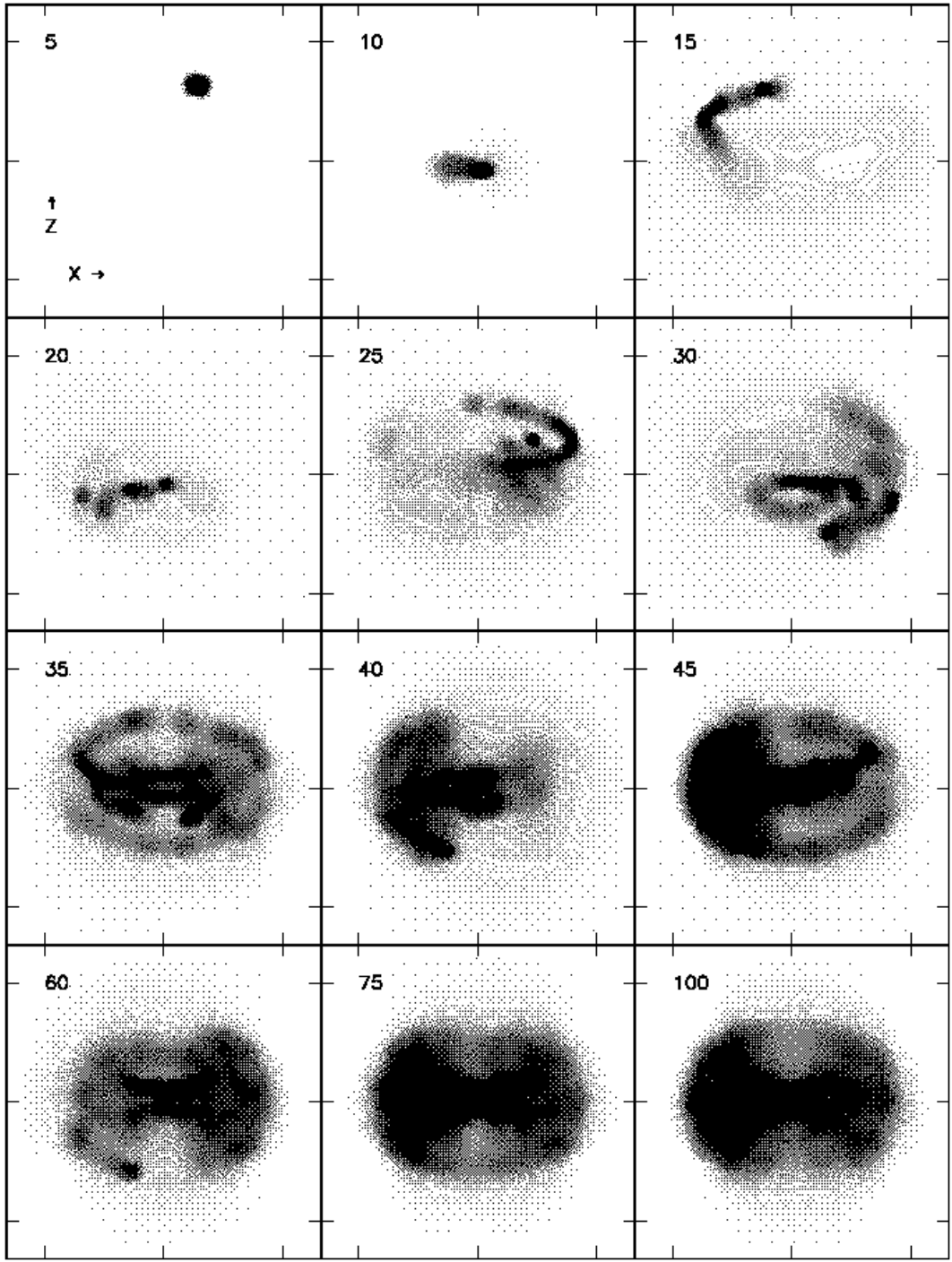}{7.in}{0.}{75}{75}{-220}{-40}
\caption{
An example of chaotic mixing in a fixed triaxial potential
(Merritt \& Valluri 1996).
An ensemble of $10^4$ isoenergetic particles move independently in
the time-dependent gravitational field generated by the
Imperfect Ellipsoid with $m_0=0.01$.
Each frame is labelled by the elapsed time in units of the period
of the long-axis orbit; the particles are seen in projection
against the $x-z$ plane.
After $\sim 100$ orbital periods, the particles have reached a
nearly stationary, coarse-grained state.
This example suggests the way in which time-independent density
components exist even in potentials that are not characterized by
integrals of the motion aside from the energy.
}
\end{figure}
  
Similar calculations in three dimensions 
were carried out in the ``imperfect ellipsoid'' potential by 
Merritt \& Valluri (1996) (Figure 11).
Ensembles of particles were evolved starting from small patches 
on the equipotential surface, initial conditions that would 
have generated box orbits in an integrable potential.
Merritt \& Valluri also found an initially rapid approach to a near-invariant 
distribution followed by a slower evolution.
Ensembles with starting points near the $y-z$ instability strip 
(Figure 3) exhibited the most rapid mixing, with e-folding times 
of only $\sim10$ crossing times in models with small core radii 
$m_0$ or with central point masses containing $\gap 0.1\%$ 
of the total mass.
Ensembles started near regions where the motion was regular 
evolved more slowly, a consequence of the ``stickiness'' of 
invariant tori.
Merritt \& Valluri suggested that chaotic mixing would tend to 
axisymmetrize at least the central regions of triaxial galaxies 
with high central concentrations or massive nuclear black holes.

Kandrup (1998a) compared the efficiency of phase mixing and chaotic 
mixing in 2D and 3D potentials.
He noted that -- for initially very localized ensembles -- the 
two processes occur at very different rates: chaotic mixing 
takes place on the Liapunov, or exponential divergence, time scale 
while the phase mixing rate falls to zero.
But phase mixing of a group of points with a finite extent can 
be much more rapid.
Furthermore the mixing rate of chaotic ensembles typically 
falls below the Liapunov rate once the trajectories separate; 
this is especially true for those stochastic orbits that are confined 
over long periods of time to restricted parts of phase space.
The effective rates of phase mixing and chaotic mixing might 
therefore be comparable in real galaxies.
Kandrup noted also that chaotic mixing in 3D potentials can occur 
at substantially different rates in different directions.

Smooth potentials are idealizations of real elliptical galaxies, 
which sometimes contain imbedded disks, shells, or other fine 
structure.
As seen by a single star, these distortions would add 
small-amplitude perturbing forces to the mean field.
Such perturbations would not be expected to have much consequence 
for either strongly chaotic or nearly regular orbits, but they
might have an appreciable effect on the evolution of weakly stochastic 
orbits, by scattering trajectories away from a trapped region into a 
region where mixing is more rapid.
Goodman \& Schwarzschild (1981) found that small perturbations had 
relatively little effect on the behavior of stochastic orbits in their 
nearly-integrable triaxial model.
Habib, Kandrup \& Mahon (1996, 1997) observed much more striking 
effects of noise on the evolution of trapped stochastic orbits 
in two dimensionial potentials.
These authors found that even very weak noise, with 
a characteristic time scale $|{{1\over v}{\delta v\over\delta 
t}}|^{-1}$ of order $10^6$ crossing times, could 
induce substantial changes in the motion of trapped stochastic
orbits over $\sim 10^2$ periods.
Merritt \& Valluri (1996) also considered the effect of noise 
on chaotic mixing in triaxial potentials.
They observed a significant enhancement in the mixing rate of 
trapped ensembles for noise with a characteristic time scale 
of $\sim 10^3$ orbital periods.
These results suggest that chaotic mixing in strongly 
time-dependent potentials, like those associated with a 
collapsing protogalaxy, might be much more efficient than in 
experiments based on fixed potentials.

\subsubsection{Mixing in Time-Dependent Potentials}

In a real galaxy, the approach to a stationary state takes place 
against the backdrop of a time-varying potential.
King (1962), H\'enon (1964), Lynden-Bell (1967) and others realized 
that relaxation 
under such conditions might be very efficient, and the latter 
author made the bold suggestion that the relaxation rate could be 
directly identified with the rate of change of a star's 
potential energy.
Lynden-Bell proposed that the time scale for
``violent relaxation'' was 
\begin{equation}
T_{vr}={3\over 4} \langle{\dot\Phi^2\over\Phi^2}\rangle ^{-1/2}
\end{equation}
with $\dot\Phi$ the time derivative of the gravitational
potential.
For a collapsing sphere, Lynden-Bell noted that $T_{vr}$ 
was similar to the collapse time, consistent with 
$N$-body simulations which showed nearly complete relaxation 
after just a few radial oscillations.

A rearrangement of stars in energy space can only take place if the
potential is time-dependent.
But in other respects, the identification of $\partial\Phi/\partial t$ 
with a relaxation rate is problematic.
Instantaneous changes in the potential imply only a relabelling 
of particle energies, not necessarily a divergence of
phase space trajectories or a redistribution of energies, which 
are also prerequisites for relaxation.
It is in fact possible to construct oscillating 
models which exhibit no tendency to relax
(\cite{log88}; \cite{sri89}; \cite{srn89}; \cite{mfr90}).
The failure of these models to reach a steady state
is a result of their being constructed  
in such a way that mixing is inhibited;
thus the distribution of particle energies does not relax
even though the energy of individual particles is constantly changing.
For instance, in Sridhar's (1989) pulsating models, 
all of the stars move with the same fixed frequencies and no mixing
can occur.
Such examples are artificial but demonstrate that mixing is 
no less a prerequisite for relaxation in time-dependent potentials 
than it is in fixed potentials.
Potential fluctuations would be expected to promote relaxation 
only to the extent that they encourage mixing.

Mixing that is driven by changes in the mean-field potential 
is self-limiting in the sense that the potential approaches 
a steady state as the mixing progresses.
The effectiveness of ``violent relaxation'' therefore depends 
strongly on how long the mixing is able to continue, hence on
the initial conditions.
A number of numerical simulations have borne this out.
Hoffman, Shlosman \& Shaviv (1979) followed the collapse and 
relaxation of a spherical galaxy starting from two initial states 
with different values of the ``virial ratio'' $|2T/W|$, where $T$ 
and $W$ are the initial kinetic and potential energies.
For $|2T/W|=0.49$ they found a deeper collapse than for 
$|2T/W=0.59|$, but the subsequent decay of the virial ratio to 
its equilibrium value of $1$ was also more rapid, thus allowing 
less time for the mixing to take place.
Van Albada and May (\cite{val82}; \cite{mav84}) 
simulated the evolution of initially ``cold'' clouds with 
irregular shapes and clumpy particle distributions.
They found that the degree of relaxation -- as measured, for 
instance, by the change in the binding energy distribution -- 
was well correlated with the initial degree of irregularity. 
They argued that clumpy initial conditions were effective at 
driving relaxation since they permitted the mixing to continue 
for a prolonged period of time.
Villumsen (1984) carried out similar experiments starting from 
spherical as well as highly flattened initial conditions.
He too noted the greater complexity of the motion in the 
nonspherical collapses and he argued that this complexity was 
necessary if the evolving galaxy was to lose memory of its 
initial state.
Similar conclusions were reached by McGlynn (1984) and by 
Londrillo, Messina \& Stiavelli (1991).

The view of relaxation in a time-dependent potential as a 
self-limiting process was incorporated into a heuristic model by 
Wiechen, Ziegler and collaborators
(\cite{wzs88}; \cite{ziw89}; \cite{szw90}; \cite{ziw90}).
These authors assumed that the degree to which an 
out-of-equilibrium stellar system relaxes could be estimated 
by comparing it to a reference system, the ``lowest energy 
state,'' defined in the following way.
Starting from the initial state, they rearranged the matter in phase 
space so as to minimize the total energy; in other 
words, they required the largest values of $f$ to coincide with the 
most negative values of $h=v^2/2 + \Phi$ and vice versa.
At the same time, they required that Liouville's theorem be satisfied, i.e. 
that the volume of each element of phase space be conserved.
Wiechen, Ziegler \& Schindler (1988) showed that such a ``lowest 
energy state'' could always be uniquely defined.
This state is however unreachable, since its energy is (by 
definition) different from that of the initial state.
But Ziegler \& Wiechen (1989) suggested that the energy difference 
between the initial and lowest-energy states
could be taken as a measure of the degree to which the 
potential would evolve before the mixing came to a halt.
They then simulated the mixing process by applying a smoothing 
operator to the initial state, 
after demonstrating that the smoothing process always 
increases the total energy by making the distribution of 
matter less compact.
They therefore chose the smoothing length such that 
the reference model corresponding to the smoothed initial distribution
had the same energy as the unsmoothed initial state.
Finally they identified this smoothed, lowest energy state with the 
equilibrium configuration.
Ziegler \& Wiechen (1989) applied their formalism to a single, 
spherical initial state and found good agreement with 
a full numerical simulation.

Nozakura (1992) applied Wiechen \& Ziegler's algorithm to the 
spherical initial states considered by van Albada (1982) in his $N$-body 
simulations.
Nozakura found that the formalism was useful for predicting the final 
state when the initial state was only moderately far from 
equilibrium, with a virial ratio $2T/|W|\gap 0.5$,
but that the agreement was poor for stronger collapses.
Ziegler, Wiechen \& Arendt (1994) argued that the mixing in 
strong collapses was ``non-uniform,'' i.e. that the appropriate 
smoothing length to apply to particles in the (final) core and 
halo should be different.
Lacking a detailed understanding of the evolution, however, they were 
unable to specify how the smoothing length should vary with position.
They noted also that the tendency of initially spherical models 
to acquire highly elongated shapes during the collapse 
(\cite{mea85}; \cite{agm90}; \cite{cah92}) 
had no natural explanation in their model.

\begin{figure}
\plotfiddle{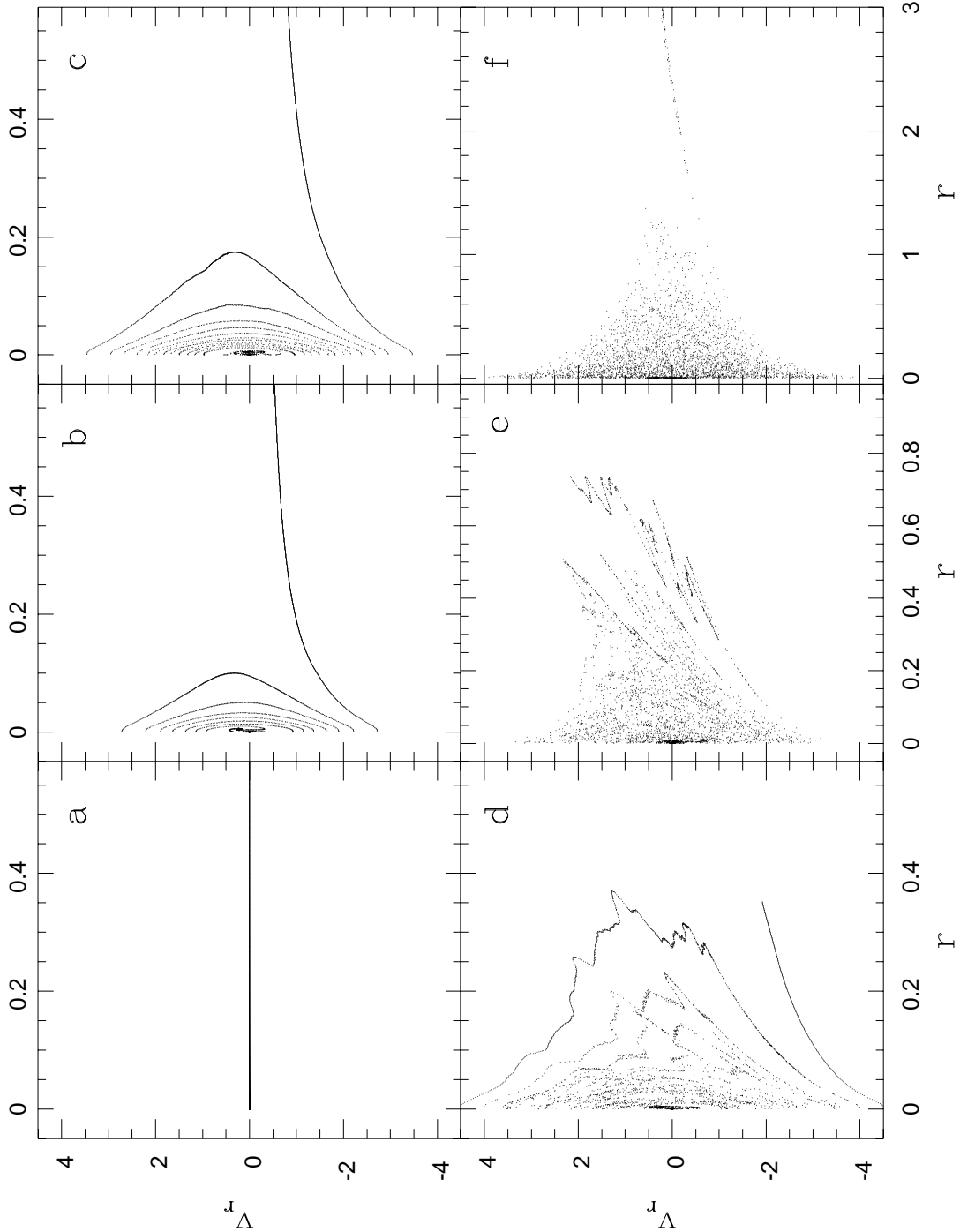}{7.in}{0.}{75}{75}{-220}{-40}
\caption{
Evolution of a spherically-symmetric system of particles in $(r,v_r)$ phase
space starting with zero velocity and $\rho\propto r^{-3/2}$
(Henriksen \& Widrow 1997).
The initial evolution is similar to
phase mixing in spite of the rapidly varying potential;
after a certain point, the motion undergoes a transition to
a more chaotic flow, allowing rapid
mixing between particles on neighboring phase-space
streams.
This example shows that the distinction made in Figure 10,
between phase mixing and chaotic mixing in a fixed potential, is
qualitatively valid in time-dependent potentials as well.
 }
\end{figure}

An interesting illustration of mixing in a time-dependent 
potential was presented by Henriksen \& Widrow (1997).
These authors simulated spherically-symmetric 
collapses of ``cold'' particle distributions with power-law density 
profiles, $\rho\sim r^{-\gamma}$, $0<\gamma<3$.
The initial evolution was found to be essentially similar to 
phase mixing, in spite of the rapidly varying potential: the 
single phase-space thread corresponding to the initial model 
became more and more tightly wound as the sphere oscillated
(Figure 12).
After a certain point, however, the motion underwent a transition to
a more complex flow, which allowed rapid
mixing between particles on neighboring phase-space 
streams.
While highly idealized, Henriksen \& Widrow's simulations provide a 
striking illustration of the way in which the efficiency of 
relaxation is tied more to the complexity of the phase-space flow
than to the rate of change of the potential.

\subsubsection{Entropy}

Mixing in a time-dependent potential allows the energies of stars 
to change, and if the mixing is very efficient or prolonged, one 
might expect the distribution of energies to approach a ``most 
probable'' state.
Much the same thing takes place in a gas, where collisions 
between molecules imply a rapid exchange of kinetic energies;
the most probable distribution of particle energies 
is obtained by maximizing the entropy, leading to the 
well-known Maxwell-Boltzmann velocity distribution.
One can carry out a similar calculation for stellar systems 
(\cite{ogo65}; \cite{lyn67}; \cite{shu78}), with the 
not-very-surprising result that the velocity distribution
should be Maxwellian in this case as well.
However the self-consistent 
model corresponding to a Maxwellian velocity distribution,
the isothermal sphere, has an infinite radius and mass.
One might interpret this to mean that 
the phase-space structure of a real galaxy should be
similar in some respects to that of an isothermal sphere,
but the entropy arguments give little indication of 
precisely what the similarities should be.

A number of authors have nevertheless attempted to
use entropy arguments to draw conclusions about the end 
state of collisionless relaxation
(\cite{thl86}; \cite{mad87}; \cite{whn87}; \cite{mat88}; 
\cite{sph92}; \cite{soa96}).
In addition to the difficulty just mentioned, 
these studies must deal with the fact that
the equilibrium state of a stellar system depends strongly on 
the initial conditions.
If the correct final state is to be singled out by an entropy
maximization, the dependence of the final state on the initial 
state must somehow be translated into a set of constraints.
Inferring the final state then becomes equivalent to specifying
these constraints.
There is currently no clear understanding of how this can be done 
and it is not even certain that the exercise would be physically 
enlightening.

An illustration of the subtleties associated with entropy
arguments is the controversy surrounding the study of
Tremaine, Lynden-Bell \& H\'enon (1986).
These authors noted, following Tolman (1938), 
that any functional of the form
\begin{equation}
S_C(f) = -\int C(f) d{\bf x} d{\bf v}
\end{equation}
is precisely conserved during collisionless evolution;
here $C(f)$ is any ``convex'' function of $f$, $d^2C/df^2\geq 0$.
The constancy of $S_C$ is a result of the detailed conservation 
of phase space volume implied by Liouville's theorem.
However if one computes $S_C$ at some time $t_2$ using a 
coarse-grained approximation to $f$, and compares it to the value 
of $S_C$ computed at an earlier time $t_1$ using the exact $f$, 
one finds
\begin{equation}
S_C(t_2) > S_C(t_1).
\label{tolman}
\end{equation}
One is tempted to conclude that $S_C$ is a 
monotonically increasing function of time and that any 
such functional can play the role of an entropy.
Such a conclusion was rejected by a number of authors
(\cite{bin87}; \cite{dej87}; \cite{kan87}; \cite{sri87})
on the following grounds.
Identifying the initial state with the fine-grained $f$ is 
equivalent to assuming that the initial state is known with 
infinite precision.
The increase in $S_C$ is then a consequence 
of the coarse-graining: at any time, the coarse-grained $S_C$ is 
greater than the fine-grained $S_C$, hence $S_C(t_2) > S_C(t_1)$.
But it does not follow that $S_C$ is monotonically increasing and 
it might well decrease.
In fact, equation (\ref{tolman}) is true irrespective of whether $t_2$ is 
greater or less than $t_1$.

Entropy increases of a physically interesting sort are always 
associated with irreversibility, that is with chaos (\cite{rue91}), 
and it is remarkable that so few discussions of entropy in the 
galactic dynamics literature have paid any attention to the 
dynamical properties of the phase-space flow.
Chaotic mixing in a fixed potential, as illustrated in Figure 11, 
represents a true entropy increase in the sense understood by 
Boltzmann, and the irreversibility observed in numerical 
simulations of ``violent relaxation'' (e.g. Figure 12)
suggests that something very similar takes place in time-dependent 
potentials as well.
It is in this direction that a fuller understanding of 
collisionless relaxation probably lies.

\subsection{Mixing-Induced Evolution}

The irreversibility of mixing flows like the one illustrated in 
Figure 11 implies a reduction in the effective number of orbits: 
all the stochastic trajectories at a given energy are gradually
replaced by a single invariant ensemble, whose shape is typically 
not well matched to that of the galaxy.
If time scales for chaotic mixing are comparable to galaxy 
lifetimes, this reduction might be expected to encourage a galaxy to 
evolve away from a triaxial shape toward a more axisymmetric one, 
in which most of the orbits are tubes that avoid the destabilizing center.

A pioneering study of mixing-induced evolution was carried out by 
Norman, May \& van Albada (1985), who performed $N$-body 
integrations of triaxial models in which a central compact object 
was slowly grown.
They found a modest degree of evolution toward more spherical 
shapes near the center.
Norman et al.'s simulations were intended to represent the effect of a 
massive nuclear black hole on the surrounding galaxy; however, 
because of computational limitations, they were forced to adopt a 
softening length for the central object that was close to the 
gravitational radius of influence $r_h\sim GM_h/\sigma^2$ of a true point 
mass.
In addition, their initial models were chosen to be close to 
axisymmetric.
As Norman et al. noted, both of these factors would tend to limit
the evolution, the first by eliminating large-angle 
scattering of trajectories by the central object, the second by 
reducing the fraction of stars on boxlike orbits that could pass 
near the center.

A natural way to increase the central density of a galaxy 
model is through the inclusion of a dissipative component which 
accumulates near the center as its energy and angular momentum are 
carried away.
$N$-body studies of galaxy formation that include such a 
component often reveal an evolution toward more 
axisymmetric shapes as the central mass grows.
Katz \& Gunn (1991) carried out such simulations 
starting from initial conditions containing either no ``gas,''
or a $10:1$ ratio of ``stellar'' to ``gas'' particles.
Initially the two components were well mixed; however the gas 
particles quickly accumulated near the center where they formed a 
disk.
In the absence of a dissipative component, the stellar particles 
in these simulations relaxed into strongly triaxial shapes, with 
typical axis ratios of $\sim 1:0.7:0.5$.
When gas was included, the final stellar distribution was 
found to always be nearly oblate-axisymmetric, with $b/a\gap 0.9$.
The short-to-long axis ratio was nearly unchanged.
Udry (1993) simulated dissipation in 
triaxial galaxy models by a number of schemes:
including an ad hoc drag force on $10\%$ of the particles;
adding a central particle containing $2.5\%$ of the stellar mass; 
or including 
a centrally concentrated distribution of particles containing 
$5\%$ or $10\%$ of the total mass.
While the simulations without ``dissipation'' resulted in 
models with a range of triaxialities, the simulations including 
dissipation evolved to nearly axisymmetric shapes, with $0.7\lap 
b/a\lap 1$.
Dubinski (1994) simulated dissipative infall inside an initially 
triaxial halo by slowly increasing the mass associated 
with a central, fixed component having either a spherical or 
disklike shape.
The final mass of this central object was $\sim5\%$ the mass of 
the surrounding halo; however its scale length was large, roughly $0.1$ 
that of the halo.
Dubinski nevertheless found a modest increase in the intermediate-to-long axis 
ratio of the halo, from $\sim 0.6$ to $\sim 0.8$, with less change in $c/a$.

Barnes (1992, 1996) reported the results of $N$-body simulations of 
mergers between disk galaxies with and without a
dissipative component.
In the absence of gas, the remnants acquired shapes that 
depended strongly on the initial orientations and impact 
parameter of the colliding disks; many of the remnants were 
strongly triaxial.
When as little as $1.5\%$ of the total mass was added in the form of 
dissipative particles, the remnants acquired much more axisymmetric
shapes.
Barnes (1996) cataloged the stellar orbits in the remnants and 
found a preponderance of short-axis tubes in the remnants 
containing gas, as expected for nearly axisymmetric potentials.
Barnes \& Hernquist (1996) confirmed these results with further 
simulations and noted that the final shapes of the remnants were 
well correlated with the depth of the central potential.
They speculated that the evolution toward oblate figures was 
driven largely by the destabilization of box orbits moving in the 
deepened potential, i.e. by chaotic mixing.

The effect of a massive central singularity on the structure of an 
initially triaxial galaxy was studied by Merritt \& Quinlan (1998) 
using an $N$-body code with much higher spatial and temporal 
resolution than in the pioneering simulation of Norman et al.
The final shape achieved by their model after growth of the 
black hole was nearly spherical at the center and close 
to axisymmetric throughout; the rate of change of the 
galaxy's shape depended strongly on the ratio $M_h/M_g$ of black 
hole mass to galaxy mass.
When $M_h/M_g$ was less than about $0.3\%$, the galaxy evolved 
in shape over $\sim10^2$ orbital periods, while increasing
$M_h/M_g$ to $3\%$ caused the galaxy to become
axisymmetric in little more than a crossing time.
Merritt \& Quinlan proposed that the rapid evolution toward
axisymmetric shapes for $M_h/M_g\gap 2.5\%$ would tend to cut off the 
supply of fuel to a growing black hole, thus imposing an upper 
limit to its mass.

Chaotic mixing induced by a central density cusp or nuclear black 
hole could have a number of other consequences for the structure 
of elliptical galaxies and bulges (\cite{mer98}).
Elliptical galaxies with nonsmooth phase space populations are 
expected to be ``boxy'' or peanut-shaped, since individual 
orbits, both boxes and tubes, are strongly dimpled when seen in 
projection (\cite{bip85}).
The fact that most elliptical galaxies have accurately elliptical 
isophotes suggests that some process has acted to smooth the
distribution of orbital turning points at every energy.
``Violent relaxation'' is one candidate for this smoothing 
process, but it is apparently not effective enough, since the 
end-products of $N$-body simulations are often strongly
peanut-shaped 
(e.g. \cite{mav85}; \cite{qug86}; \cite{udr93}; \cite{zwa94}).
An alternative mehcanism is chaotic mixing induced by a central 
black hole; Merritt \& Quinlan (1998) noted that the 
growth of a central mass concentration converts boxy, triaxial 
systems into axisymmetric ones with accurately elliptical 
isophotes.
This result may explain the observed correlation between boxiness and 
kinematical measures of triaxiality (\cite{kob96}).
On a deeper level, the distribution function of an axisymmetric 
galaxy formed in this way would be biased toward forms for $f$ 
that are as nearly constant as possible, i.e. which do not 
depend strongly on a third integral.
Detailed modelling of M32, a galaxy that is old compared to 
expected time scales of chaotic mixing, in fact suggests a best-fit 
$f\approx f(E,L_z)$ (\cite{vdm98}).

\section{COLLECTIVE INSTABILITIES}

The chaotic mixing discussed in the previous section is driven 
by the exponential instability of stochastic motion; the effect of such
mixing is typically to erase a density perturbation
by spreading particles in phase space.
However it is possible for a density perturbation to grow
by inducing motion that reinforces the original overdensity.
\footnote{Distinguishing between the two sorts of instability
in a numerical simulation can sometimes be difficult;
see the discussion of bending instabilities in \S 6.3.} 
Such collective instabilities typically require the unperturbed
motion to be highly correlated and they have been most thoroughly 
studied in thin disks, which are subject to a variety of unstable 
modes when sufficiently ``cold'' (\cite{sew93}).
In elliptical galaxies, where the stellar motions are more nearly
random in direction, a perturbation in the density might be 
expected to rapidly attenuate as the stars move along their 
respective orbits.
However it turns out that the motion in a variety of 
physically reasonable models is sufficiently correlated to induce 
growing modes.

The simplest model for an elliptical galaxy is a sphere in which the 
stellar velocities are isotropic, $f=f(E)$.
Antonov (1960, 1962) established a necessary and sufficient 
condition for linear stability of spherical systems satisfying 
$df/dE<0$, in the form of a complicated variational principle.
He went on to derive a number of simpler, sufficient conditions 
for stability.
The most important of these are:

\begin{quote}
I. A spherical system with $f=f(E)$ and $df/dE<0$ is stable to 
all non-radial (i.e. non-spherically-symmetric) perturbations.
\end{quote}

\begin{quote}
II. A spherical system with $f=f(E)$, $df/dE<0$ and 
$d^3\rho/d\Phi^3\le 0$ is stable to all perturbations.
\end{quote}

\noindent
Antonov was able to show that the family of ``stellar dynamical 
polytropes'' defined by 
\begin{equation}
f(E) = f_0(E_0-E)^{n-3/2}, \ \ \ \ E\le E_0,
\label{poly}
\end{equation}
is stable for $n\ge 3/2$, i.e. for all values of $n$ such that 
$df/dE\le 0$.
Antonov's second theorem was generalized by Doremus, Feix \& 
Baumann (1971), Sygnet et al. (1984) and Kandrup \& Sygnet (1985) 
to include any spherical system with $df/dE<0$.
This result may be used to verify that many of the 
isotropic models that resemble real galaxies are linearly stable.

A number of attempts have been made to generalize Antonov's 
theorems to spherical systems with anisotropic velocities, 
$f=f(E,L^2)$.
One such generalization follows directly from Theorem II above: 
stability to radial perturbations is guaranteed if 
$\left({\partial\over\partial E^*}\right)^2 \int_{E^*}^0 f(E,L^2) 
{dE\over (E-E^*)}<0$ (\cite{dem88}).
Theorems of an even more general nature have been proposed, 
but their validity is not always clear.
Dor\'emus \& Feix (1973) and Gillon, Doremus \& Baumann (1976) 
analyzed a ``water bag'' model for spherical stellar systems, in 
which the distribution function is represented as a set of bags of 
incompressible fluid; a perturbation is interpreted as a change 
in the volumes of the bags rather than as a change in the numerical value of 
the distribution function.
They found stability to radial perturbations when 
$\partial f/\partial E<0$ and stability to all perturbations when 
$\partial f/\partial L<0$ as well.
The former result was confirmed by Kandrup \& Sygnet (1985) 
and Perez \& Aly (1996),
but the latter result would seem to definitely contradict numerical 
studies of the radial-orbit instability described below (\S 6.2);
in fact Perez \& Aly were able to find an error in the Gillon et 
al. proof.
A criterion proposed by Hjorth (1994) for stability of a certain 
class of anisotropic models is also inconsistent with numerical 
results (\cite{mez97}).

As an alternative to proving theorems on stability,
one can use physical intuition to search for particular, unstable models.
This approach has been extremely successful, leading to the 
identification of several types of unstable behavior 
in models of hot stellar systems.
Results on four such instabilities are reviewed below, roughly in 
the order in which their importance was first recognized.
The first is a pulsational instability of spherical systems, 
first reported by H\'enon (1973).
The second is the radial-orbit instability, postulated by Antonov 
(1973) and observed by Polyachenko (1981).
Third are bending instabilities, described as early as 1966 by 
Toomre in the context of disks, and observed in a number of 
$N$-body studies starting around 1990.
Finally, instabilities in slowly-rotating oblate models with
nearly circular orbits are discussed; these instabilities were first 
noticed in disk models and later seen in oblate spheroids.

\subsection{H\'enon's instability}

Antonov's proofs leave open the question of the stability of 
systems with anisotropic or non-monotonic distribution functions.
H\'enon (1973) carried out the first systematic search for 
instabilities in anisotropic spherical models.
He found that the polytropic models defined by equation (\ref{poly}) 
were stable to spherically symmetric modes, with the possible exception 
of the $n=1/2$ model, which appeared to oscillate at a level 
slightly above the noise. 
(The $n=1/2$ model is extreme in the sense that all stars have 
exactly the same energy.)
Barnes, Goodman \& Hut (1986) later established the stability of these 
models to non-spherical modes as well.
H\'enon went on to test the stability to radial perturbations of the 
``anisotropic polytropes'' defined by
\begin{equation}
f(E,L^2) = f_0L^{-2m}(E_0-E)^{n-3/2}.
\label{anpoly}
\end{equation}
Models generated from equation (\ref{anpoly}) have velocity 
ellipsoids with fixed axis ratios 
$\sigma_r^2/\sigma_t^2=(1-m)^{-1}$, where $\sigma_r$ and 
$\sigma_t$ are the 1D, radial and tangential components of the 
velocity dispersion tensor.
H\'enon found that the oscillatory instability that seemed to be 
present in the isotropic model with $n=1/2$ became stronger as 
the velocity ellipsoid was made more prolate.
He mapped out the region of instability in the $(m,n)$-plane and 
found that models with $n$ as large as $\sim 1$ could be unstable 
when the velocity ellipsoid was sufficiently elongated.

Barnes, Goodman \& Hut (1986) reanalyzed the stability of the 
anisotropic polytropes using a more sophisticated $N$-body code.
They confirmed H\'enon's results and noted that the instability 
boundary for radial modes could be approximated by $n-m=1/2$, 
suggesting instability even for $n=3/2$ when the orbits were 
fully radial.
Barnes et al. noted that the stability boundary coincided roughly with the 
condition that the distribution of radial velocity have two 
peaks, and suggested that the ``two-stream'' nature of the 
velocity distribution was a necessary condition for this instability.
They noted further that $\partial f/\partial E>0$ was a requirement
for a spherical model to have such a double-peaked 
distribution.
Neither H\'enon nor Barnes et al. discussed in detail the 
structure of their unstable models after they had reached a steady 
state, although the major effect of the instability seemed to be 
a radial spreading of stars.

The two conditions for instability established by these studies
-- $\partial f/\partial E>0$, and radially-elongated velocity 
ellipsoids -- imply an equilibrium density profile that diverges near the 
center, roughly as a power law.
Until recently, such a density dependence was considered unphysical, 
but it is now known that early-type galaxies always have power-law
density cusps (\cite{mef95}; \cite{geb96}).
Further work on H\'enon's instability would therefore be useful.

\subsection{The radial-orbit instability}

Antonov (1973) suggested that a spherical model constructed from
purely radial orbits would be unstable to clumping of particles around any
radius vector.
\footnote{Antonov's proof would seem to establish only the instability of 
the motion of a single particle and not necessarily the existence of a
growing mode.}
The instability was verified numerically by Polyachenko (1981), 
who followed the evolution of a 200-particle radial-orbit model into a bar.
Barnes (1985) rediscovered the instability while using an $N$-body
code to test the stability of H\'enon's generalized polytropes; 
he observed growing barlike distortions in radially anisotropic
models with a wide range 
of $m$'s and $n$'s.

The instability is similar to one first described by Lynden-Bell 
(1979) in the context of radially-hot disks.
In spherical systems with well-behaved central potentials, 
elongated orbits are nearly closed, making slightly less than two radial 
oscillations for each circulation in angle.
Such an orbit acts like a slowly precessing ellipse, or rod 
(\cite{pol89}); the precession rate tends to zero for low 
$L$.
In the presence of a barlike perturbation to the potential, 
an elongated orbit experiences a torque which causes it to precess more 
rapidly toward the bar; after passing through the bar, the orbit 
loses angular momentum and precesses more slowly.
The result is an enhancement of the density 
in the neighborhood of the bar.

The strength of the instability, and its effect on the shapes of 
initially spherical models, were explored in a number of 
$N$-body studies.
Merritt \& Aguilar (1985) examined the stability of models with 
Dehnen's density law, $\gamma=2$, and three different types of 
distribution function.
They noted that the behavior of the instability depended sensitively 
on the low-$L$ dependence of $f$.
In models where the phase-space density was divergent at low $L$
(as in H\'enon's anisotropic polytropes), 
the growth rate of the instability and the final ellipticity of 
the model were found to be gradually increasing functions of the 
anisotropy.
Even very mildly anisotropic models constructed from such 
distribution functions were found to be unstable.
Barnes, Goodman \& Hut (1986) reported similar 
behavior in their $N$-body study of H\'enon's polytropes.
Merritt \& Aguilar also investigated a family of models 
generated from a distribution function of the form $f=f(Q), 
Q=E+L^2/2r_a^2$, 
in which the phase-space density is constant on spheroids in velocity 
space and everywhere finite.
The latter models are characterized by a central region, $r\lap 
r_a$, in which the motion is nearly isotropic and an outer 
envelope where the motion becomes increasingly radial (\cite{osi79};
\cite{mer85}).
Merritt \& Aguilar found that the instability appeared suddenly 
in these models when $r_a$ fell below a critical value, 
roughly $0.3$ times the half-mass radius.
The neutrally-stable model from this family had a global anisotropy ratio 
$2T_r/T_t\approx 2.3$, 
with $T_r$ and $T_t$ the kinetic energies in radial and 
tangential motions respectively.

Much of the subsequent work on the radial orbit instability has 
focussed on models with $f=f(Q)$.
Meza \& Zamorano (1997) extended the Merritt \& Aguilar
study to include Dehnen models with a variety of cusp slopes $\gamma$.
They found a mild dependence of the critical anisotropy ratio $2T_r/T_t$
on $\gamma$, from $\sim 2.0$ when $\gamma=2$ (slightly less than 
the value found by Merritt \& Aguilar for the same family) to 
$\sim 2.6$ for $\gamma=0$.
May \& Binney (1986) used an adiabatic deformation technique to 
evaluate the stability of models with $f=f(Q)$ and the isochrone 
density law.
They estimated a critical value of $\sim 2.2$ for the anisotropy 
ratio.
Dejonghe \& Merritt (1988) found $(2T_r/T_t)_{crit}\approx 1.9$ 
in $f=f(Q)$ models with the Plummer density law.
Bertin \& Stiavelli (1989) used an $N$-body code to test the 
stability of models with distribution functions of the form 
$f\propto |E|^{3/2}\exp (aQ)$, designed to represent galaxies 
that form via radial collapse (\cite{stb85};
\cite{mtj89}).
They found $(2T_r/T_t)_{crit}\approx 1.9$.
Stiavelli \& Sparke (1991) derived a very similar criterion for 
instability in models with $f\propto |E|^{3/2}\exp(aE)/(1+cL^2)$.
Perez et al. (1996) noted the presence of the instability in 
$f=f(Q)$ models with the density profiles of isotropic polytropes.
However they did not look closely at the critical parameter 
values defining the onset of the instability.
Allen, Palmer \& Papaloizou (1990) also noted the existence of the instability 
in an $N$-body study of modified polytropes but did not attempt 
to establish the stability boundary.

The existence of slowly-growing modes in models with very small 
global anisotropies, reported by 
Merritt \& Aguilar (1985) and Barnes, Goodman \& Hut (1986),
calls into question the usefulness of diagnostics like
$2T_r/T_t$ for judging instability.
Palmer \& Papaloizou (1987) showed that instability to barlike 
modes is in fact guaranteed in models where $f$ is unbounded at 
small $L$, regardless of the degree of velocity anisotropy.
These results prompted an analysis of the stability of a 
model of the giant elliptical galaxy M87 constructed by Newton \& 
Binney (1984); their model is close to isotropic except near the 
center, where $f$ increases rapidly at small $L$.
The model was found to be mildly unstable to a barlike mode
(\cite{mer87b}).

An alternative to $N$-body techniques when evaluating the 
stability of equilibrium models is Kalnajs's (1972) matrix 
algorithm, which yields numerical expressions for the normal modes 
and their growth rates.
Kalnajs's algorithm is in principle superior to $N$-body 
techniques when searching for the exact stability boundary in a 
family of models, but in practice the eigenvalue equation 
can be difficult to solve with accuracy when the growth rate is small.
Polyachenko \& Shukhman (1984) pioneered this approach in the 
study of spherical systems, deriving the lowest-order modes for a 
family of nearly-homogeneous models derived by Osipkov.
They found instability for $2T_r/T_t\gap 1.6$.
Results for two additional families -- H\'enon's generalized 
polytropes, equation (\ref{anpoly}), and a family of models based 
on Plummer's density law -- were reported by Fridman \& 
Polyachenko (1984).
For the first family, a critical anisotropy ratio $\sim 1.4$ was 
quoted; however Palmer \& Papaloizou (1987) proved that all 
radially-anisotropic models from this family would be unstable.
For the second family, Fridman \& Polyachenko found 
$(2T_r/T_t)_{crit}\approx 1.5$;
\footnote{
In a reanalysis, Polyachenko (1985) found a critical 
ratio of slightly more than 2.}
Dejonghe \& Merritt (1988), in an $N$-body study, 
found instability in this family for $(2T_r/T_t)_{crit}\gap 
1.9$.

Some of these inconsistencies can be attributed to the 
difficulties of implementing Kalnajs's technique.
In a very convincing study, Saha (1991) derived the lowest-order 
unstable models for the family of anisotropic, $f=f(Q)$ 
isochrone models first analyzed by May \& Binney (1986).
Saha found instability for $2T_r/T_t>1.4$, substantially smaller 
than May \& Binney's estimate of 2.2; this result suggests that
spherical models may be unstable over a much wider range of 
anisotropies than the $N$-body studies indicate.
Saha (1992) went on to analyze a second family of isochrone models 
with a two-parameter distribution function; the first parameter 
$r_a$ defined the size of the isotropic core, while the second 
parameter $\beta_{\infty}$ fixed the degree of anisotropy at 
large radii.
Saha found a critical anisotropy ratio that ranged from 
$2T_r/T_t \approx 2.3$ in models whose envelopes contained purely 
radial orbits
to $\sim 1.8$ in models where the asymptotic anisotropy was 
$\sigma_r/\sigma_t\approx 1.6$.
Weinberg (1991) derived the unstable modes for $f=f(Q)$ models 
with the density profiles of isotropic Michie-King models; the 
latter constitute a one-parameter sequence characterized by the 
value of the dimensionless central potential $W_0$ (King 1966).
Weinberg demonstrated instability for low $r_a$ in models with 
several different values of $W_0$.
For $W_0=5$, he found a critical anisotropy radius of $r_a\approx 
1.2$ in units of the King core radius.
Bertin et al. (1994) evaluated the stability of the Stiavelli-Bertin 
(1985) models and found $(2T_r/T_t)_{crit}\approx 1.6$, 
substantially less than in the earlier $N$-body study of  
Bertin \& Stiavelli (1989).

These normal-mode calculations suggest that spherical models with 
$2T_r/T_t\gap 2.3$ are generally unstable, and that models with 
distribution functions that increase rapidly toward small $L$ can 
be unstable for much lower values of the mean anisotropy.
A model with $2T_r/T_t=2.3$ has an average ratio of radial to 
tangential velocity dispersions of $\sigma_r/\sigma_t\approx 
1.5$; thus, one might expect to never see more extreme 
anisotropies in spherical galaxies.
One of the few galaxies for which the anisotropy has been 
convincingly measured is M87 (\cite{vdm94}; \cite{meo97}),
where $\sigma_r/\sigma_t\approx 1.5$ over a wide range in 
radius.

The radial-orbit instability would be expected to persist in 
axisymmetric models with radially-elongated velocity ellipsoids.
Levison, Duncan \& Smith (1990) reported barlike modes in highly 
flattened ($c/a\approx 0.4$) oblate models when the radial 
velocity dispersion exceeded $\sim 0.7$ times the velocity 
disperision perpendicular to the meridional plane.
This result suggests that the radial-orbity instability might be 
very effective in flattened models; it would be useful to verify 
the result in an independent study.

Elliptical galaxies often contain central mass concentrations 
which could strongly influence the elongated orbits that 
drive the radial-orbit instability.
Palmer \& Papaloizou (1988) discussed the effect of a 
central point mass on the instability of models with distribution 
functions peaked near $L=0$.
They noted that a central point mass greatly increases the precession 
rate of low-$L$ orbits; at a given energy, there is a minimum 
precession rate corresponding to orbits with a certain nonzero 
$L$.
If this minimum precession rate is larger, for most orbits, than 
the rate at which a barlike mode would grow in the model without a 
central point mass, they argued that the instability would be 
inhibited.
Preliminary $N$-body simulations suggested a critical mass 
of the central object of roughly $0.1\%$ the mass of the model, 
similar to the masses inferred for the nuclear black holes in a 
number of galaxies.
Given the likely ubiquity of black holes in elliptical galaxies,
more work along the lines of Palmer \& Papaloizou's would be valuable.

Polyachenko (1981) noted that collapse from cold 
initial conditions tends to produce equilibrium systems with 
strong radial anistropies, and he pointed out that such systems
would be unstable to nonspherical modes.
He went on to verify (\cite{pol85}, 1992) the evolution of 
initially cold and spherical $N$-body clouds into bars.
A number of subsequent studies have shown that a wide range of 
initial conditions can produce collapses where the radial-orbit 
instability is active.
Merritt \& Aguilar (1985) found that intially spherical clouds 
with an $r^{-1}$ density profile evolved into triaxial bars when 
$(2T/|W|)_0$ was less than about 0.2, with $T$ and $W$ the total 
kinetic and potential energies of the initial configuration.
The final elongation was found to increase smoothly with 
decreasing initial ``temperature''; the coldest initial 
conditions produced models with a final axis ratio of $\sim 2.5:1$.
The instability was also found to reduce the central 
concentration of the relaxed models.
Min \& Choi (1989) also found a critical virial ratio of $\sim 
0.2$ in collapses starting from homogeneous initial conditions.
Aguilar \& Merritt (1990) extended their earlier work to 
flattened and rotating initial conditions.
They found that collapses with $(2T/|W|)_0\lap 0.1$ produced 
final shapes that were nearly uncorrelated with initial 
shapes; this characteristic final shape was approximately prolate 
with a $2:1$ axis ratio.
The minimum amount of rotation required to inhibit the bar 
instability was found to be $\lambda\approx 0.1$, with $\lambda$ 
the standard dimensionless spin parameter.
Cannizzo \& Hollister (1992) explored the dependence of the final 
shape of a collapsing cloud on its initial density profile.
They distributed particles according to $\rho\propto r^{-n}$ 
initially with $0<n<2.5$ and set the virial ratio to a very low 
value, $(2T/|W|)_0\approx 0.01$.
Their final configurations were approximately prolate, with axis 
ratios that varied from $\sim 1.8:1$ for $n=2.5$ to $\sim 1.3:1$ 
for $n=0$.
\footnote{Cannizzo \& Hollister adopted a relatively large 
value for the interparticle softening length in their simulations, 
and this fact may explain the less extreme elongations found by them 
than by other authors who adopted similar initial conditions.}
Theis \& Spurzem (1998) carried out a series of direct-summation 
$N$-body collapse simulations using large numbers of particles 
($N\approx 3\times 10^4$) and small softening lengths.
Particles were initially distributed according to Plummer's 
law, with $2T/W\approx 0.04$.
The final models were prolate/triaxial and very elongated, with 
$c/a\approx 0.4-0.5$.

As a number of authors have noted, the final elongations of these 
models are close to the maximum value permitted by the bending 
instability discussed below.
Polyachenko (1992) suggested that the distribution of elliptical galaxy
axis ratios might be determined by the competition between these 
two instabilities.
However recent determinations of the frequency function of 
elliptical galaxy intrinsic shapes (\cite{trm95}, 1996; 
\cite{ryd96})
find a peak near $c/a\approx 0.8$, too round to be convincingly 
explained in this way.

\subsection{Bending instabilities}

Toomre (1966) showed that a thin stellar sheet has a tendency to buckle when
the stars have large random velocities along the plane and are constrained to
remain in a single thin layer as the sheet bends.
\footnote{Toomre has pointed out that the names ``firehose'' and 
``hose-pipe'' are not really appropriate for this 
instability, which has more in common with the Kelvin-Helmholtz 
instability that occurs when two fluids slide past one another, 
or with beads sliding along an oscillating string (Parker 1958).} 
The constraining force comes from the vertical self-gravity of the sheet,
which is large if the sheet is thin.
Stars moving across a bend are forced to oscillate vertically as they pursue
their unperturbed horizontal motions, and the bend will grow if the
gravitational restoring forces from the perturbation are too weak to provide
the vertical acceleration required.
The thin-sheet dispersion relation is (Toomre 1966)
\begin{equation}
\omega^2 = 2\pi G\Sigma k - \sigma_x^2 k^2
\label{toom}
\end{equation}
where $\sigma_x$ is the velocity dispersion parallel to the sheet
and $\Sigma$ is its surface density.
The first term, which arises from the perturbed gravity, is stabilizing,
while the second term, due to the centrifugal force that the stars exert on
the sheet, is destabilizing.
Because the stabilizing force decreases less
rapidly with the wavelength of the bend than does the destabilizing
force ($\lambda^{-1}$ vs. $\lambda^{-2}$), short-wavelength
perturbations are the most unstable.
For sufficiently long wavelengths $\lambda>\lambda_J=\sigma_x^2/G\Sigma$, the
gravitational restoring force dominates and the sheet is stable.
Bending instabilities are precisely complementary, in this sense, to the
Jeans instability in the plane, which is stabilized at wavelengths
$\lambda<\lambda_J$.
Toomre's analysis, which was based on the moment equations, 
was shown by Mark (1971) and Kulsrud, Mark \& Caruso (1971) to 
be valid for slabs of arbitrary vertical structure whenever the 
wavelength of the perturbation greatly exceeds the thickness of 
the slab.

At wavelengths shorter than the actual vertical thickness of the layer,
Toomre argued that the bending would once again be stabilized.
Since the thickness of a slab scales as $\sigma_z^2/G\Sigma$, 
with $\sigma_z$ the vertical dispersion, Toomre noted that both 
long- and short-wavelength perturbations should be stabilized 
when $\sigma_z/\sigma_x$ was sufficiently large.
He estimated a critical value of $0.30$ for this ratio.
Polyachenko \& Shukhman (1977) found a very similar value for the 
critical anisotropy in a homogenous slab, for which the bending 
takes the form of surface corrugations.
Araki (1985) computed the exact linear normal modes of a 
finite-thickness slab with an anisotropic Gaussian velocity distribution.
He found that bending at all wavelengths was stabilized when the ratio of
vertical to horizontal velocity dispersions exceeded $\sim 0.293$, 
remarkably close to Toomre's earlier estimate.
At the instability threshold, the single, neutrally-unstable 
mode has a wavelength of about 1.2 in units of $\lambda_J$.

The elongation of a pressure-supported galaxy with the Toomre-Araki 
critical anisotropy is approximately $1:15$, 
which would seem to suggest that bending modes are
unlikely to be important in structuring elliptical galaxies.
But a number of subsequent studies revealed instability
in much rounder models.
Polyachenko \& Shukhman (1979) and Vandervoort (1991) 
calculated the exact linear normal modes for constant-density 
stellar spheroids, both oblate and prolate.
In models where the distribution function favored 
radial motions, both studies found critical axis ratios of $\sim 
1:3$, not $\sim1:15$.
Merritt \& Hernquist (1991) carried out an $N$-body study of 
inhomogeneous prolate models; they also found a critical axis 
ratio of $0.3-0.4$.

\begin{figure}
\plotfiddle{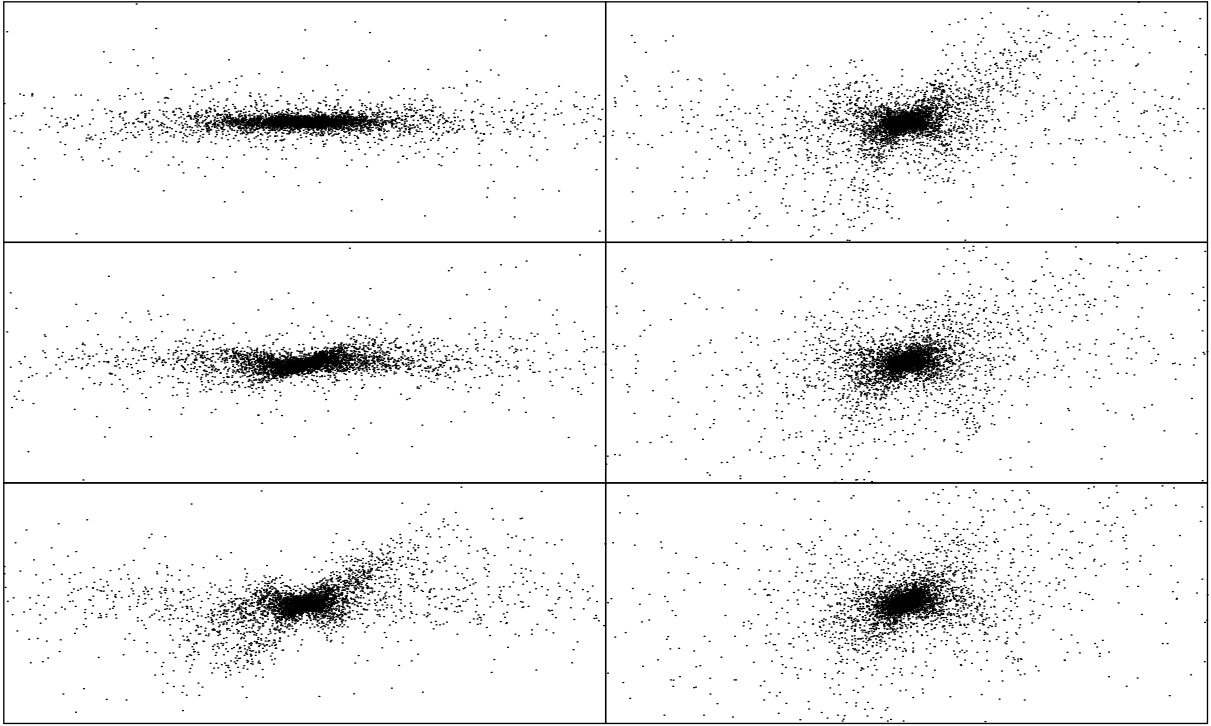}{7.in}{0.}{140}{140}{-360}{-600}
\caption{
Bending instability in a prolate, ``shell-orbit'' model
(Merritt \& Hernquist 1991).
The initial model has axis ratio $c/a=0.1$; frames are separated
by $\sim 4$ periods of the long-axis orbit.
The dominant mode is an antisymmetric bend that induces a
long-lived peanut shape.
Prolate models with elongations more extreme than $\sim 3:1$ are
found to be unstable to bending modes.
 }
\end{figure}

The discrepancy was explained in the following way (\cite{mes94}).
In finite or inhomogenous models, the gravitational restoring 
force from the bend -- represented by the first,
stabilizing term in equation (\ref{toom}) -- is substantially weaker than 
in infinite sheets and slabs.
In fact, bending modes in thin systems with realistic density
profiles are often not stabilized by gravity even at the longest 
wavelengths.
However bending modes can still be stabilized in such systems, 
for the following reason.
Stars in a finite-thickness system oscillate vertically
with an unperturbed frequency $\kappa_z$; like any oscillator, the phase of
a star's response to the imposed bending depends entirely on whether the
forcing frequency $kv_x$ is greater than or less than its natural frequency.
If $kv_x>\kappa_z$ for most stars, the overall density response to the
perturbation will produce a potential opposite to that imposed by the bend
and the disturbance will be damped.
Merritt \& Sellwood showed that this mechanism was responsible 
for the short-wavelength stabilization found by Toomre (1966) and 
Araki (1985) in the infinite slab.
In a finite or inhomogeneous stellar spheroid, where the 
gravitational restoring force from the bend is much smaller than 
in an infinite slab, almost all of the stabilization is due to
this out-of-phase damping.
Since a typical star in such a system feels a 
vertical forcing frequency from a long-wavelength bend that is 
roughly twice the frequency $\omega_x$ of its orbital motion along the 
long axis, stability to global modes requires that the forcing frequency 
be less than $\omega_z$, the frequency of orbital motion along the short axis.
The resulting condition $2\omega_x>\omega_z$ predicts stability for a
pressure-supported, homogeneous prolate spheroid rounder than $\sim 1:2.94$,
in excellent agreement with the normal-mode calculations of 
and Fridman \& Polyachenko (1984) and Vandervoort (1991).

Fridman \& Polychenko (1984) went on to suggest that the absence of
elliptical galaxies flatter than $\sim {\rm E}6$ 
was due to bending instabilities.
The work summarized above is consistent with their hypothesis, at least
to the extent that elliptical galaxies are pressure-supported systems 
in which the stellar motions are largely radial.
Merritt \& Sellwood (1994) pointed out that oblate galaxies with 
azimuthally biased motions can remain stable even when
somewhat flatter; indeed, Hunter \& Toomre (1969) showed that even 
perfectly thin disks are stable to bending modes when all the stars orbit 
in one direction, since the bending wave is free to travel with 
the stars.
Sellwood \& Merritt (1994) followed the nonlinear evolution of 
oblate models with various initial velocity distributions and 
found that models where a modest fraction of the pressure was 
tangentially directed could reach equilibrium when still rather 
thin.
They suggested that such models were reasonable representations of 
S0 galaxies like NGC 4550 that appear to contain two, nearly equal 
counter-rotating streams of stars (\cite{rgk92}; \cite{mek94}).

Bending instabilities may also be relevant to the formation 
of galactic bulges.
Rapidly-rotating disks are unstable to the formation of bars; 
early indications that these bars might thicken with time 
(\cite{cos81}; \cite{com90}) were spectacularly confirmed 
by Raha et al. (1991) who carried out three-dimensional 
simulations revealing a violent buckling instability.
Raha et al. argued that they were witnessing a coherent bending 
mode driven by the inequality between the stellar velocity 
dispersions along and perpendicular to the bar.
The end result in their simulations was a thick, peanut-shaped 
sub-system very 
similar in appearance to the ``boxy bulges'' seen in some disk 
galaxies (\cite{sha87}).

An alternative explanation for the thickening of 
$N$-body bars was proposed by Combes et al. (1990) and Pfenniger \& 
Friedli (1991), who suggested that the bending was due to the 
vertical instability of the $x_1$, bar-supporting orbits.
This instability first occurs at a $2:1$ bifurcation, which 
generates a family of stable banana orbits 
perpendicular to the bar; as discussed above (\S 4.2.2), 
the vertical banana orbits appear to be heavily populated in 3D 
$N$-body bars.
However the evidence is strongly against the hypothesis that the 
$2:1$ bifurcation is responsible for the vertical thickening.
The buckling of $N$-body bars is observed to be extremely coherent, 
particularly in initially thin bars, which bend almost into a V-shape 
(Raha et al. 1991).
Friedli \& Pfenniger (1990) found that the vertical thickening could 
be essentially eliminated by imposing reflection symmetry about 
the disk plane, which is the expected behavior for a collective mode 
but which has no natural explanation in terms of orbital 
instability.
Merritt \& Hernquist's (1991) observation of bending modes in 
integrable prolate models is also inconsistent 
with the Combes et al. hypothesis, since these models do not 
support banana orbits.
The strong dependence of the bending rate on model thickness, 
seen in a number of studies, is also most naturally explained in 
terms of a collective instability.
Merrifield (1996) noted that the condition $2\omega_x<\omega_z$ 
derived by Merritt \& Sellwood (1994) for instability to 
large-scale bending modes also implies the existence of a 
$2:1$ vertical resonance, a coincidence that may explain the 
early confusion about the origin of the bending.

\subsection{Instabilities in radially-cold models}

A disk constructed purely from circular orbits is locally 
Jeans-unstable and clumps into rings (Toomre 1964); this behavior 
is independent of the net rotation velocity, that is, the 
fraction of stars that rotate clockwise versus counterclockwise.
These ringlike modes can be stabilized either by making the disk 
hot or by thickening the disk into a spheroid, for which the 
potential disturbance caused by a given perturbation in surface 
density is weaker than in a disk (Vandervoort 1970).
Bishop (1988) reported the ring instability in a
radially-cold oblate model with an axis ratio of $1:4$.
Merritt \& Stiavelli (1990) found a critical axis ratio of $\sim 
0.4$ for these unstable modes in the same family of 
thin-orbit models.
De Zeeuw \& Schwarzschild (1991) analyzed the same models yet a third 
time, using an adiabatic deformation technique, and 
estimated a critical axis ratio of $\sim 0.33$.
The effect of radial pressure on the ring instability was 
addressed by Sellwood \& Valluri (1997), who considered a family 
of two-integral oblate models.
The radial pressure in a two-integral model increases with 
decreasing elongation, due to the equality of radial and vertical 
velocity dispersions; this implies a rapid stabilization of 
ringlike modes as the model is made rounder.
Sellwood \& Valluri found stability to axisymmetric modes at an 
axis ratio of $\sim 1:4$ in their models.
Ringlike modes in axisymmetric models were also reported by 
Sellwood \& Merritt (1994) and Robijn (1995).

\begin{figure}
\plotfiddle{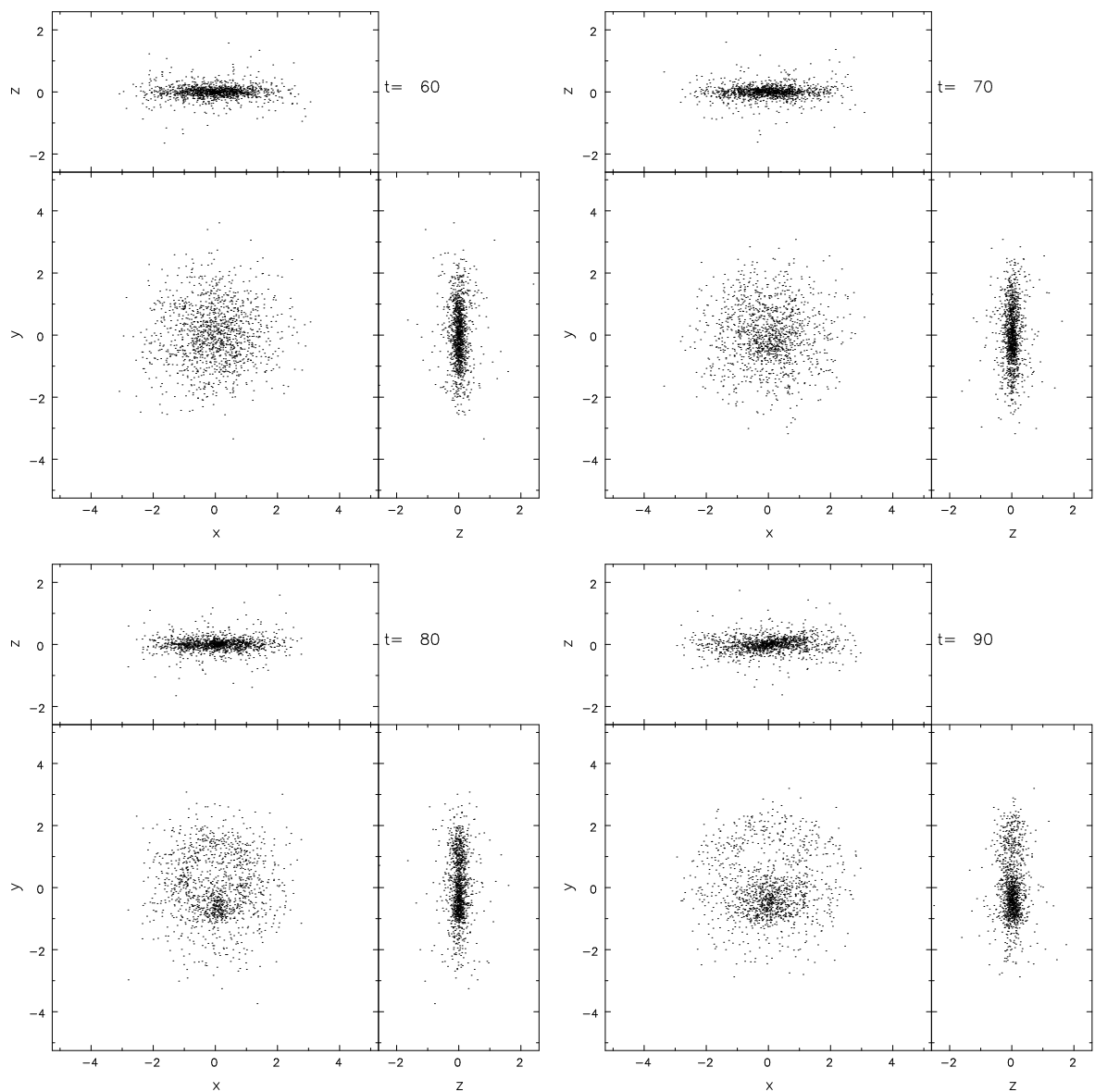}{7.in}{-90.}{170}{170}{-1000}{550}
\caption{
Lopsided instability in a nonrotating oblate model from the
Kuzmin-Kutuzov (1962) family.
The initial model has $f=f(E,L_z^2)$ and an axis ratio
$c/a\approx 0.1$; hence the stellar motions are nearly circular,
with very little radial pressure.
Modes like these persist in two-integral oblate models as round
as $c/a\approx 0.2$, and in even rounder models when the velocity
ellipsoid is biassed toward circular motions.
 }
\end{figure}

The prevalence of $m=1$ or lop-sided modes in slowly-rotating disks 
suggests that lop-sided modes would be the 
dominant ones in axisymmetric models whose rotation was too 
slow to encourage the formation of a bar.
Nonrotating oblate models constructed from thin orbits ($\S 3.2$) 
are in fact unstable to lopsided modes regardless of their axis 
ratio (\cite{mes90}).
Levison, Duncan \& Smith (1990) reported lopsided modes in 
nonrotating, axisymmetric oblate models with axis ratio of $\sim 
0.4$; the instability vanished when the radial velocity 
dispersion exceeded 
$\sim 0.4$ times the tangential velocity dispersion.
Sellwood \& Merritt (1994) also reported lopsided modes in their 
flattened, counter-rotating models and found that fairly large 
radial pressures were required for stability.
In their study of two-integral oblate models, Sellwood \& Valluri 
(1997) found instability to lopsided modes for $c/a\lap 0.2$; in 
other words, the radial pressure of two-integral models 
rapidly stabilizes the modes as the models are made rounder.
Sellwood \& Valluri found that the lopsided modes became less 
important as the net rotation of their models was increased, but 
they persisted even in models with net angular momentum 90\% that 
of a maximally rotating model.
Robijn (1995) carried out a normal-mode analysis of 
Kuz'min-Kutuzov models with little radial pressure and 
confirmed the Merritt \& Stiavelli result that such models 
were unstable to lopsided modes regardless of axis ratio.
He found that the growth rate of lopsided modes was more 
strongly affected by increasing the radial velocity dispersion 
than by adding net rotation.

The persistence of unstable modes even in nearly spherical, 
radially-cold models suggests that precisely spherical galaxies 
composed of circular orbits might be unstable.
Bisnovatyi-Kogan, Zel'dovich \& Fridman (1968) showed that the 
circular-orbit sphere is always stable to radial modes, i.e. to 
clumping into spherical shells.
However Fridman \& Polyachenko (1984, Vol. 1, p. 179) showed that 
a circular-orbit sphere with a radially-increasing density, 
$d\rho/dr > 0$, is generally unstable to 
non-spherically-symmetric modes.
Barnes, Goodman \& Hut (1986) explored such instabilities in 
their $N$-body study of H\'enon's generalized polytropes.
For $m<0$, equation (\ref{anpoly}) implies a preponderance of 
nearly-circular orbits and a density profile that peaks at 
nonzero $r$.
Barnes et al. found that all models with $m\lap -1/2$ exhibited 
quadrupole oscillations with growing amplitude, reaching 
stability only after a substantial rearrangement of matter.

\subsection{The ``tumbling bar instability''}

Papaloizou, Palmer \& Allen (1991) and Allen, Papaloizou \& Palmer 
(1992) reported barlike instabilities in nearly spherical models 
with very small amounts of rotation.
Additional arguments in favor of this ``tumbling bar 
instability'' were presented by Palmer (1994a,b).
However, the instability appears not to exist.
Sellwood \& Valluri (1997) tested the 
stability of the Allen et al. models, using precisely the same 
files of initial particle coordinates but a different $N$-body code.
No traces were found of growing modes, nor did Sellwood \& 
Valluri find any unstable barlike distortions in another family 
of slowly-rotating, axisymmetric models.
The evolution seen in the Allen et al. study was traced
to an improper treatment of variable time steps in their $N$-body 
code.

\subsection{Instabilities of triaxial models}

The instabilities discussed above in spherical and axisymmetric
models would be expected to
persist in non-axisymmetric models with appropriate orbital 
populations.
Smith \& Miller (1982) were the first to search for 
dynamical instabilities in triaxial models, using 
an $N$-body code to evolve a realization of 
Schwarzschild'd (1979) model.
Although they witnessed significant evolution in the shape of the 
model -- the peanut-shaped isophotes of the initial configuration 
became much more nearly elliptical -- Smith \& Miller concluded 
that there was no evidence for exponentially-growing modes.
De Zeeuw \& Schwarzschild (1989) used an adiabatic 
deformation technique to look for unstable barlike modes in 
Statler's (1987) set of triaxial models based on the Perfect 
Ellipsoid.
They confirmed the existence of a barlike instability for triaxial 
models in which box orbits were heavily populated.
The instability appeared to vanish in triaxial models that were 
sufficiently elongated.

\bigskip

The preparation of this review was supported by NSF grants 
AST 93-18617 and AST 96-17088
and by NASA grant NAG 5-2803.
I am very much indebted to W. Dehnen, P. Saha, J. Sellwood, M. Valluri and 
R. van der Marel who read substantial portions of the manuscript 
and made many suggestions for improvement.
J. Perez and V. Polyachenko also made helpful comments.
My understanding of the issues discussed here has benefitted 
enormously from discussions with A. Bahri, H. Kandrup, J. Binney, 
G. Contopoulos, O. Gerhard, J. Laskar, J. Meiss, A. Moser, 
Y. Papaphilippou, and Y. Sinai.

\end{document}